\documentclass[pdflatex,sn-mathphys-num]{sn-jnl}
\usepackage[utf8]{inputenc}
\usepackage{amsmath}
\usepackage{amscd}
\usepackage{amssymb}
\usepackage{amsfonts}
\usepackage{graphicx}
\usepackage{mathsfbf}
\usepackage{isomath}
\usepackage{xcolor}
\usepackage{tikz-cd}
\usepackage{subcaption}

\usepackage{multirow}%
\usepackage{amsthm}%
\usepackage{mathrsfs}%
\usepackage[title]{appendix}%
\usepackage{textcomp}%
\usepackage{manyfoot}%
\usepackage{booktabs}%
\usepackage{algorithm}%
\usepackage{algorithmicx}%
\usepackage{algpseudocode}%
\usepackage{listings}%
\usepackage{nicematrix}
\usepackage{makecell}
\usepackage{tabularray}

\usepackage{csquotes}

\newcommand\defeq{\mathrel{\stackrel{\makebox[0pt]{\mbox{\normalfont\tiny def}}}{=}}}

\numberwithin{equation}{section}

\begin{document}

\title[Harmonic fields and the mechanical response of a cellular monolayer to ablation]
{Harmonic fields and the mechanical response of a cellular monolayer to ablation}

\author*[1]{\fnm{Oliver E.} \sur{Jensen}}\email{oliver.jensen@manchester.ac.uk}

\author[1]{\fnm{Christopher K.} \sur{Revell}}\email{christopher.revell@manchester.ac.uk}

\affil*[1]{\orgdiv{Department of Mathematics}, \orgname{University of Manchester}, \orgaddress{\street{Oxford Road}, \city{Manchester}, \postcode{M13 9PL}, \country{UK}}}

\abstract{Multicellular tissues, such as the epithelium coating a developing embryo, often combine complex tissue shapes with heterogeneity in the spatial arrangement of individual cells.  Discrete approximations, such as the cell vertex model, can accommodate these geometric features, but techniques for analysis of such models are underdeveloped. Here, we express differential operators defined on a network representing a monolayer of confluent cells in a framework inspired by discrete exterior calculus, considering scalar fields defined over cell vertices and centres and vector fields defined over cell edges.  We achieve this by defining Hodge stars, wedge products and musical isomorphisms that are appropriate for a disordered monolayer for which cell edges and links between cell centres are not orthogonal, as is generic for epithelia.  We use this framework to evaluate the harmonic vector field arising in an ablated planar monolayer, demonstrating an approximate 1/\textit{r} scaling of the upper bound of the field's amplitude, where \textit{r} is the distance from the ablation.  Using a vertex model that incorporates osmotic effects, we then calculate the mechanical response of a monolayer in a jammed state to ablation.  Perturbation displacements exhibit long-range coherence, monopolar and quadrupolar features, and an approximate 1/\textit{r} near-hole upper-bound scaling, implicating the harmonic field.  The upper bounds on perturbation stress amplitudes scale approximately like 1/\textit{r}$^2$, a feature relevant to long-range mechanical signalling.
}


\maketitle


\section{Introduction}\label{sec1}

The relationship between geometric structure and biomechanical function is of central interest in the study of multicellular tissues.  For example, the epithelium that lines internal organs or coats embryos is formed from confluent cells with approximately polygonal apical faces.  This tight packing has an important barrier function but also regulates the mechanical environment of individual cells, influencing their response to mechanical cues \cite{heisenberg2013, debelly2022}.  An epithelium can undergo a phase change from a solid to a fluidized state, whereby a subtle change of the cells' material properties leads to a dramatic transition from a jammed to an unjammed configuration \cite{atia2018, mao2024}, promoting cell mobility.  Topological defects in the arrangement of cells may organise some aspects of morphogenesis \cite{hoffmann2022, vafa2022} and defect movement (via cell neighbour exchanges) is intrinsic to the plasticity of epithelial tissues.  These factors motivate the development of multiscale modelling approaches that can relate microstructure to tissue-level phenomena. 

The vertex model provides a powerful and popular computational framework with which to simulate cell mechanics at the tissue scale \cite{weliky1990, farhadifar2007, fletcher2014}.  An epithelium coating a surface is represented geometrically through the location of the vertices of its polygonal cells. The vertex model describes the dynamic evolution of such a monolayer as a flow (of vertices over a manifold) down a gradient of mechanical energy, interspersed with topological changes of the cell network (via neighbour exchange, division, extrusion or intercalation).  To connect this individual-based model to more conventional approaches, formal homogenization techniques can be used to derive continuum-level descriptions, for example when the cellular microstructure has a regular periodic organisation \cite{murisic2015}.  In general, however, upscaling techniques rely on one or more \textit{ad hoc} assumptions \cite{ishihara2017, fielding2022, hernandez2022} that may only partially capture important microstructural features.  Discrete calculus offers an alternative route to bridge the gap to the continuum level, by formulating descriptions of mechanical behaviour in a language that mirrors continuum descriptions while retaining complete microstructural information \cite{jensen2020}.  The spectral properties of continuum differential operators, which underlie many solution methods at the macroscale, are then replaced by spectral properties of discrete operators \cite{jensen2022, cowley2024}.

Discrete differential operators can be defined over polygonal meshes using the principles of mimetic finite differences \cite{alexa2011, lipnikov2014}.  Exploiting this approach, we derived operators over a  primal planar network of cells, and a dual network of triangles connecting adjacent cell centres \cite{jensen2022}, such that the operators act on scalar fields defined on cell vertices and cell centres, and on vector fields defined on cell edges and links between cell centres.  An alternative set of operators emerging naturally (via cell area changes) in the vertex model \cite{cowley2024} are appropriate for scalar fields defined on cell centres and vector fields defined on cell vertices, complementing operators defined in \cite{degoes2020} that are appropriate for scalar fields defined on vertices and vector fields on cells.  In each case, one can identify Laplacian operators (expressed as matrices) that are discretizations of the continuum $\nabla^2$ operator over the network provided by the cells themselves, the cotan Laplacian being one well-known example \cite{alexa2011}.  In its standard formulation, in which cell mechanical energy includes a contribution from cell perimeters, the vertex model also incorporates more exotic Laplacian operators, which do not appear to have a direct relationship with $\nabla^2$, that regulate the evolution of a cell monolayer \cite{cowley2024}.  

Below, we express the geometric operators identified in \cite{jensen2022} in a framework inspired by discrete exterior calculus (DEC).  As well as strengthening the theoretical foundations of existing results, this allows the development of a wider repertoire of geometric tools with which to analyse discrete mechanical models.  We identify exterior derivatives, sharp and flat operators, and wedge products, with which standard operations of vector calculus can be expressed \cite{grady2010, perot2014, crane2018, wang2023}.  This has been undertaken previously for cellular networks with suitable symmetries \cite{desbrun2005}, and related methods have been exploited to address a range of problems in mechanics \cite{yavari2008, jensen2020, srinivasa2021, boom2022}. Delaunay triangulation and Voronoi tessellation are popular geometric models that together ensure orthogonality between cell edges and links between cell centres.  While this can  be a useful approximation in many circumstances, this symmetry can be violated in epithelia \cite{jensen2020}.  One feature that distinguishes our task from existing studies is the requirement to avoid imposing edge--link orthogonality.  A price to be paid is an increase in the number of distinct operators \cite{jensen2022}.  

Our prior study \cite{jensen2022} addressed simply-connected monolayers, and exploited Helmholtz--Hodge decomposition to recover the scalar stress potentials corresponding to a field of equilibrium forces acting at vertices, thereby revealing so-called couple stresses acting in the neighbourhood of cell vertices.  A common (albeit invasive) experimental approach for stress inference in epithelia is to measure the response to an ablation (or wounding) of a small region of a monolayer \cite{rauzi2010, liang2016, kong2019, gomez2020, babu2024, villeneuve2025}.  The self-healing capacity of an epithlieum after injury is of major biological significance, involving biochemical signalling (\hbox{e.g.} via calcium \cite{oconnor2021} and chemoattractants that drive an inflammatory response \cite{weavers2016}), mechanical signalling (mediated by mechanosensors such as PIEZO1 and YAP/TAZ \cite{pena2024}) and inducing a mechanical response (including `purse-string' formation around a hole, fluidization in surrounding cells \cite{tetley2019} and directed cell migration \cite{lim2024}).  YAP/TAZ is also implicated in regulation of cell volume and of cell tension via levels of apical myosin \cite{perez2018}.  From a mathematical perspective, introducing a hole in a domain is significant because, as we shall demonstrate, the change in topology creates a so-called harmonic field (lying in the kernel of a Laplace--de Rahm operator), which captures in geometric terms part of the response to formation of the hole; we recall the continuous harmonic solution for a punctured linearly-elastic disc in Appendix~\ref{app:disc}.  We evaulate discrete harmonic fields here, and use them to interpret the mechanical impact of ablation, highlighting the remarkably coherent multipolar features of displacement fields and algebraic scaling properties of stress and displacement fields.  With mobile chemical factors in mind, we also show how the vertex model can be adapted to incorporate osmotic effects, allowing biochemical processes to influence effective cell mechanical properties, \hbox{e.g.} through macromolecular crowding \cite{zhou2009, urbanska2024}.  However we do not embark on simulations of the wider wound-healing response, instead referring the reader to studies such as \cite{lee2011, tetley2019, mosaffa2020, bai2023, babu2024, almada2025}.

This study straddles some traditionally distant disciplines, which can lead to confusion over terminology and potentially unfamiliar notation.  The term `vector' will be reserved for `traditional' vectors in $\mathbb{R}^2$ or $\mathbb{R}^3$ having a physically interpretable length and orientation.  The summation convention is avoided, and it will be convenient to express some linear operators explicitly in terms of the bases over which they act rather than as matrices.  We will also simplify terminology and notation introduced in \cite{jensen2022} that was inspired in part by conventions established in mimetic finite differences.  In particular, we distinguish the primary two-dimensional (2D) differential operators grad and curl, which form an exact sequence ($\mathrm{curl}\,\circ\mathrm{grad}=0$), from their respective adjoints (under suitable inner products) $-\mathrm{div}$ and rot, satisfying $-\mathrm{div}\,\circ\mathrm{rot}=0$.  Discrete fields defined over cells and vertices (so-called cochains, analogues of differential forms) will typically have two scalar components, labelled by $\parallel$ and $\perp$ (denoting an association with projections of a vector field onto directions parallel or perpendicular to edges or links).  Because edges and links need not be orthogonal, we will discuss the rotated operators cograd, cocurl, corot and codiv.  We will show how the rotated operators on the primal cell network resemble, but are generally distinct from, the unrotated operators on the dual (triangulated) network.  In \cite{jensen2022}, fields labelled with $\parallel$ and $\perp$ were treated separately; here they are handled in a unified way as elements of  2-component covector fields.  

We will consider a model of a 
planar epithelium defined over a flat 2D manifold $\mathcal{M}$ embedded in $\mathbb{R}^3$.  
Cells are defined in terms of vertices, edges and faces lying in $\mathcal{M}$.  In the language of algebraic topology, such objects are respectively 0-chains, 1-chains and 2-chains, and functions defined over them are cochains.  While it is common to define an $m$-cochain over an $m$-chain, here we retain the flexibility to define $n$-cochain-valued $m$-cochains, where $n$ and $m$ may differ.  As suggested above, we focus in particular on 1-cochain-valued $m$-cochains, represented by two scalar components (labelled with $\parallel$ and $\perp$) defined over $m$-chains for $m=0,1,2$.  Accordingly, the Hodge stars and wedge products that we deploy differ from (but complement) those proposed by other authors (\hbox{e.g.} \cite{desbrun2005}).

The first aim of the present work is therefore to recast operators defined in \cite{jensen2022} in the language of DEC, accommodating the requirement for edges and links not to be orthogonal.  Thus in Sec.~\ref{sec:prdr} we define $\mathrm{d}$, $\wedge$, $\star$, $\sharp$, and $\flat$ and the spaces over which they act.  This allows us in Sec.~\ref{sec:do} to write gradients as $(\mathrm{d}\phi)^\sharp$, curls as $(\star\mathrm{d}\mathsfbf{b}^\flat)^\sharp$, rots as $(\star\mathrm{d}\mathsf{f}^\flat)^\sharp$ and divergences as $\star\, \mathrm{d}\star \mathsfbf{b}^\flat$, for suitable discrete fields (cochains) $\phi$, $\mathsfbf{b}$, and $\mathsf{f}$.  This treatment allows construction of the associated Laplacian operators (\ref{sec:lap}), in particular Laplace--de Rahm operators defined over edges and links of the monolayer. Our second aim is to exploit Helmholtz--Hodge decomposition and a bespoke computational tool \cite{Revell_DiscreteCalculus_jl} to investigate networks containing one or more holes (Sec.~\ref{sec:hh}).  We show in Sec.~\ref{sec:vm} how discrete differential operators facilitate the inclusion of osmotic effects in the vertex model, and then apply Helmholtz--Hodge decomposition to the rotated force potential of equilibrium monolayers \cite{jensen2020} to compute the associated stress potentials (Sec.~\ref{sec:abl}).  While this has been pursued previously for simply-connected monolayers \cite{jensen2022}, here we calculate the discrete harmonic fields of ablated monolayers (Sec.~\ref{sec:hm}) and use these to evaluate stress potentials (Sec.~\ref{sec:ablation}) interpret stress and displacement fields (Sec.~\ref{sec:abph}). Readers interested in the more physical aspects of ablation will find relevant results in Sec.~\ref{sec:abph}.  

\section{Model and methods}
\label{sec:model}

This section develops a DEC-inspired framework suitable for cellular monolayers before returning to the vertex model in Sec.~\ref{sec:vm}.  We begin in Sec.~\ref{sec:network} by establishing the basic geometric and topological features of the framework on which we will construct differential operators.  

\subsection{Network properties}
\label{sec:network}

\begin{figure}
\begin{center}
\includegraphics[width=0.95\textwidth]{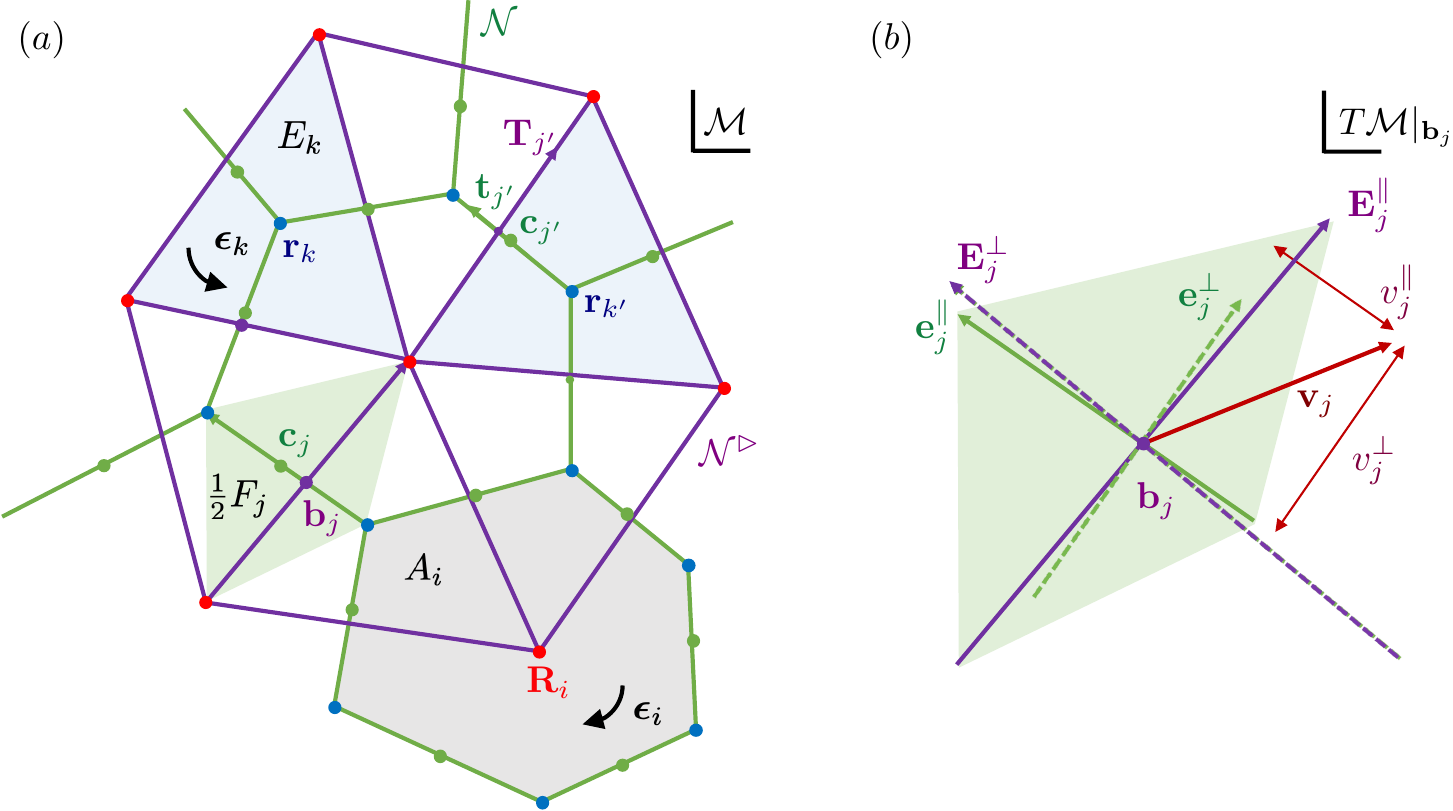}
\end{center}
\caption{(a) Schematic diagram illustrating the primal network $\mathcal{N}$ (green cell edges $\mathbf{t}_{j'}$) and the dual network $\mathcal{N}^\rhd$ (purple links $\mathbf{T}_{j'}$).  Cell vertices ($\mathbf{r}_k$ and $\mathbf{r}_{k'}$, blue dots) are associated with triangle areas ($E_k$, shaded blue).  Edge centroids ($\mathbf{c}_j$ and $\mathbf{c}_{j'}$, green dots), and edge-link intersections ($\mathbf{b}_j$, purple dots) are associated with quadrilateral areas ($\tfrac{1}{2}F_j$, shaded green).  Cell centres ($\mathbf{R}_i$, red dots) are associated with cell areas ($A_i$, shaded grey).  Cell orientation $\boldsymbol{\epsilon}_i$ and the opposite triangle orientation $\boldsymbol{\epsilon}_k$ are indicated. (b) The tangent plane $T\mathcal{M}\vert_{\mathbf{b}_j}$, showing basis vectors $\mathbf{e}_j^\parallel\equiv \mathbf{t}_j$, $\mathbf{E}_j^\parallel\equiv \mathbf{T}_j$, rotated vectors $\mathbf{e}_j^\perp$, $\mathbf{E}_j^\perp$ and the projection $\{v_j^\parallel,v_j^\perp\}^\top$ of a vector $\mathbf{v}_j$ onto $\mathbf{e}_j^\parallel$ and $\mathbf{e}_j^\perp$.  
}
\label{fig:Figure1schematic}
\end{figure}

We represent a cell monolayer as a set of confluent polygons.  We use $i=1,\dots,N_c$ to label cells or cell centres, $j=1,\dots, N_e$ to label cell edges and links between cell centres and $k=1,\dots,N_v$ to label cell vertices or triangles spanned by cell centres.  The primal network $\mathcal{N}$ is a polygonal tiling (a simplicial complex) of cells; the dual network $\mathcal{N}^\rhd$ is the triangulation connecting adjacent cell centres (Fig.~\ref{fig:Figure1schematic}a).  We consider either a simply-connected network, for which $N_v -N_e+N_c =1$ (viewing the monolayer as a topological disk), or allow for $n_h$ internal holes, in which case $N_v-N_e+N_c=1-n_h$.   
We define $\mathcal{V}$, $\mathcal{E}$ and $\mathcal{F}$ to be the vector spaces containing $0$-chains (vertices), $1$-chains (edges) and $2$-chains (cell faces) of the primal network; these are spanned respectively by bases $\mathsf{q}_k$, $\mathsf{q}_j$ and $\mathsf{q}_i$, for $k=1,\dots, N_v$, $j=1,\dots,N_e$ and $i=1,\dots, N_c$.   The dual network is built from vector spaces $\mathcal{C}$ (cell centres), $\mathcal{L}$ (links) and $\mathcal{T}$ (triangles), spanned respectively by $\mathsf{q}_i$, $\mathsf{q}_j$ and $\mathsf{q}_k$; $\mathcal{V}$, $\mathcal{E}$ and $\mathcal{F}$ are isomorphic respectively to $\mathcal{T}$, $\mathcal{L}$ and $\mathcal{C}$.

Orientations are assigned to all elements of each network and are encoded in signed incidence matrices $A_{jk}$, $B_{ij}$ \cite{grady2010} mapping between bases $\mathsf{q}_k$, $\mathsf{q}_j$ and $\mathsf{q}_i$.  Following \cite{desbrun2005, jensen2020}, we ensure that orientations assigned to the dual network are consistent with those assigned (arbitrarily) to the primal network.
We choose $\boldsymbol{\epsilon}_i$ (the $2\times 2$ matrix describing a $\pi/2$ rotation) to represent clockwise orientations of all cells and $\boldsymbol{\epsilon}_k=-\boldsymbol{\epsilon}_i$ to represent anticlockwise orientations of all triangles.  The topology of both networks is then fully specified by matrix operators $\mathsf{A}=\sum_{j,k}A_{jk}\mathsf{q}_j\otimes\mathsf{q}_k$ and $\mathsf{B}=\sum_{i,j}B_{ij}\mathsf{q}_i\otimes\mathsf{q}_j$, satisfying \cite{grady2010}
\begin{equation}
 \mathsf{B}\mathsf{A}=\mathsf{0}.
 \label{eq:ba}
\end{equation}
This fundamental relationship arises because $\mathsf{A}^\top$ and $\mathsf{B}^\top$ are boundary operators on $\mathcal{N}$ and the boundary of any set of cells has no boundary (for example, there are no vertices connected to a single edge), so that $\mathsf{A}^\top \mathsf{B}^\top=\mathsf{0}$.  $C_{ik}=\tfrac{1}{2}\sum_j \vert B_{ij}\vert \,\vert A_{jk} \vert $ defines the face-vertex adjacency matrix $\mathsf{C}$ and $Z_i=\sum_k C_{ik}$ gives the number of vertices per cell.

\begin{figure}
    \centering
    \includegraphics[width=\textwidth]{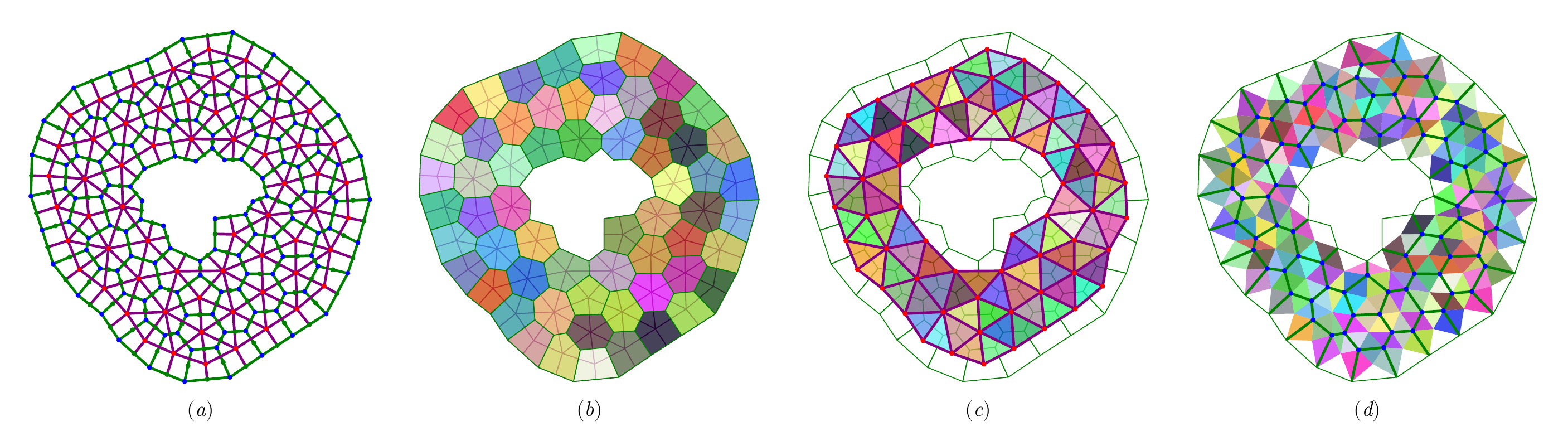}
    \caption{An ablated monolayer showing network construction and representations of geometric quantities.  (a) Primal network $\mathcal{N}$ of cell edges (green lines) and dual network $\mathcal{N}^\rhd$ of cell links (purple lines), as shown in detail in Fig.~\ref{fig:Figure1schematic}(a).  Links are bounded by cell centres  (red dots) or, at the monolayer periphery, edge centroids (green dots).
    (b) Randomly coloured polygons show cell areas $A_i$, overlaid on $\mathcal{N}$ and $\mathcal{N}^\rhd$. (c) The reduced dual network $\hat{\mathcal{N}}^\rhd$ (purple lines) lacks links to peripheral edge midpoints; randomly coloured polygons show internal triangle areas $E_k$, overlaid on $\mathcal{N}$.  (d) The reduced primal network $\hat{\mathcal{N}}$ (green lines) lacks peripheral edges and peripheral vertices; blue dots show the vertices of $\hat{\mathcal{N}}$; randomly coloured polygons show quadrilateral areas $\tfrac{1}{2}F_j$ at non-peripheral edges.  In $\hat{\mathcal{N}}$, edges normal to the periphery do not terminate in a vertex. 
    }
    \label{fig:Figure2constructors}
\end{figure}

The networks lie on a flat, oriented 2D Riemannian manifold $\mathcal{M}$ that is embedded in $\mathbb{R}^3$.   Vertices at $\mathbf{r}_k$ ($k=1,\dots,N_v$) and cell centres at $\mathbf{R}_i$ ($i=1,\dots,N_c$) lie in $\mathcal{M}$.  
Lengths and areas are evaluated using the metric associated with $\mathcal{M}$.  Evolution of vertex $k$ takes place in the tangent space $T \mathcal{M}\vert_{\mathbf{r}_k}\subset \mathbb{R}^2$.  The union of such spaces over the network is the tangent bundle $\Gamma(T\mathcal{M}_{\mathcal{V}})$.  
We will consider discrete vector fields defined over edges and links, sitting in the tangent bundles $\Gamma(T\mathcal{M}_{\mathcal{E}})$ and $\Gamma(T\mathcal{M}_{\mathcal{L}})$.  It is convenient to define each bundle as the union of tangent spaces $T\mathcal{M}\vert_{\mathbf{b}_j}$, where $\mathbf{b}_j\in \mathcal{M}$ denotes the intersection of edge $j$ with link $j$ (Fig.~\ref{fig:Figure1schematic}a,b). (These tangent spaces align with $\mathcal{M}$ because it is flat, but it is helpful to distinguish them for conceptual purposes.) Special provision is made for cells at the periphery $\partial \mathcal{N}$ of an isolated monolayer, as illustrated in Fig.~\ref{fig:Figure2constructors}; we will not need to include peripheral edges in $\Gamma(T\mathcal{M}_{\mathcal{E}})$ or links connected to peripheral edges in $\Gamma(T\mathcal{M}_{\mathcal{L}})$.

As illustrated in Fig.~\ref{fig:Figure1schematic}(a), cell centres are defined as vertex centroids $\mathbf{R}_i=Z_i^{-1}\sum_k C_{ik}\mathbf{r}_k$, and cell edges connecting adjacent vertices and links connecting adjacent cells are respectively
\begin{equation}
    \mathbf{t}_j={\textstyle\sum_k} A_{jk}\mathbf{r}_k,\quad \mathbf{T}_j={\textstyle \sum_j} B_{ij}\mathbf{R}_i.
    \label{eq:edgelink}
\end{equation}
Oriented cell faces are $A_i\boldsymbol{\epsilon}_i$, with area $A_i$; oriented triangle faces are $E_k\boldsymbol{\epsilon}_k$, with area $E_k$.  We define $F_j$ as the area of the parallelogram spanned by $\mathbf{t}_j$ and $\mathbf{T}_j$ so that quadrilaterals with area $\frac{1}{2}F_j$ tile the monolayer (Figs~\ref{fig:Figure1schematic}a and \ref{fig:Figure2constructors}d).  Centroids of each edge are defined by $\mathbf{c}_j=\tfrac{1}{2}\sum_k \vert A_{jk}\vert \mathbf{r}_k$; these are distinct in general from $\mathbf{b}_j$ (Fig.~\ref{fig:Figure1schematic}a).  

\subsubsection{Vector spaces on the primal and dual networks}

Under a scalar-valued natural pairing $\langle \cdot \vert \cdot \rangle$ \cite{desbrun2005, wang2023}, each basis of the spaces defined over $\mathcal{N}$ and $\mathcal{N}^\rhd$ induces a basis $\mathsf{q}_k^*$, $\mathsf{q}_j^*$ and $\mathsf{q}_i^*$ in one of the dual spaces 
\begin{subequations}
\label{eq:vsp}
\begin{align}
    \Omega_0^0(\mathcal{N})&\equiv \mathcal{V}^*, & \Omega_0^1(\mathcal{N})&\equiv \mathcal{E}^*, &   \Omega_0^2(\mathcal{N})&\equiv \mathcal{F}^*,\\
\Omega_0^2(\mathcal{N}^\rhd)&\equiv \mathcal{T}^*, & \Omega_0^1(\mathcal{N}^\rhd)&\equiv \mathcal{L}^*,  &  \Omega_0^0(\mathcal{N}^\rhd)&\equiv \mathcal{C}^*.
\end{align}
\end{subequations}
These spaces hold scalar-valued cochains.  To explain notation, we define $\Omega_n^m(\mathcal{N})$ to hold $n$-cochain-valued $m$-cochains over the network $\mathcal{N}$.  Thus for a network $\mathcal{N}$ confined to the 2D manifold $\mathcal{M}$, for $m=0,1,2$, $\Omega_0^m(\mathcal{N})$ and $\Omega_2^m(\mathcal{N})$ hold scalars and elements of $\Omega^m_1(\mathcal{N})$ (covectors) have two scalar components.  The dual bases of the spaces (\ref{eq:vsp}) satisfy 
\begin{equation}
\langle \mathsf{q}_k^*\vert \mathsf{q}_{k'}\rangle=\delta_{kk'}, \quad
\langle \mathsf{q}_j^*\vert \mathsf{q}_{j'}\rangle=\delta_{jj'}, \quad
\langle \mathsf{q}_i^*\vert \mathsf{q}_{i'}\rangle=\delta_{ii'}.
\end{equation}
Thus, for a 0-chain $\mathsf{f}\in \mathcal{V}$ and a scalar-valued 0-cochain $\phi\in\Omega_0^0(\mathcal{N})$, we can write $\mathsf{f}=\sum_k\langle \mathsf{q}^*_k\vert\mathsf{f}\rangle \mathsf{q}_k\equiv \sum_k f_k\mathsf{q}_k$ and ${\phi}=\sum_k\langle \phi\vert \mathsf{q}_k\rangle \mathsf{q}_k^*\equiv \sum_k \phi_k \mathsf{q}_k^*$. The pairing is given explicitly in this case by $\langle \cdot \vert \cdot \rangle:\Omega_0^0(\mathcal{N})\times\mathcal{V}\rightarrow \mathbb{R}$ where
\begin{equation}
    \langle \phi \vert \mathsf{f} \rangle =\left\langle {\textstyle{\sum_k}} \phi_k \mathsf{q}_k^* \vert
{\textstyle{\sum_{k'}}} f_{k'} \mathsf{q}_{k'} \right\rangle={\textstyle{\sum_k}}\phi_k f_k.
\label{eq:pairing}
\end{equation}
When the chain $\mathsf{f}$ is an indicator function defining a set of vertices, (\ref{eq:pairing}) can be interpreted as an integral of $\phi$ over the chain $\mathsf{f}$.  $\mathsf{A}^\top$ and $\mathsf{B}^\top$ act as boundary operators, while their adjoints with respect to the natural pairing, $\mathsf{A}^*\equiv \sum_{jk}A_{jk}\mathsf{q}_j^*\otimes \mathsf{q}_k^*$ and $\mathsf{B}^*\equiv \sum_{ij}B_{ij} \mathsf{q}_i^*\otimes \mathsf{q}_j^*$, act as difference (or coboundary) operators acting on cochains.  Thus
\begin{equation}
    \langle \mathsf{A}^* \phi \vert \mathsf{g}\rangle={\textstyle\sum_{j,k}}  g_j A_{jk} \phi_k=\langle\phi  \vert \mathsf{A}^\top\mathsf{g}\rangle, \quad
    \langle \mathsf{B}^* \psi \vert \mathsf{h}\rangle={\textstyle\sum_{i,j}} h_i B_{ij} \psi_j=\langle \psi \vert \mathsf{B}^\top\mathsf{h}\rangle,
    \label{eq:apg}
\end{equation}
for $\mathsf{g}\in \mathcal{E}$, $\mathsf{h}\in\mathcal{F}$, $\phi\in\Omega_0^0(\mathcal{N})$, $\psi\in\Omega_0^1(\mathcal{N})$.  Eq.~(\ref{eq:apg}a) shows that the integral of $\mathsf{A}^*\phi$ along a path specified by edges $\mathsf{g}$ is equivalent to $\phi$ evaluated at the vertices bounding the path.  Eq.~(\ref{eq:apg}b) shows that the integral of $\mathsf{B}^*\psi$ over a patch of cells specified by the chain $\mathsf{h}$ is equivalent to $\psi$ evaluated around the cell edges bounding the patch.

$\mathsf{A}^*$ and $\mathsf{B}^*$ inherit from (\ref{eq:ba}) the properties $\mathsf{B}^*\mathsf{A}^*=\mathsf{0}$ and $\mathsf{A}^{*\top}\mathsf{B}^{*\top}=\mathsf{0}$, forming the exact sequences
\begin{subequations}
\begin{equation}
\label{eq:es1}
\begin{tikzcd}
\Omega_0^0(\mathcal{N})  \arrow[r, "\mathsf{A}^{*}"] 
& \Omega_0^1(\mathcal{N}) \arrow[r, "\mathsf{B}^{*}"] 
& \Omega_0^2(\mathcal{N}).
\end{tikzcd}
\end{equation}
and
\begin{equation}
 \label{eq:es2}
 \begin{tikzcd}
\Omega_0^2(\mathcal{N}^\rhd)  
& \Omega_0^1(\mathcal{N}^\rhd) \arrow[l, "\mathsf{A}^{*\top}"] 
& \Omega_0^0(\mathcal{N}^\rhd) \arrow[l, "\mathsf{B}^{*\top}"].
\end{tikzcd}
\end{equation}
\end{subequations}
Furthermore, for $\psi\in\Omega_0^1(\mathcal{N})$, Helmholtz--Hodge decomposition \cite{bhatia2012, lim2019} implies that 
\begin{subequations}
\label{eq:hh}
\begin{equation}
\psi=\mathsf{A}^*\phi+\mathsf{B}^{*\top}\theta +\mathsf{x}
\end{equation}
for some $\phi\in\Omega_0^0(\mathcal{N})$, $\theta\in\Omega_0^2(\mathcal{N})$ and $\mathsf{x}\in\Omega_0^1(\mathcal{N})$, where 
\begin{equation}
    (\mathsf{A}^*\mathsf{A}^{*\top}+\mathsf{B}^{*\top}\mathsf{B}^*)\mathsf{x}=\mathsf{0}, \quad \mathsf{A}^{*\top}\psi=\mathsf{A}^{*\top}\mathsf{A}^*\phi, \quad \mathsf{B}^*\psi =\mathsf{B}^*\mathsf{B}^{*\top}\theta.
\end{equation}
 \end{subequations}
Eq.~(\ref{eq:hh}a) illustrates how $\Omega_0^1(\mathcal{N})$ can be partitioned into the orthogonal subspaces $\mathrm{im}(\mathsf{A}^*)$ of dimension $N_v-1$ (the so-called cut space), $\mathrm{im}(\mathsf{B}^{*\top})$ of dimension $N_c$ (the so-called cycle space) and $\mathrm{ker}(\mathsf{A}^*\mathsf{A}^{*\top}+\mathsf{B}^{*\top}\mathsf{B}^*)=\mathrm{ker}(\mathsf{A}^{*\top})\cap \mathrm{ker}(\mathsf{B}^*)=\mathrm{ker}(\mathsf{A}^{*\top})/\mathrm{im}(\mathsf{B}^{*\top})$ \cite{lim2019} with dimension equal to the number of holes $n_h$ in the monolayer.  $\mathrm{ker}(\mathsf{A}^{*\top})$ and $\mathrm{im}(\mathsf{A}^*)$ have dimensions $n_h+N_c$ and $N_v-1$ respectively, summing to $N_e=N_v-1+N_c+n_h$; $\mathrm{ker}(\mathsf{B}^{*})$ and $\mathrm{im}(\mathsf{B}^{*\top})$ have dimensions $N_v-1+n_h$ and $N_c$ respectively, also summing to $N_e$.  The cycle space contains all closed paths around cell edges.  Analogous representations to (\ref{eq:hh}) follow for scalar-valued cochains defined on $\Omega_0^1(\mathcal{N}^\rhd)$, exploiting (\ref{eq:es1}, \ref{eq:es2}).  Below, we will extend the decomposition (\ref{eq:hh}a), which is based solely on topological information, by incorporating appropriate metric information to describe vectors defined on edges and links.  Differences between combinatorial Laplacians, such as $\mathsf{A}^{*\top}\mathsf{A}^*$, $\mathsf{B}^*\mathsf{B}^{*\top}$ and $\mathsf{A}^*\mathsf{A}^{*\top}+\mathsf{B}^{*\top}\mathsf{B}^*$ in (\ref{eq:hh}b), and metric-dependent Laplacians are discussed in \cite{ribando2024}.  

For isolated monolayers of interest here (\hbox{e.g.} Fig.~\ref{fig:Figure2constructors}), suitable boundary conditions must be applied to the potentials $\phi$ and $\theta$ in (\ref{eq:hh}) (and their analogues).  We explain in Appendix~\ref{app:bc1} how this can be accommodated by use of modified forms of the incidence matrices, $\hat{\mathsf{A}}$ and $\hat{\mathsf{B}}$ in (\ref{eq:abred}), that suppress contributions from peripheral edges and peripheral vertices while satisfying $\hat{\mathsf{B}}\hat{\mathsf{A}}=\mathsf{0}$.  These are defined over reduced networks $\hat{\mathcal{N}}$ and $\hat{\mathcal{N}}^\rhd$ that lack peripheral edges and vertices and links to peripheral edges respectively (Fig.~\ref{fig:Figure2constructors}c,d).  We proceed by defining functions over these reduced networks.

\subsection{Operators on networks of cells}
\label{sec:prdr}

Exterior calculus uses an economical notation whereby individual symbols can have multiple interpretations, depending on the object on which they act and the spaces in which these objects sit.  Below, we will identify instances of the exterior derivative $\mathrm{d}$, musical isomorphisms ($\sharp$, $\flat$), wedge product ($\wedge$), interior product ($\iota$) and Hodge star ($\star$), chosen to be consistent with operators defined in \cite{jensen2022}.  Notationally, we will distinguish vectors (in bold font) that describe positions or orientations ($\mathbf{r}_k\in \mathcal{M}$, $\mathbf{R}_i\in\mathcal{M}$; $\mathbf{t}_j\in T\mathcal{M}\vert_{\mathbf{b}_j}$, $\mathbf{T}_j\in T\mathcal{M}\vert_{\mathbf{b}_j}$, etc.) from 1-cochain-valued cochains (in sans serif) having two scalar components labelled with $\parallel$ or $\perp$.  We will use vectors locally parallel ($\parallel$) to $\mathbf{t}_j$ or $\mathbf{T}_j$ ($j=1,\dots,N_e$), and vectors orthogonal to them ($\perp$) within $T\mathcal{M}\vert_{\mathbf{b}_j}$,  as local bases for vector fields (Fig.~\ref{fig:Figure1schematic}b).  Accordingly, we define the (vector) space $\mathcal{P}\subset\mathbb{R}^2$ with covector basis $\{\mathsf{p}^\parallel,\mathsf{p}^\perp\}^\top$ that holds components of 1-cochain-valued cochains; we shall call such objects $\mathcal{P}$-valued cochains.  We then extend the definition of cochain spaces (\ref{eq:vsp}) so that $\mathcal{P}$-valued $m$-cochains sit within the spaces
\begin{equation}
\Omega_1^m(\hat{\mathcal{N}})\equiv \Omega_0^m(\hat{\mathcal{N}})\times \mathcal{P}, \quad
\Omega_1^m(\hat{\mathcal{N}}^\rhd)\equiv \Omega_0^m(\hat{\mathcal{N}}^\rhd)\times \mathcal{P}, \quad m=0,1,2.
\end{equation}
Thus, over the primal reduced network $\hat{\mathcal{N}}$, a label $\parallel$ [or $\perp$] that appears on cochain elements that are defined over vertices ($m=0$) or cell faces ($m=2$) signifies that the cochain element is associated with the projection of a vector field onto $\parallel$ [or $\perp$] basis vectors in the tangent bundle $\Gamma(TM_{\mathcal{E}})$.

It is convenient to embed orthogonality in $\mathcal{P}$-space within the natural pairing (\ref{eq:pairing}), which we extend by defining $\langle\cdot \vert\cdot\rangle_{\mathcal{P}}$ to satisfy, for $1\leq i,i'\leq N_c$ 
\begin{subequations}
\begin{align}
    \langle \mathsf{p}^\parallel \mathsf{q}_i^*\vert \mathsf{p}^\perp \mathsf{q}_{i'}\rangle_{\mathcal{P}} &= 0, & 
        \langle \mathsf{p}^\perp \mathsf{q}_i^*\vert \mathsf{p}^\parallel \mathsf{q}_{i'}\rangle_{\mathcal{P}} &= 0, \\
    \langle \mathsf{p}^\parallel \mathsf{q}_i^*\vert \mathsf{p}^\parallel \mathsf{q}_{i'}\rangle_{\mathcal{P}} &= \delta_{ii'}, & 
        \langle \mathsf{p}^\perp \mathsf{q}_i^*\vert \mathsf{p}^\perp \mathsf{q}_{i'}\rangle_{\mathcal{P}} &= \delta_{ii'},
\end{align}
\end{subequations}
and likewise for $j$ and $k$.  Then, for $\phi\in \Omega_1^0(\hat{\mathcal{N}})$ and $\mathsf{f}\in 
\mathcal{V}\times\mathcal{P}$, where 
\begin{subequations}
\begin{align}
\mathsf{f}\equiv \{\mathsf{f}^\parallel,\mathsf{f}^\perp\}^\top &\equiv \mathsf{f}^\parallel \mathsf{p}^\parallel+\mathsf{f}^\perp\mathsf{p}^\perp\equiv{\textstyle\sum_k}(f_k^\parallel\mathsf{p}^\parallel+f_k^\perp \mathsf{p}^\perp)\mathsf{q}_k
\equiv {\textstyle\sum_k}f_k\mathsf{q}_k,\\
\phi\equiv \{\phi^\parallel,\phi^\perp\}^\top &\equiv \phi^\parallel \mathsf{p}^\parallel+\phi^\perp\mathsf{p}^\perp\equiv{\textstyle\sum_k}(\phi_k^\parallel\mathsf{p}^\parallel+\phi_k^\perp \mathsf{p}^\perp)\mathsf{q}_k^*
\equiv {\textstyle\sum_k}\phi_k\mathsf{q}_k^*,
\end{align}
\end{subequations}
the pairing (\ref{eq:pairing}) is extended such that 
\begin{equation}
   \langle \phi\vert \mathsf{f}\rangle_{\mathcal{P}} =\langle\{\phi^\parallel,\phi^\perp\}\vert \{\mathsf{f}^\parallel,\mathsf{f}^\perp\}^\top\rangle_{\mathcal{P}} =\langle \phi^\parallel\vert \mathsf{f}^\parallel\rangle + \langle \phi^\perp \vert \mathsf{f}^\perp\rangle={\textstyle \sum_k}(\phi_k^\parallel f_k^\parallel +\phi_k^\perp f_k^\perp).
   \label{eq:f2}
\end{equation}
$\Omega_1^1 (\hat{\mathcal{N}})$ and $\Omega_1^1(\hat{\mathcal{N}}^\rhd)$ are dual to $\Gamma(T\mathcal{M}_{\mathcal{E}})$ and $\Gamma(T\mathcal{M}_{\mathcal{L}})$ respectively (in a sense to be clarified below) and so can be considered as cotangent bundles.  

We introduce Hodge stars $\star$ to connect the sequences (\ref{eq:es1}, \ref{eq:es2}) on the reduced primal and dual networks, so that
\begin{equation}    
\label{eq:spacen}
\begin{tikzcd}[row sep=large]
\Omega_{2-n}^2(\hat{\mathcal{N}}^\rhd) \arrow[d, "\star_{2-n,2}^\rhd" description, shift left=2.5ex]
& \arrow[l, "\mathsf{A}_n^{*\top}"] \Omega_{2-n}^1(\hat{\mathcal{N}}^\rhd) \arrow[d, "\star_{2-n,1}^\rhd" description, shift left=2.5ex]  
& \arrow[l, "\mathsf{B}_n^{*\top}"] \Omega_{2-n}^0(\hat{\mathcal{N}}^\rhd) \arrow[d, "\star_{2-n,0}^\rhd" description, shift left=2.5ex] \\
\Omega_n^0(\hat{\mathcal{N}}) \arrow[u, "\star_{n,0}" description, shift left=2.5ex] \arrow[r, "\mathsf{A}_n^{*}"]   
& \Omega_n^1(\hat{\mathcal{N}}) \arrow[r, "\mathsf{B}_n^{*}"] \arrow[u, "\star_{n,1}" description, shift left=2.5ex]
& \Omega_n^2(\hat{\mathcal{N}}) \arrow[u, "\star_{n,2}" description, shift left=2.5ex],  
\end{tikzcd} \qquad (n=0,1,2).
\end{equation}
As (\ref{eq:spacen}) illustrates, Hodge stars are distinguished by two subscripts, so that
\begin{equation}
\label{eq:hodef}
    \star_{n,m}:\Omega_n^m(\hat{\mathcal{N}})\rightarrow\Omega_{2-n}^{2-m}(\hat{\mathcal{N}}^\rhd), \quad \star_{n,m}^\rhd:\Omega_n^m(\hat{\mathcal{N}}^\rhd)\rightarrow\Omega_{2-n}^{2-m}(\hat{\mathcal{N}}).
\end{equation}
The first subscript on $\star_{n,m}$ denotes the cochain-value $n$ of its argument; the second denotes the underlying space from which it acts.  Specific definitions are provided below.  In (\ref{eq:spacen}), we set $\mathsf{A}_0^*\equiv \hat{\mathsf{A}}^*$, $\mathsf{A}_2^*\equiv \hat{\mathsf{A}}^*$, $\mathsf{B}_0^*\equiv \hat{\mathsf{B}}^*$, $\mathsf{B}_2^*\equiv \hat{\mathsf{B}}^*$ (consistent with \ref{eq:es1} and \ref{eq:es2}) and define $\mathsf{A}_1^*$ and $\mathsf{B}_1^*$ shortly.  We shall pay particular attention to $\mathcal{P}$-valued cochains, with $n=1$, extending (\ref{eq:spacen}) to
\begin{equation}    
\label{eq:dr}
\begin{tikzcd}[row sep=large]
& \Gamma(T\mathcal{M}_{\mathcal{L}}) \arrow[d, "\flat" description, shift right=1ex] 
& \mathcal{C}\times\mathcal{P}\arrow[d, "\flat" description, shift right=1ex] \\
\Omega_1^2(\hat{\mathcal{N}}^\rhd) 
\arrow[d, shift left=2.5ex, "\star_{1,2}^\rhd" description] 
& \arrow[l, "\mathsf{A}_1^{*\top}"] \Omega_1^1(\hat{\mathcal{N}}^\rhd) \arrow[d, "\star_{1,1}^\rhd" description, shift left=2.5ex] \arrow[u, shift right=1ex, "\sharp" description] 
& \arrow[l, "\mathsf{B}_1^{*\top}"] \Omega_1^0(\hat{\mathcal{N}}^\rhd) \arrow[d, "\star_{1,0}^\rhd" description, shift left=2.5ex] \arrow[u, "\sharp" description, shift right=1ex]\\
\Omega_1^0(\hat{\mathcal{N}}) \arrow[u, "\star_{1,0}" description, shift left=2.5ex] \arrow[r, "\mathsf{A}_1^{*}"] \arrow[d, "\sharp" description, shift right=1ex]  
& \Omega_1^1(\hat{\mathcal{N}}) \arrow[r, "\mathsf{B}_1^{*}"] \arrow[d, "\sharp" description, shift right=1ex] \arrow[u, "\star_{1,1}" description, shift left=2.5ex]
& \Omega_1^2(\hat{\mathcal{N}})  \arrow[u, "\star_{1,2}" description, shift left=2.5ex] 
\\
\hat{\mathcal{V}}\times\mathcal{P} \arrow[u, shift right=1ex, "\flat" description] 
& \Gamma(T\mathcal{M}_{\mathcal{E}}) \arrow[u, shift right=1ex, "\flat" description]
& 
\end{tikzcd},
\end{equation}
which provides the framework over which we build differential operators.  We define $\hat{\mathcal{V}}$ in (\ref{eq:dr}) to be $\mathcal{V}$ restricted to internal vertices (as illustrated in Fig.~\ref{fig:Figure2constructors}(d)). The sharp and flat operators appearing in (\ref{eq:dr}) will be defined as we proceed.  Maps between spaces, introduced below, are summarised in Table~\ref{tab:maps}.   When edges and links are not orthogonal, we will see that $\star_{1,1}^\rhd$ in (\ref{eq:dr}) differs from $-(\star_{1,1})^{-1}$, although we will impose that 
\begin{equation}
\star_{1,0}^{-1}=-\star_{1,2}^\rhd, \quad \star_{1,2}^{-1}=-\star_{1,0}^\rhd.
\label{eq:h012}
\end{equation}
As explained below, the $-$ sign arises from the action of $\star$ on the value-leg of $n=1$ cochains, which involves a rotation in $\mathcal{P}$-space.

\begin{table}
    \centering
    \begin{tabular}{|c|c| c| }
    \hline 
    Maps & Spaces & Representation  \\
    \hline
$\mathsf{A}_1^*$ & $\Omega_1^0(\hat{\mathcal{N}})\rightarrow \Omega_1^1(\hat{\mathcal{N}})$ 
& $\sum_{j,k}A_{jk}\mathsf{q}_j^*\otimes\mathsf{q}_k^*\otimes \mathsf{I}_{\mathcal{P}}$  \\
    $\mathsf{B}_1^*$ & $\Omega_1^1(\hat{\mathcal{N}}) \rightarrow \Omega_1^2(\hat{\mathcal{N}})$ & $\sum_{i,j}B_{ij}\mathsf{q}_i^*\otimes\mathsf{q}_j^*\otimes \mathsf{I}_{\mathcal{P}}$  \\
$\flat$ & $\mathcal{V}\times\mathcal{P}\rightarrow \Omega_1^0(\hat{\mathcal{N}})$ & $\sum_{k,k'}\delta_{kk'}\mathsf{q}_k^*\otimes\mathsf{q}_{k'}\otimes \mathsf{I}_{\mathcal{P}}$ \\
$\flat$ & $\Gamma(T\mathcal{M}_{\mathcal{E}})\rightarrow \Omega_1^1(\hat{\mathcal{N}})$ & $\sum_{j,j'}\delta_{jj'}\{\mathbf{e}_{j\parallel}\cdot, \mathbf{e}_{j\perp}\cdot\}\mathsf{q}_j^*\otimes\mathsf{q}_{j'}\otimes \mathsf{I}_{\mathcal{P}}$ \\
$\flat$ & $\mathcal{C}\times\mathcal{P}\rightarrow \Omega_1^0(\hat{\mathcal{N}}^\rhd)$ & $\sum_{i,i'}\delta_{ii'}\mathsf{q}_i^*\otimes\mathsf{q}_{i'}\otimes \mathsf{I}_{\mathcal{P}}$ \\
$\flat$ & $\Gamma(T\mathcal{M}_{\mathcal{L}})\rightarrow \Omega_1^1(\hat{\mathcal{N}}^\rhd)$ &  $\sum_{j,j'}\delta_{jj'}\{\mathbf{E}_{j\parallel}\cdot, \mathbf{E}_{j\perp}\cdot\}\mathsf{q}_j^*\otimes\mathsf{q}_{j'}\otimes \mathsf{I}_{\mathcal{P}}$ \\
\hline
$\sharp$ & $\Omega_0^1\rightarrow \mathcal{E}$ & $\sum_{j,j'}\delta_{j j'}(1/t_j)\mathsf{q}_j\otimes\mathsf{q}_{j'}$  \\
$\sharp$ & $\Omega_1^1(\hat{\mathcal{N}})\rightarrow\Gamma(T\mathcal{M}_{\mathcal{E}})$ & $\sum_{j,j'} \delta_{jj'}\{\mathbf{e}_{j\parallel},\mathbf{e}_{j\perp}\} \mathsf{q}_j\otimes \mathsf{q}_j^* $  \\
$\sharp$ & $\Omega_1^1(\hat{\mathcal{N}}^\rhd)\rightarrow\Gamma(T\mathcal{M}_{\mathcal{L}})$ & $\sum_{j,j'} \delta_{jj'}\{\mathbf{E}_{j\parallel},\mathbf{E}_{j\perp}\} \mathsf{q}_j\otimes \mathsf{q}_j^* $ \\
$\sharp$ & $\Omega_1^0(\hat{\mathcal{N}}) \rightarrow\mathcal{V}\times\mathcal{P}$ & $\sum_{k,k'} \delta_{kk'} \mathsf{q}_k\otimes \mathsf{q}_k^* \otimes \mathcal{I}_{\mathcal{P}} $  \\
$\sharp$ & $\Omega_1^0(\hat{\mathcal{N}}^\rhd)\rightarrow\mathcal{C}\times\mathcal{P}$ & $\sum_{k,k'} \delta_{kk'} \mathsf{q}_k\otimes \mathsf{q}_k^* \otimes \mathcal{I}_{\mathcal{P}} $  \\
\hline
$\star_{0,0}$
& $\Omega_0^0(\hat{\mathcal{N}})\rightarrow\Omega_2^2(\hat{\mathcal{N}}^\rhd) $& $\sum_{k,k'} \delta_{k k'} E_k \mathsf{q}_k^*\otimes \mathsf{q}_{k'}^*$  \\
$\star_{0,0}^\rhd$ 
& $\Omega_0^0(\hat{\mathcal{N}}^\rhd)\rightarrow\Omega_2^2(\hat{\mathcal{N}})$& $\sum_{i,i'} \delta_{i i'} A_i \mathsf{q}_i^*\otimes \mathsf{q}_{i'}^*$  \\
\hline
$\star_{1,0}$ & $\Omega_1^0(\hat{\mathcal{N}}) \rightarrow \Omega_1^2(\hat{\mathcal{N}}^\rhd)$ &$\sum_{k,k'} \delta_{k k'} E_k \mathsf{q}_k^*\otimes \mathsf{q}_{k'}^*\otimes \boldsymbol{\epsilon}_{\mathcal{P}}$ \\
$\star_{1,0}^\rhd$ & $\Omega_1^0(\hat{\mathcal{N}}^\rhd)\rightarrow \Omega_1^2(\hat{\mathcal{N}})$& $\sum_{i,i'} \delta_{i i'} A_i\mathsf{q}_i^*\otimes \mathsf{q}_{i'}^*\otimes \boldsymbol{\epsilon}_{\mathcal{P}}$ \\
\hline
$\star_{0,1}$ & $\Omega_0^1(\hat{\mathcal{N}})\rightarrow \Omega_2^1(\mathcal{N^\rhd})$ & $\sum_{j,j'} \delta_{j j'} (F_j/t_j^2) \mathsf{q}_j^*\otimes \mathsf{q}_{j'}^*$ \\
$\star_{1,1}$ & $\Omega_1^1(\hat{\mathcal{N}})\rightarrow \Omega_1^1(\hat{\mathcal{N}}^\rhd)$ & $\sum_{j,j'}\delta_{j j'}(F_j/t_j^2) \mathsf{q}_j^*\otimes \mathsf{q}_{j'}^*\otimes \boldsymbol{\epsilon}_{\mathcal{P}}$ \\
$\star_{2,1}$ & $\Omega_2^1(\hat{\mathcal{N}})\rightarrow \Omega_0^1(\mathcal{N^\rhd})$ & $\sum_{j,j'}\delta_{j j'}(F_j/t_j^2) \mathsf{q}_j^*\otimes \mathsf{q}_{j'}^*$ \\
\hline
$\star_{0,1}^\rhd$ & $\Omega_0^1(\mathcal{N^\rhd})\rightarrow \Omega_2^1(\hat{\mathcal{N}})$ & $\sum_{j,j'}\delta_{j j'}(F_j/T_j^2) \mathsf{q}_j^*\otimes \mathsf{q}_{j'}^*$ \\
$\star_{1,1}^\rhd$ & $\Omega_1^1(\hat{\mathcal{N}}^\rhd)\rightarrow \Omega_1^1({\mathcal{N}})$ & $\sum_{j,j'}\delta_{j j'}(F_j/T_j^2)\mathsf{q}_j^*\otimes \mathsf{q}_{j'}^*\otimes \boldsymbol{\epsilon}_{\mathcal{P}}$  \\
$\star_{2,1}^\rhd$ & $\Omega_2^1(\mathcal{N^\rhd})\rightarrow \Omega_0^1(\hat{\mathcal{N}})$ & $\sum_{j,j'}\delta_{j j'}(F_j/T_j^2) \mathsf{q}_j^*\otimes \mathsf{q}_{j'}^*$\\
\hline
    \end{tabular}
    \caption{Definitions of maps, the spaces over which they act and their explicit representation in terms of relevant bases.}
    \label{tab:maps}
\end{table}

We will require suitable inner products in order to derive differential operators.  Inner products are defined here in terms of $\wedge$ and $\star$; thus we define these operators, along with $\mathrm{d}$, $\sharp$ and $\flat$, in the following subsections, before addressing differential operators in Sec.~\ref{sec:do}. 

\subsubsection{The exterior derivative}

We define the exterior derivative $\mathrm{d}$ in (\ref{eq:dr}) as $\mathsf{A}_1^*:\Omega_1^0(\hat{\mathcal{N}})\rightarrow \Omega_1^1(\hat{\mathcal{N}})$, $\mathsf{B}_1^*:\Omega_1^1(\hat{\mathcal{N}})\rightarrow \Omega_1^2(\hat{\mathcal{N}})$, or the transposes $\mathsf{B}_1^{*\top}:\Omega_1^0(\hat{\mathcal{N}}^\rhd)\rightarrow \Omega_1^1(\hat{\mathcal{N}}^\rhd)$, $\mathsf{A}_1^{*\top}:\Omega_1^1(\hat{\mathcal{N}}^\rhd)\rightarrow \Omega_1^2(\hat{\mathcal{N}}^\rhd)$, where
\begin{equation}
\mathsf{A}_1^*\equiv \hat{\mathsf{A}}^*\otimes\mathsf{I}_{\mathcal{P}}, \quad 
\mathsf{B}_1^*\equiv \hat{\mathsf{B}}^*\otimes \mathsf{I}_{\mathcal{P}},\quad
\mathsf{I}_{\mathcal{P}}\equiv \mathsf{p}^\parallel\otimes \mathsf{p}^\parallel + \mathsf{p}^\perp\otimes \mathsf{p}^\perp.
\end{equation}
Thus $\mathsf{I}_{\mathcal{P}}$ is represented by the $2\times 2$ identity matrix in the $\{\mathsf{p}^\parallel,\mathsf{p}^\perp\}^\top$ basis.  Eq.~(\ref{eq:ba}) ensures that $\mathrm{d}$ is nilpotent, via $\mathsf{B}_1^*\mathsf{A}_1^*=(\hat{\mathsf{B}}^*\otimes\mathsf{I}_{\mathcal{P}})(\hat{\mathsf{A}}^*\otimes\mathsf{I}_{\mathcal{P}})=(\hat{\mathsf{B}}^*\hat{\mathsf{A}}^*)\otimes \mathsf{I}_{\mathcal{P}}=\mathsf{0}$; likewise $\mathsf{A}_1^{*\top}\mathsf{B}_1^{*\top}=\mathsf{0}$.  As illustrated in (\ref{eq:dr}), $\mathrm{d}$ maps $\mathcal{P}$-valued 0-cochains to $\mathcal{P}$-valued 1-cochains, and $\mathcal{P}$-valued 1-cochains to $\mathcal{P}$-valued 2-cochains, over the reduced primal network $\hat{\mathcal{N}}$ and its dual $\hat{\mathcal{N}}^\rhd$.  Use of the reduced networks anticipates the implementation of boundary conditions at $\partial\mathcal{N}$.

\subsubsection{Sharp and flat operators}

Maps between $\hat{\mathcal{V}}\times\mathcal{P}$ and $\Omega_1^0(\hat{\mathcal{N}})$ (see (\ref{eq:dr})), for $\mathsf{f}\equiv \sum_k\{f_k^\parallel, f_k^\perp\}^\top\mathsf{q}_k\in \hat{\mathcal{V}}\times\mathcal{P}$ and $\phi\equiv \sum_k\{\phi_k^\parallel, \phi_k^\perp\}^\top\mathsf{q}_k^* \in\Omega_1^0(\hat{\mathcal{N}})$, involve a change of basis and are defined by $\mathsf{f}^\flat=\sum_k \{f_k^\parallel,f_k^\perp\}^\top \mathsf{q}_k^*$ and $\phi^\sharp=\sum_k \{\phi_k^\parallel,\phi_k^\perp\}^\top \mathsf{q}_k$, so that $(\mathsf{f}^\flat)^\sharp=\mathsf{f}$ and $(\phi^\sharp)^\flat=\phi$.   Using (\ref{eq:f2}), these induce metrics 
\begin{subequations}
\begin{align}
\langle\mathsf{f}^\flat \vert \mathsf{f}\rangle_{\mathcal{P}}&=\langle\mathsf{f}^{\parallel\flat} \vert \mathsf{f}^\parallel\rangle +\langle\mathsf{f}^{\perp\flat} \vert \mathsf{f}^\perp\rangle
={\textstyle \sum_k} (f_k^{\parallel2}+f_k^{\perp 2}),\\
\langle\phi \vert \phi^\sharp\rangle_{\mathcal{P}}&=\langle\phi^\parallel \vert \phi^{\parallel\sharp}\rangle+\langle\phi^\perp \vert \phi^{\perp\sharp}\rangle={\textstyle\sum_k} (\phi_k^{\parallel 2}+\phi_k^{\perp 2}).
\end{align}
\end{subequations}
Analogous $\sharp$ and $\flat$ operators connect $\mathcal{C}\times\mathcal{P}$ and $\Omega_1^0(\hat{\mathcal{N}}^\rhd)$ (Table~\ref{tab:maps}).

To span tangent bundles, we introduce spatial basis vectors in $\Gamma(T\mathcal{M}_{\mathcal{E}})$ and $\Gamma(T\mathcal{M}_{\mathcal{L}})$.  The contravariant and  covariant bases aligned to edge $j$ and link $j$ at $\mathbf{b}_j$ are defined respectively
, using (\ref{eq:edgelink}), as 
\begin{subequations}
    \begin{align}
    \mathbf{e}_{j\parallel}&={\mathbf{t}_j}/{t_j^2}, & 
    \mathbf{e}_{j\perp}&={\boldsymbol{\epsilon}_i\mathbf{t}_j}/{t_j^2}, & 
     \mathbf{e}_j^{\parallel}&=\mathbf{t}_j, &
 \mathbf{e}_j^\perp&=\boldsymbol{\epsilon}_i\mathbf{t}_j, \\
\mathbf{E}_{j\parallel}&={\mathbf{T}_j}/{T_j^2}, &
    \mathbf{E}_{j\perp}&={\boldsymbol{\epsilon}_k\mathbf{T}_j}/{T_j^2}, & 
     \mathbf{E}_j^{\parallel}&=\mathbf{T}_j, &
     \mathbf{E}_j^\perp&=\boldsymbol{\epsilon}_k\mathbf{T}_j,
\end{align}
\label{eq:cocobases}
\end{subequations}
so that
\begin{subequations}
\label{eq:loceu}
    \begin{align}
     \mathbf{e}_j^{\parallel} \cdot \mathbf{e}_{j\parallel} &= 1, & 
      \mathbf{e}_j^{\parallel} \cdot \mathbf{e}_{j\perp} &= 0, &
\mathsfbf{g}_j&\equiv \mathbf{e}_j^\parallel \otimes \mathbf{e}_{j\parallel}+\mathbf{e}_j^\perp \otimes \mathbf{e}_{j\perp}, \\
     \mathbf{E}_j^{\parallel} \cdot \mathbf{E}_{j\parallel} &= 1, & 
      \mathbf{E}_j^{\parallel} \cdot \mathbf{E}_{j\perp} &= 0, &
\mathsfbf{G}_j&\equiv \mathbf{E}_j^\parallel \otimes \mathbf{E}_{j\parallel}+\mathbf{E}_j^\perp \otimes \mathbf{E}_{j\perp}.
\end{align}
\end{subequations}
Here the dot product exploits the local Euclidean metric $\mathsfbf{g}_j$, $\mathsfbf{G}_j$ of the manifold $\mathcal{M}$.   Orientations are chosen such that $\mathbf{e}_j^\perp$ [$\mathbf{E}_j^\perp$] aligns (reasonably closely) with $\mathbf{E}_j^\parallel$ [$\mathbf{e}_j^\parallel$] (Fig.~\ref{fig:Figure1schematic}b), allowing the definition of the area associated with edges and links (Fig.~\ref{fig:Figure1schematic}a)
\begin{equation}
\label{eq:fj}
    F_j=\mathbf{T}_j\cdot {\boldsymbol{\epsilon}}_i\mathbf{t}_j=\mathbf{t}_j\cdot{\boldsymbol{\epsilon}_k}\mathbf{T}_j.
\end{equation}
Care is needed when edges and links are not orthogonal ($\mathbf{e}_j^\parallel\cdot \mathbf{E}_j^\parallel \neq 0$), as is generic for real epithelia \cite{jensen2020}.  

A basis for vectors defined on edges in $\Gamma(T\mathcal{M}_{\mathcal{E}})$ is provided by $\mathsf{q}_j\left\{\mathbf{e}_{j\parallel}, \mathbf{e}_{j\perp}\right\}^\top$, $j=1,\dots,N_e$.  Thus a typical element of $\Gamma(T\mathcal{M}_{\mathcal{E}})$ can be written $\mathsfbf{v}=\sum_j\mathbf{v}_j\mathsf{q}_j$, where $\mathbf{v}_j=\langle \mathsf{q}_j^*\vert \mathsfbf{v} \rangle$ and 
\begin{equation}
\label{eq:bj}
\mathbf{v}_j=(\mathbf{v}_j \cdot \mathbf{e}_j^\parallel)\mathbf{e}_{j\parallel} + (\mathbf{v}_j \cdot \mathbf{e}_j^\perp)\mathbf{e}_{j\perp}\equiv v_j^\parallel \mathbf{e}_{j\parallel}+v_j^\perp\mathbf{e}_{j\perp}.
\end{equation}
Similarly, a vector $\mathsfbf{V}\in\Gamma(T\mathcal{M}_{\mathcal{L}})$ has components $V_j^\parallel=\mathbf{V}_j\cdot\mathbf{E}_j^\parallel$ and $V_j^\perp=\mathbf{V}_j\cdot\mathbf{E}_j^\perp$ ($j=1,\dots,N_e$).
For a $\mathcal{P}$-valued 1-cochain $\psi\in\Omega_1^1(\hat{\mathcal{N}})$, we define the sharp operator to be a projection onto the contravariant basis in $\Gamma(T\mathcal{M}_{\mathcal{E}})$, \begin{equation}
\psi^\sharp=(\{\psi^\parallel,\psi^\perp\}^{\top})^{\sharp}={\textstyle{\sum_j}}( \psi_j^\parallel \mathbf{e}_{j\parallel} + \psi_j^\perp \mathbf{e}_{j\perp} )\mathsf{q}_j.
\end{equation}
For $\mathsfbf{v}$ in $\Gamma(T\mathcal{M}_{\mathcal{E}})$ we define the flat operator to be the $\mathcal{P}$-valued 1-cochain obtained by contraction with the covariant basis
\begin{equation}
\mathsfbf{v}^\flat={\textstyle{\sum_j}} \{ (\mathbf{v}_j\cdot\mathbf{e}_j^{\parallel}), (\mathbf{v}_j\cdot\mathbf{e}_j^{\perp})\}^\top \mathsf{q}_j^*
=\left\{ {\textstyle{\sum_j}} {v}_j^{\parallel} \mathsf{q}_j^*,{\textstyle{\sum_j}}  {v}_j^{\perp} \mathsf{q}_j^*\right\}^\top  \equiv \left\{\mathsf{v}^\parallel, \mathsf{v}^\perp \right\}^\top\in \Omega_1^1(\hat{\mathcal{N}}).  
\end{equation}
Thus $(\mathsfbf{v}^\flat)^\sharp=\mathsfbf{v}$ and $(\psi^\sharp)^\flat=\psi$.  This induces the metric 
\begin{align}
    \langle \mathsfbf{v}^\flat\vert \mathsfbf{v}\rangle_{\mathcal{P}} &= 
    \left\langle {\textstyle{\sum_j}} \{ (\mathbf{v}_j\cdot\mathbf{e}_j^{\parallel}), (\mathbf{v}_j\cdot\mathbf{e}_j^{\perp})\} \mathsf{q}_j^* \bigg\vert  {\textstyle{\sum_{j'}}} \mathbf{v}_{j'} \mathsf{q}_{j'}\right\rangle_{\mathcal{P}} \nonumber \\
     & \equiv
    \left\langle {\textstyle{\sum_j}} \left[ (\mathbf{v}_j\cdot\mathbf{e}_j^{\parallel})\mathbf{e}_{j\parallel}\cdot + (\mathbf{v}_j\cdot\mathbf{e}_j^{\perp})\mathbf{e}_{j\perp}\cdot  \right]  \mathsf{q}_j^* \bigg\vert {\textstyle{\sum_{j'}}} \mathbf{v}_{j'} \mathsf{q}_{j'}\right\rangle_{\mathcal{P}} \nonumber \\
&=    {\textstyle\sum_j}  \left[(\mathbf{v}_j\cdot\mathbf{e}_j^{\parallel})(\mathbf{v}_j\cdot\mathbf{e}_{j \parallel})+(\mathbf{v}_j\cdot\mathbf{e}_j^{\perp})(\mathbf{v}_j\cdot\mathbf{e}_{j \perp})\right]={\textstyle\sum_j}\mathbf{v}_j^T \mathsfbf{g}_j \mathbf{v}_j={\textstyle\sum_j}\vert \mathbf{v}_j\vert^2.
\end{align}
Under the same definition, the metric induced by the sharp operation is 
\begin{align}
\langle \psi\vert \psi^\sharp \rangle_{\mathcal{P}} &= \left\langle{\textstyle \sum_j}(\psi_j^\parallel, \psi_j^\perp)\mathsf{q}_j^*\bigg\vert {\textstyle{\sum_{j'}}}(\psi^\parallel_j\mathbf{e}_{j\parallel}\mathsf{q}_j+(\psi^\perp_j\mathbf{e}_{j\parallel}\mathsf{q}_j) \right\rangle_{\mathcal{P}} \nonumber \\
& = 
\left\langle {\textstyle \sum_j}(\psi_j^\parallel \mathbf{e}_{j\parallel}\cdot +\psi_{j\perp} \mathbf{e}_j^\perp\cdot) \mathsf{q}_j^* \bigg\vert {\textstyle{\sum_{j'}}}(\psi^\parallel_j\mathbf{e}_{j\parallel}\mathsf{q}_j+(\psi^\perp_j\mathbf{e}_{j\parallel}\mathsf{q}_j) \right\rangle_{\mathcal{P}} \nonumber \\
&= {\textstyle{\sum_j}} \psi_j^{\parallel 2} \vert\mathbf{e}_{j\parallel}\vert^2 + \psi_j^{\perp 2} \vert \mathbf{e}_{j\perp}\vert^2={\textstyle\sum_j} (\psi_j^{\parallel 2}+\psi_j^{\perp 2})/t_j^2.
\end{align}
This is reasonable when interpreted as the magnitude of a covector $\psi$.   Equivalent definitions of $\sharp$ and $\flat$ connect $\Psi\in \Omega_1^1(\hat{\mathcal{N}}^\rhd)$ and $\Gamma(T\mathcal{M}_{\mathcal{L}})$ (Table~\ref{tab:maps}).

\subsubsection{Wedge products}

We define the wedge product $\wedge:\Omega^1_1(\hat{\mathcal{N}})\times\Omega^1_1(\hat{\mathcal{N}})\rightarrow \Omega^1_2(\hat{\mathcal{N}})$ between vectors $\mathsfbf{v}\in \Gamma(T\mathcal{M}_{\mathcal{E}})$ and $\mathsfbf{w}\in \Gamma(T\mathcal{M}_{\mathcal{E}})$, with $\mathsfbf{v}^\flat=\{\mathsf{v}^\parallel,\mathsf{v}^\perp\}^\top$ and $\mathsfbf{w}^\flat=\{\mathsf{w}^\parallel,\mathsf{w}^\perp\}^\top$, as the 2-cochain-valued 1-cochain (\hbox{i.e.} the scalar field defined on edges)
\begin{equation}
\label{eq:wedge1}
    \mathsfbf{v}^\flat\wedge \mathsfbf{w}^\flat\equiv {\textstyle \sum_j}({v}_j^\parallel {w}_j^\perp-{v}_j^\perp {w}_j^\parallel)\mathsf{q}_j^*={\textstyle\sum_j}\{v_j^\parallel,v_j^\perp\}(-\boldsymbol{\epsilon}_{\mathcal{P}})\{w_j^\parallel,w_j^\perp\}^\top \mathsf{q}_j^*, 
\end{equation}  
where
\begin{equation}
 \boldsymbol{\epsilon}_{\mathcal{P}}\equiv \mathsf{p}^\perp \otimes \mathsf{p}^\parallel-\mathsf{p}^\parallel\otimes \mathsf{p}^\perp=\left(\begin{matrix}
        0&-1\\1&0
    \end{matrix}\right).
\end{equation}
Clearly, $\mathsfbf{v}^\flat \wedge \mathsfbf{v}^\flat=\mathsf{0}$ and $\mathsfbf{v}^\flat \wedge \mathsfbf{w}^\flat=-\mathsfbf{w}^\flat \wedge \mathsfbf{v}^\flat$.  A similar definition holds for for $\mathsfbf{V}$ and $\mathsfbf{W}$ in $\Gamma(T\mathcal{M}_{\mathcal{L}})$, with $\wedge$ again acting via $-\boldsymbol{\epsilon}_\mathcal{P}$.   The wedge product between $\phi\in \Omega_0^1(\hat{\mathcal{N}})$ and $\mathsfbf{v}\in \Gamma(T\mathcal{M}_{\mathcal{E}})$ is defined as
\begin{equation}
\label{eq:wedge7}
    \mathsfbf{v}^\flat\wedge\phi=\phi\wedge\mathsfbf{v}^\flat={\textstyle\sum_j}\phi_j(-{\boldsymbol{\epsilon}}_{\mathcal{P}})\{v_j^\parallel,v_j^\perp\}^\top\mathsf{q}_j^*={\textstyle\sum_j}\phi_j\{v_j^\perp,-v_j^\parallel\}^\top\mathsf{q}_j^*
\end{equation}
implementing so-called graded anticommutivity of $\wedge$.  

Cochain elements of $\Omega_n^m(\hat{\mathcal{N}})$ have two legs (adopting terminology of \cite{kanso2007} and others describing bundle-valued forms), one relating to the value $n$ of the cochain and one relating to the underlying chain $m$.  In (\ref{eq:wedge1}, \ref{eq:wedge7}), $\wedge$ acts on the value leg, mapping between (rather than within) the sequences illustrated in (\ref{eq:spacen}).  Additional wedge products can be defined that act on the $m$-leg \cite{desbrun2005}, {\hbox{i.e.}} within sequences with fixed $n$, such as
\begin{subequations}
\label{eq:altwedge}
    \begin{align}
\mathsf{a}\tilde{\wedge}\mathsf{b}&\equiv
    {\textstyle\sum_{i, j,j',k}} A_{jk} B_{ij'} \vert A_{j'k} \vert \vert B_{ij}\vert  a_j b_{j'} \mathsf{q}_k^* \in \Omega_0^2(\hat{\mathcal{N}}^\rhd),\\
\mathsf{u}\tilde{\wedge}\mathsf{v}&\equiv 
    {\textstyle\sum_{i, j,j',k}} 
    B_{ij} A_{j'k}\vert B_{ij'} \vert \vert A_{jk}\vert u_j v_{j'} \mathsf{q}_i^* \in \Omega_0^2(\hat{\mathcal{N}}),
\end{align}
\end{subequations}
for 
$\mathsf{a}\in\Omega_0^1(\hat{\mathcal{N}}^\rhd)$, 
$\mathsf{b}\in\Omega_0^1(\hat{\mathcal{N}}^\rhd)$,
$\mathsf{u}\in\Omega_0^1(\hat{\mathcal{N}})$, 
$\mathsf{v}\in\Omega_0^1(\hat{\mathcal{N}})$.  
For example, defining $\mathsf{T}_x=\sum_j (\mathbf{T}_j\cdot \hat{\mathbf{x}}) \mathsf{q}_j^*\in \Omega_0^1(\hat{\mathcal{N}}^\rhd)$ and $\mathsf{T}_y=\sum_j (\mathbf{T}_j\cdot \hat{\mathbf{y}}) \mathsf{q}_j^*\in\Omega_0^1(\hat{\mathcal{N}}^\rhd)$ where $\hat{\mathbf{x}}$ and $\hat{\mathbf{y}}$ are Cartesian unit vectors, then
\begin{equation}
    \mathsf{T}_y\tilde{\wedge}\mathsf{T}_x =6\mathsf{E}\equiv 6{\textstyle\sum_k} E_k\mathsf{q}_k^*.
\end{equation}
Thus $\tilde{\wedge}:\Omega^1_0(\hat{\mathcal{N}}^\rhd)\times\Omega^1_0(\hat{\mathcal{N}}^\rhd)\rightarrow \Omega^2_0(\hat{\mathcal{N}}^\rhd)$ can be used to evaluate the area of a triangle $E_k$.  We mention the more traditional wedge product (\ref{eq:altwedge}) which increases the value of $m$ (see \cite{desbrun2005}), to emphasise the distinction with (\ref{eq:wedge1}), which increases the value of $n$; only the latter is used in what follows.

\subsubsection{Hodge star operators}

For given $n$, Hodge stars connect the spaces of cochains (\ref{eq:spacen}) by introduction of metric information, expressed as areas $A_i$, $F_j$ and $E_k$ and lengths $t_j$ and $T_j$ (Fig.~\ref{fig:Figure1schematic}a).  Recall that, excluding the periphery, the monolayer is tiled by cells through $A_i$, triangles via $E_k$, or the quadrilaterals spanned by edges and links via $\tfrac{1}{2}F_j$ (Fig.~\ref{fig:Figure2constructors}).  For later reference, these quantities are gathered into matrix operators
\begin{subequations}
\label{eq:mops}
\begin{align}
\hat{\mathsf{E}}&\equiv {\textstyle\sum_{k,k'}} E_k\delta_{k k'} \mathsf{q}_k^*\otimes\mathsf{q}_{k'}^*
,& \mathsf{H}&\equiv {\textstyle\sum_{i,i'}} A_i\delta_{i i'} \mathsf{q}_i^*\otimes\mathsf{q}_{i'}^*
,\\ \hat{\mathsf{T}}_e&\equiv {\textstyle\sum_{j,j'} (t_j^2/F_j)\delta_{j j'} \mathsf{q}_j^*\otimes\mathsf{q}_{j'}^*}
,& \hat{\mathsf{T}}_l&\equiv {\textstyle\sum_{j,j'} (T_j^2/F_j)\delta_{j j'} \mathsf{q}_j^*\otimes\mathsf{q}_{j'}^*}
.
\end{align}
\end{subequations}
From (\ref{eq:fj}), $\hat{\mathsf{T}}_e=\hat{\mathsf{T}}_l^{-1}$ only when edges and links are orthogonal.  As Hodge stars are defined over the reduced networks $\hat{\mathcal{N}}$ and $\hat{\mathcal{N}}^\rhd$, the hats on $\hat{\mathsf{E}}$, $\hat{\mathsf{T}}_e$ and $\hat{\mathsf{T}}_l$ denote exclusion of peripheral vertices, edges and links respectively in the sums in (\ref{eq:mops}).

Recalling (\ref{eq:hodef}), the operator $\star_{1,0}:\Omega_1^0(\hat{\mathcal{N}})\rightarrow \Omega_1^2(\hat{\mathcal{N}}^\rhd)$ is defined, for a $\mathcal{P}$-valued 0-cochain $\phi$, by 
\begin{equation}
\label{eq:star10}
\star_{1,0} \phi\equiv 
{\textstyle \sum_k} E_k  \mathsf{q}_k^*\boldsymbol{\epsilon}_{\mathcal{P}}\phi_k =
{\textstyle \sum_k} E_k  \mathsf{q}_k^*\{-\phi_k^\perp, \phi_k^\parallel \}^\top. 
\end{equation}
A rotation in $\mathcal{P}$-space is included to accommodate the underlying transition from edges on $\hat{\mathcal{N}}$ to links on $\hat{\mathcal{N}}^\rhd$.  
Likewise $\star_{1,0}^\rhd:\Omega_1^0(\hat{\mathcal{N}}^\rhd)\rightarrow \Omega_1^2(\hat{\mathcal{N}})$ is defined for a $\mathcal{P}$-valued 0-cochain $\Phi$ by $\star_{1,0}^\rhd \Phi\equiv 
\sum_i A_i  \mathsf{q}_i^*\boldsymbol{\epsilon}_{\mathcal{P}}\Phi_i$.  We distinguish $\star_{1,1}:\Omega_1^1(\hat{\mathcal{N}})\rightarrow\Omega_1^1(\hat{\mathcal{N}}^\rhd)$, defined by 
\begin{equation}
\label{eq:star1}
\star_{1,1} \psi\equiv 
{\textstyle\sum_j}  \mathsf{q}_j^* (F_j/t_j^2)\boldsymbol{\epsilon}_{\mathcal{P}}\psi_j,    
\end{equation}
when using the bases $\mathbf{e}_j$ in (\ref{eq:cocobases}a), from $\star_{1,1}^\rhd:\Omega_1^1(\hat{\mathcal{N}}^\rhd)\rightarrow\Omega_1^1(\hat{\mathcal{N}})$, defined by $\star_{1,1}^\rhd \Psi\equiv 
\sum_j  \mathsf{q}_j^* (F_j/T_j^2)\boldsymbol{\epsilon}_{\mathcal{P}}\Psi_j$, when using the bases $\mathbf{E}_j$ in (\ref{eq:cocobases}b).  When acting on scalars defined over edges or links, $\star_{0,1}$, $\star_{2,1}$ and $\star_{0,1}^\rhd$, $\star_{2,1}^\rhd$ lack the rotation $\boldsymbol{\epsilon}_{\mathcal{P}}$ (Table~\ref{tab:maps}).

The analogue of $\star_{1,0}$ in (\ref{eq:star10}) acting on scalar-valued 0-cochains, $\star_{0,0}:\Omega_0^0(\hat{\mathcal{N}})\rightarrow \Omega_2^2(\hat{\mathcal{N}}^\rhd)$, identifies internal triangle areas via
\begin{equation}
    \star_{0,0}\hat{\mathsf{1}}_v^\flat={\textstyle\sum_k} E_k\mathsf{q}_k^*,
\end{equation}
where $\hat{\mathsf{1}}_v\in \hat{\mathcal{V}}$ is the chain identifying all vertices of $\hat{\mathcal{N}}$.  The equivalent `top form' for cells is given by $\star_{0,0}^\rhd:\Omega_0^0(\hat{\mathcal{N}}^\rhd)\rightarrow \Omega_2^2(\hat{\mathcal{N}})$ via $\star_{0,0}^\rhd\mathsf{1}_c^\flat={\textstyle\sum_i} A_i\mathsf{q}_i^*$, with $\mathsf{1}_c\in\mathcal{C}$.

We define the interior product $\iota$ in Appendix~\ref{app:interior} and illustrate how standard scalar and vector products can be recovered, exploiting a duality between $\iota$ and $\wedge$ that is mediated by $\star$.

\subsubsection{Inner products}

We can finally define inner products on $\Gamma(T\mathcal{M}_{\mathcal{E}})$ and $\Gamma(T\mathcal{M}_{\mathcal{L}})$ respectively, by combining the natural pairing with the wedge product and Hodge star as 
\begin{subequations}
\label{eq:inner1}
\begin{align}
[\mathsfbf{v},\mathsfbf{w}]_{\hat{\mathcal{E}}}\defeq & \langle \mathsfbf{v}^\flat \wedge \star \mathsfbf{w}^\flat \vert \hat{\mathsf{1}}_e\rangle={\textstyle\sum_j F_j (\mathbf{v}_j\cdot\mathbf{w}_j}), \\
[\mathsfbf{V},\mathsfbf{W}]_{\hat{\mathcal{L}}}\defeq & \langle \mathsfbf{V}^\flat \wedge \star \mathsfbf{W}^\flat \vert \hat{\mathsf{1}}_l\rangle={\textstyle\sum_j F_j (\mathbf{V}_j\cdot\mathbf{W}_j}),
\end{align}
\end{subequations}
where the chain $\hat{\mathsf{1}}_e\in \hat{\mathcal{E}}$ [$\hat{\mathsf{1}}_l\in\hat{\mathcal{L}}$] identifies every edge in $\hat{\mathcal{N}}$ [link in $\hat{\mathcal{N}}^\rhd$].  Clearly $[\mathsfbf{w},\mathsfbf{w}]_{\hat{\mathcal{E}}}=\sum_j F_j \vert\mathbf{w}_j\vert^2\geq 0$ for any $\mathsfbf{w}\in \Gamma(T\mathcal{M}_{\mathcal{E}})$; similarly, $[\mathsfbf{W},\mathsfbf{W}]_{\hat{\mathcal{L}}}\geq 0$ for any $\mathsfbf{W}\in \Gamma(T\mathcal{M}_{\mathcal{L}})$.  The half-weights $\tfrac{1}{2}F_j$ are illustrated in Fig.~\ref{fig:Figure2constructors}(d).  Using (\ref{eq:star10}), we also define an inner product on $\hat{\mathcal{V}}\times\mathcal{P}$ as
\begin{subequations}
    \begin{align}
\label{eq:inner0}
    [\mathsf{f},\mathsf{g}]_{\hat{\mathcal{V}}}& \defeq \langle \mathsf{f}^\flat \wedge \star\mathsf{g}^\flat \vert \hat{\mathsf{1}}_v\rangle =\langle \{\mathsf{f}^\parallel,\mathsf{f}^\perp\}^\top\wedge \star_{1,0} \{\mathsf{g}^\parallel,\mathsf{g}^\perp\}^\top\vert \hat{\mathsf{1}}_v \rangle \nonumber \\
    &  ={\textstyle \sum_k} \{f_k^\parallel, f_k^\perp\}^\top \wedge \{-g_k^\perp,g_k^\parallel\}^\top E_k ={\textstyle\sum_k} (f_k^\parallel g_k^\parallel+f_k^\perp g_k^\perp)E_k.
\end{align}
The weights $E_k$ are illustrated in Fig.~\ref{fig:Figure2constructors}(c). Again it is clear that $[\mathsf{f},\mathsf{f}]_{\hat{\mathcal{V}}}=\sum_k [(f_k^{\parallel})^2+(f_k^\perp)^2]E_k\geq 0$ for any $\mathsf{f}\in \hat{\mathcal{V}}\times\mathcal{P}$.  Likewise we define an inner product on $\mathcal{C}\times\mathcal{P}$ as
\begin{equation}
\label{eq:inner2}
    [\mathsf{u},\mathsf{v}]_{\mathcal{C}}\defeq \langle \mathsf{u}^\flat \wedge \star_{1,0}^\rhd \mathsf{v}^\flat \vert \mathsf{1}_c \rangle ={\textstyle \sum_i A_i (u_i^\parallel v_i^\parallel + u_i^\perp v_i^\perp)},
\end{equation}
\end{subequations}
where the chain $\mathsf{1}_c\in\mathcal{C}$ identifies every cell centre in $\hat{\mathcal{N}}^\rhd$.  This ensures that $[\mathsf{u},\mathsf{u}]_{\mathcal{C}}\geq 0$ for any $\mathsf{u}\in \mathcal{C}$.  The weights $A_i$ are illustrated in Fig.~\ref{fig:Figure2constructors}(a).  The inner products (\ref{eq:inner1})--(\ref{eq:inner2}) match those used in \cite{jensen2022} to define discrete derivatives, except that boundary conditions are here specified more precisely through the use of the reduced networks $\hat{\mathcal{N}}$ and $\hat{\mathcal{N}}^\rhd$.

\subsection{Differential operators}
\label{sec:do}

Armed with $\mathrm{d}$, $\star$, $\sharp$ and $\flat$, we can now define differential operators, illustrated schematically in Fig.~\ref{fig:Figure3extderivsW}, using standard definitions within DEC.  Details are given in Appendix~\ref{app:evalop}.  Briefly, for $\phi\equiv \{\phi^\parallel,\phi^\perp\}^\top\in \Omega_1^0(\hat{\mathcal{N}})$, we evaluate grad as $(\mathrm{d}\phi)^\sharp$.  This is a vector in $\Gamma(T\mathcal{M}_{\mathcal{E}})$, and is the vector sum of $\mathrm{grad}^v\phi^\parallel$ parallel to cell edges and $\mathrm{grad}^v\phi^\perp$ orthogonal to edges; see (\ref{eq:grad}a).  For $\mathsfbf{b}\in\Gamma(T\mathcal{M}_{\mathcal{E}})$, we evaluate $\mathrm{curl}\,\mathsfbf{b}$ using $(\star\mathrm{d}\mathsfbf{b}^\flat)^\sharp$.   Its two scalar components in $\mathcal{C}\times\mathcal{P}$ are $\mathrm{curl}^c\,\mathsfbf{b}$ (a circulation around cells) and $\mathrm{cocurl}^c\,\mathsfbf{b}$ (interpretable as a divergence); see (\ref{eq:curl}a).  Equivalent primary operators are defined in Appendix~\ref{app:primary} for fields defined over the dual network.  Then, exploiting the inner products (\ref{eq:inner1}-\ref{eq:inner2}), we define the adjoint (codifferential) operators $-\mathrm{div}$ and rot over each network using 
\begin{subequations}
\label{eq:adj}
    \begin{align}
    [\mathsfbf{v},\mathrm{grad}\,\phi]_{\hat{\mathcal{E}}}&=[(-\mathrm{div}\,\mathsfbf{v})^\sharp,\phi^\sharp]_{\hat{\mathcal{V}}}, &
    [\mathsfbf{V},\mathrm{grad}\,\Phi]_{\hat{\mathcal{L}}}&=[(-\mathrm{div}\,\mathsfbf{V})^\sharp,\Phi^\sharp]_{\mathcal{C}}, \\
        [\mathsf{u},\mathrm{curl}\,\mathsfbf{v}]_{\mathcal{C}}&=[\mathrm{rot}\,\mathsf{u},\mathsfbf{v}]_{\hat{\mathcal{E}}},&
        [\mathsf{U},\mathrm{curl}\,\mathsfbf{V}]_{\hat{\mathcal{V}}}&=[\mathrm{rot}\,\mathsf{U},\mathsfbf{V}]_{\hat{\mathcal{L}}},
\end{align}
\end{subequations}
for any $\mathsfbf{v}\in\Gamma(T\mathcal{M}_{\mathcal{E}})$, $\mathsfbf{V}\in\Gamma(T\mathcal{M}_{\mathcal{L}})$, $\phi\in \Omega_1^0(\hat{\mathcal{N}})$, $\Phi\in \Omega_1^0(\mathcal{N^\rhd})$, $\mathsf{u}\in\mathcal{C}\times\mathcal{P}$ and $\mathsf{U}\in\hat{\mathcal{V}}\times\mathcal{P}$.  By employing $\hat{\mathsf{A}}$ and $\hat{\mathsf{B}}$, as defined in (\ref{eq:abred}), over networks $\hat{\mathcal{N}}$ (lacking peripheral vertices) and $\hat{\mathcal{N}}^\rhd$ (lacking peripheral links), we impose effective Dirichlet boundary conditions on $\phi$, $\mathsf{U}$ and effective Neumann conditions on $\mathsf{u}$ and $\Phi$, avoiding the requirement to evaluate $-\mathrm{div}\,\mathsfbf{v}$ and $\mathrm{curl}\,\mathsfbf{V}$ at peripheral vertices, and  avoiding boundary contributions in (\ref{eq:adj}).  Thus for $\mathsfbf{b}\in\Gamma(T\mathcal{M}_{\mathcal{E}})$, we evaluate $-\mathrm{div}\,\mathsfbf{b}$ (adjoint to grad) as $\star\,\mathrm{d}\star\mathsfbf{b}^\flat$; its two scalar components (in $\Omega_1^0(\hat{\mathcal{N}})$) are interpretable as a divergence ($-\mathrm{codiv}^v\,\mathsfbf{b}$) and a circulation around cells ($-\mathrm{div}^v\,\mathsfbf{b}$); see (\ref{eq:div}a).  Likewise rot, adjoint to curl, maps an element $\mathsf{f}=\{\mathsf{f}^\parallel,\mathsf{f}^\perp\}^\top\in \mathcal{C}\times\mathcal{P}$ to vectors in $\Gamma(T\mathcal{M}_{\mathcal{E}})$ in the form $(\star\,\mathrm{d}\,\mathsf{f}^\flat)^\sharp$, yielding a sum of vectors parallel to edges ($\mathrm{rot}^c\,\mathsf{f}^\perp$) and orthogonal to edges ($\mathrm{corot}^c\,\mathsf{f}^\parallel)$; see (\ref{eq:rot}a).  Equivalent operators are derived also for the dual network in Appendix~\ref{app:derived}.

\begin{figure}
\centering
\includegraphics[width=0.8\textwidth]{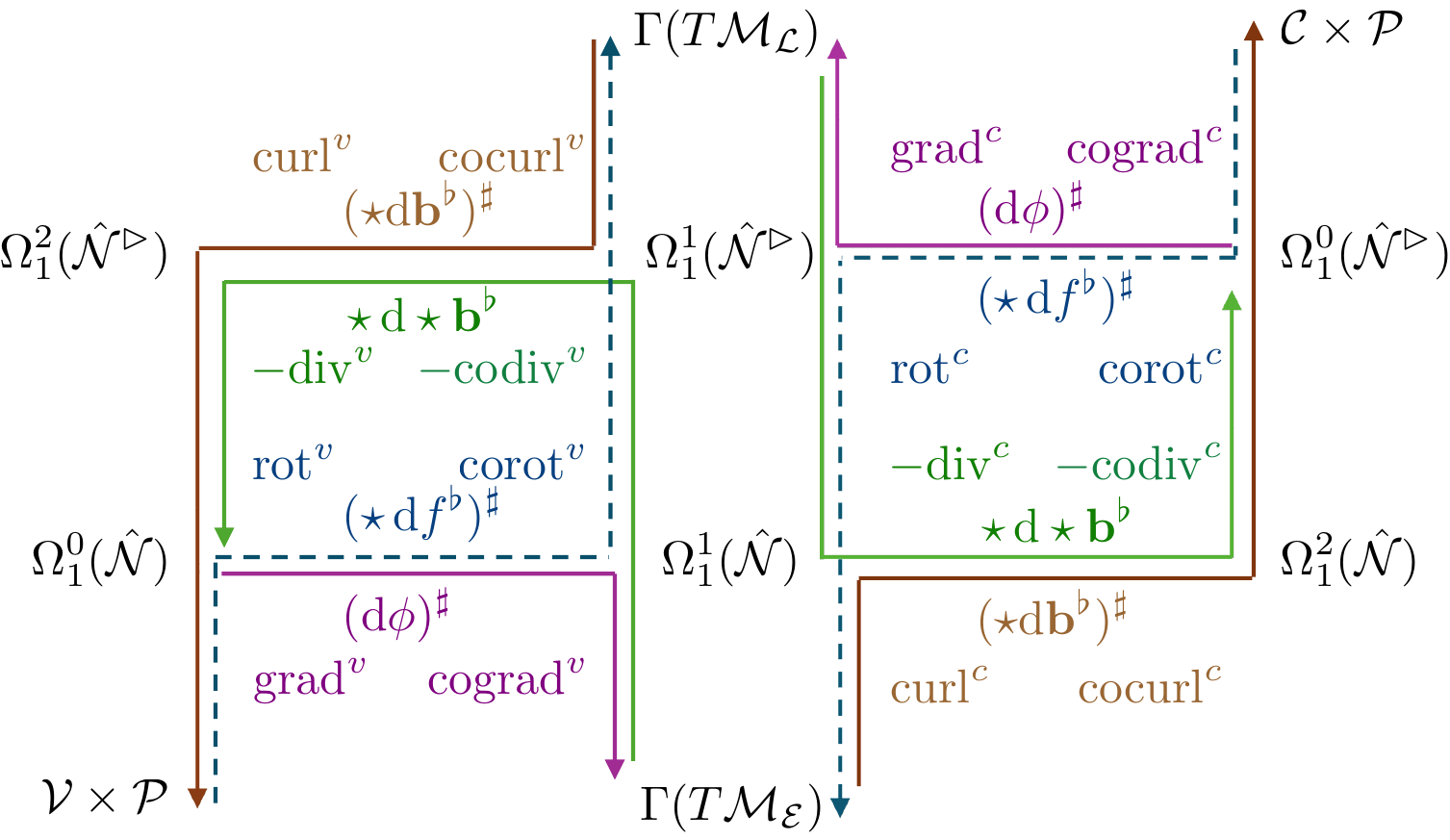}
    \caption{DEC-inspired operators constructed from the maps shown in (\ref{eq:dr}). Operators acting on vectors parallel to edges and links are paired with those perpendicular to edges and links, to form grad (\ref{eq:grad}), curl (\ref{eq:curl}), $-$div (\ref{eq:div}) and rot (\ref{eq:rot}) defined over the primal and dual networks.  Coloured arrows show the corresponding maps between spaces.  Labels of the same colour show the compact DEC representation plus the related operator components, listed below in Table~\ref{tab:diffOpDef}. Operators on the left-hand-side [right-hand-side] of the diagram have superscript $v$ [$c$], with $\mathrm{d}$ being $\mathsf{A}_1^*$ or $\mathsf{A}_1^{*\top}$ [$\mathsf{B}_1^*$ or $\mathsf{B}_1^{*\top}$].   $\mathrm{grad}$ is magenta; $\mathrm{curl}$ is brown.  The corresponding codifferentials $-{\mathrm{div}}$ and ${\mathrm{rot}}$ are green and blue (dashed) respectively.}
    \label{fig:Figure3extderivsW}
\end{figure}

We define operators in general terms in Appendix~\ref{app:evalop}, summarising them in Table~\ref{tab:diffOpDef}.  This construction confirms that the operators introduced in \cite{jensen2022} can be expressed within a DEC-like framework (some notational changes are summarised in Table~\ref{tab:diffOpDef}) and that the 16 core operators (Fig.~\ref{fig:Figure3extderivsW}) sit in four classes: four gradients of the form $(\mathrm{d}\phi)^\sharp$; four curls of the form $(\star\mathrm{d}\mathbf{b}^\flat)^\sharp$; four rots of the form $(\star\mathrm{d}\mathsf{f}^\flat)^\sharp$; and four divergences of the form $\star\,\mathrm{d}\,\star\mathbf{b}^\flat$.

\subsection{Laplacians}
\label{sec:lap}

The relationship (\ref{eq:adj}) allows us to construct positive-definite Laplacian operators $-\mathrm{div}\,\circ\,\mathrm{grad}$ and $\mathrm{curl}\,\circ\,\mathrm{rot}$ satisfying
\begin{subequations}
\begin{align}
[(-\mathrm{div}\,\circ\,\mathrm{grad}\,\phi)^\sharp,\phi^\sharp]_{\hat{\mathcal{V}}}&=[\mathrm{grad}\,\phi,\mathrm{grad}\,\phi]_{\hat{\mathcal{E}}}\geq 0, \\
[(-\mathrm{div}\,\circ\,\mathrm{grad}\,\Phi)^\sharp,\Phi^\sharp]_{\mathcal{C}}&=[\mathrm{grad}\,\Phi,\mathrm{grad}\,\Phi]_{\hat{\mathcal{L}}}\geq 0, \\
[\mathrm{curl}\,\circ\,\mathrm{rot}\,\mathsf{u},\mathsf{u}]_{\mathcal{C}}&=[\mathrm{rot}\,\mathsf{u},\mathrm{rot}\,\mathsf{u}]_{\hat{\mathcal{E}}}\geq 0, \\
[\mathrm{curl}\,\circ\,\mathrm{rot}\,\mathsf{U},\mathsf{U}]_{\hat{\mathcal{V}}}&=[\mathrm{rot}\,\mathsf{U},\mathrm{rot}\,\mathsf{U}]_{\hat{\mathcal{L}}}\geq 0, 
\end{align}
\end{subequations}
for any $\phi\in\Omega_1^0(\hat{\mathcal{N}})$, $\Phi\in\Omega_1^0(\hat{\mathcal{N}}^\rhd)$, $\mathsf{u}\in \mathcal{C}\times\mathcal{P}$, $\mathsf{U}\in\hat{\mathcal{V}}\times\mathcal{P}$.  The four scalar Laplacians, acting on $\parallel$ and $\perp$ components of $\phi\in\Omega_1^0(\hat{\mathcal{N}})$, $\Phi\in \Omega_1^0(\hat{\mathcal{N}}^\rhd)$, $\mathsf{U}\in \hat{\mathcal{V}}\times\mathcal{P}$, $\mathsf{u}\in \mathcal{C}\times\mathcal{P}$ are, making use of (\ref{eq:h012}),
\begin{subequations}
\label{eq:laps}
\begin{align}
 -{\mathrm{div}} \circ \mathrm{grad}\, \phi & =\star\, \mathrm{d}\star \mathrm{d}\phi =\star_{1,0}^{-1} \, \mathsf{A}_1^{*\top}\star_{1,1}\mathsf{A}_1^*\phi { ~=\mathsf{L}_{\mathcal{V}}\otimes\mathsf{I}_{\mathcal{P}}\phi},\\
 -{\mathrm{div}} \circ \mathrm{grad}\, \Phi & = \star\, \mathrm{d}\star \mathrm{d}\Phi =(\star_{1,0}^\rhd)^{-1} \, \mathsf{B}_1^{*}\star_{1,1}^\rhd\mathsf{B}_1^{*\top} \Phi{ ~=\mathsf{L}_{\mathcal{C}}\otimes\mathsf{I}_{\mathcal{P}}\Phi},  \\
 ({\mathrm{curl}} \circ {\mathrm{rot}} \,\mathsf{u})^\flat& =\star\, \mathrm{d}\star \mathrm{d} \mathsf{u}^{\flat}  =\star_{1,2} \, \mathsf{B}_1^{*}\star_{1,1}^{-1} \mathsf{B}_1^{*\top} \mathsf{u}^{\flat}{ ~=\mathsf{L}_{\mathcal{F}}\otimes\mathsf{I}_{\mathcal{P}}\mathsf{u}}, \\
  ({\mathrm{curl}} \circ {\mathrm{rot}} \,\mathsf{U})^\flat& = \star\, \mathrm{d}\star \mathrm{d} \mathsf{U}^{\flat} =\star_{1,2}^\rhd \, \mathsf{A}_1^{*\top}(\star_{1,1}^\rhd)^{-1} \mathsf{A}_1^{*} \mathsf{U}^{\flat}{ ~=\mathsf{L}_{\mathcal{T}}\otimes\mathsf{I}_{\mathcal{P}} \mathsf{U}},
\end{align}
\end{subequations}
where, using (\ref{eq:mops}), we recover matrix operators introduced in \cite{jensen2022}
\begin{subequations}
\label{eq:lvctf}
\begin{align}
  \mathsf{L}_\mathcal{V}&\equiv {\textstyle \sum_{k,k'}}\{\hat{\mathsf{E}}^{-1}\hat{\mathsf{A}}^\top\hat{\mathsf{T}}_e^{-1}\hat{\mathsf{A}}\}_{k,k'} \mathsf{q}_k^*\otimes \mathsf{q}_{k'}^* ,
 & \mathsf{L}_\mathcal{C}&\equiv {\textstyle \sum_{i,i'}}\{\mathsf{H}^{-1}\hat{\mathsf{B}}\hat{\mathsf{T}}_l^{-1}\hat{\mathsf{B}}^\top\}_{i,i'} \mathsf{q}_i^*\otimes \mathsf{q}_{i'}^*, \\
   \mathsf{L}_\mathcal{T}&\equiv {\textstyle \sum_{k,k'}}\{\hat{\mathsf{E}}^{-1}\hat{\mathsf{A}}^\top\hat{\mathsf{T}}_l\hat{\mathsf{A}} \}_{k,k'} \mathsf{q}_k\otimes \mathsf{q}_{k'}
 & \mathsf{L}_\mathcal{F} &\equiv {\textstyle \sum_{i,i'}}\{\mathsf{H}^{-1}\hat{\mathsf{B}}\hat{\mathsf{T}}_e\hat{\mathsf{B}}^\top\}_{i,i'} \mathsf{q}_i\otimes \mathsf{q}_{i'}.
\end{align}
\end{subequations}
Each of $\mathsf{L}_{\mathcal{C}}$ and $\mathsf{L}_{\mathcal{F}}$ has a zero eigenvalue with eigenvector $\mathsf{1}_c$.  

Laplacians defined on edges or links via $-\mathrm{grad}\,\circ\,\mathrm{div}\,+ \mathrm{rot}\,\circ\,\mathrm{curl}\,$ take the form $(\mathsfbf{L}_{\mathcal{E}} \mathsfbf{b}^\flat)^\sharp$ and $(\mathsfbf{L}_{\mathcal{L}} \mathsfbf{B}^\flat)^\sharp$, where $\mathsfbf{L}_{\mathcal{E}}:\Omega_1^1(\hat{\mathcal{N}})\rightarrow \Omega_1^1(\hat{\mathcal{N}})$ and $\mathsfbf{L}_{\mathcal{L}}:\Omega_1^1(\hat{\mathcal{N}}^\rhd)\rightarrow \Omega_1^1(\hat{\mathcal{N}}^\rhd)$ are
\begin{subequations}
\label{eq:veclaps}
    \begin{align}
\mathsfbf{L}_{\mathcal{E}}&=    \mathsf{A}_1^* \star_{1,0}^{-1} \mathsf{A}_1^{*\top}\star_{1,1}
+\star_{1,1}^{-1}\mathsf{B}_1^{*\top}\star_{1,2}\mathsf{B}_1^*, \\
\mathsfbf{L}_{\mathcal{L}}&=    \mathsf{B}_1^{*\top} (\star_{1,0}^\rhd)^{-1} \mathsf{B}_1^{*}\star_{1,1}^\rhd
+(\star_{1,1}^\rhd)^{-1}\mathsf{A}_1^{*}\star_{1,2}^\rhd\mathsf{A}_1^{*\top},
\end{align}
again using (\ref{eq:h012}).  We can write the operators as
\begin{align}
    \mathsfbf{L}_{\mathcal{E}}&={\textstyle\sum_{j,j'}}\left\{\hat{\mathsf{A}} \hat{\mathsf{E}}^{-1}\hat{\mathsf{A}}^\top\hat{\mathsf{T}}_e^{-1}+\hat{\mathsf{T}}_e\hat{\mathsf{B}}^\top \mathsf{H}^{-1}\hat{\mathsf{B}}\right\}_{j,j'}\mathsf{q}_j^*\otimes\mathsf{q}_{j'}^*\otimes \mathsf{I}_{\mathcal{P}}, \\
    \mathsfbf{L}_{\mathcal{L}}&={\textstyle\sum_{j,j'}}\left\{\hat{\mathsf{B}}^\top \mathsf{H}^{-1}\hat{\mathsf{B}}\hat{\mathsf{T}}_l^{-1}+\hat{\mathsf{T}}_l\hat{\mathsf{A}} \hat{\mathsf{E}}^{-1}\hat{\mathsf{A}}^\top\right\}_{j,j'}\mathsf{q}_j^*\otimes\mathsf{q}_{j'}^*\otimes \mathsf{I}_{\mathcal{P}}.
\end{align}
\end{subequations}
These are self-adjoint, so that for $\mathsfbf{b}\in\Gamma(\mathcal{M}_\mathcal{E})$ and $\mathsfbf{v}\in\Gamma(\mathcal{M}_\mathcal{E})$,
\begin{align}
    [(\mathsfbf{L}_{\mathcal{E}} \mathsfbf{b}^\flat)^\sharp,\mathsfbf{v}]_{\hat{\mathcal{E}}}& =\langle (\mathsfbf{L}_{\mathcal{E}} \mathsfbf{b}^\flat) \wedge \star_{1,1} \mathsfbf{v}^\flat\vert\mathsf{1}_e\rangle\nonumber \\
    &={\textstyle \sum_{j,j'}} \mathsf{b}_j^\top \left\{\hat{\mathsf{T}}_e^{-1} \hat{\mathsf{A}}\hat{\mathsf{E}}^{-1}\hat{\mathsf{A}}^\top\hat{\mathsf{T}}_e^{-1} +\hat{\mathsf{B}}^\top\mathsf{H}^{-1}\hat{\mathsf{B}}\right\}_{jj'}\mathsf{v}_{j'}=[ \mathsfbf{b},(\mathsfbf{L}_{\mathcal{E}}\mathsfbf{v^\flat)^\sharp}]_{\hat{\mathcal{E}}}
\end{align}
and likewise $[(\mathsfbf{L}_{\mathcal{L}} \mathsfbf{B}^\flat)^\sharp,\mathsfbf{V}]_{\hat{\mathcal{L}}}=[ \mathsfbf{B},(\mathsfbf{L}_{\mathcal{L}}\mathsfbf{V^\flat)^\sharp}]_{\hat{\mathcal{L}}}$ for $\mathsfbf{B}\in\Gamma(\mathcal{M}_\mathcal{L})$ and $\mathsfbf{V}\in\Gamma(\mathcal{M}_\mathcal{L})$.  Furthermore
\begin{align}
\label{eq:ker}
    [(\mathsfbf{L}_{\mathcal{E}} \mathsfbf{b}^\flat)^\sharp,\mathsfbf{b}]_{\hat{\mathcal{E}}}&={\textstyle \sum_{j,j'}} {b}_j^{\parallel\top} \{\hat{\mathsf{T}}_e^{-1} \hat{\mathsf{A}}\hat{\mathsf{E}}^{-1}\hat{\mathsf{A}}^\top\hat{\mathsf{T}}_e^{-1}\}_{jj'}{b}^\parallel_{j'} +{\textstyle \sum_{j,j'}}{b}_j^{\parallel\top}\{\hat{\mathsf{B}}^\top\mathsf{H}^{-1}\hat{\mathsf{B}}\}_{jj'}{b}^\parallel_{j'} \nonumber \\
&\quad +{\textstyle \sum_{j,j'}} {b}_j^{\perp\top} \{\hat{\mathsf{T}}_e^{-1} \hat{\mathsf{A}}\hat{\mathsf{E}}^{-1}\hat{\mathsf{A}}^\top\hat{\mathsf{T}}_e^{-1}\}_{jj'}{b}^\perp_{j'} +{\textstyle \sum_{j,j'}}{b}_j^{\perp\top}\{\hat{\mathsf{B}}^\top\mathsf{H}^{-1}\hat{\mathsf{B}}\}_{jj'}{b}^\perp_{j'} \nonumber \\
&=[(-\mathrm{div}\,\mathsfbf{b})^\sharp,(-\mathrm{div}\,\mathsfbf{b})^\sharp]_{\hat{\mathcal{V}}} +[\mathrm{curl}\,\mathsfbf{b},\mathrm{curl}\,\mathsfbf{b}]_{\hat{\mathcal{E}}}\geq 0,
\end{align}
using (\ref{eq:magcurl}) and (\ref{eq:magdiv}).
Both sums in (\ref{eq:ker}) are non-negative.  Thus $\mathsfbf{b}^\flat\in\mathrm{ker}(\mathsfbf{L}_{\mathcal{E}})$ implies that $\mathsfbf{b}\in\mathrm{ker}(\mathrm{div})$ and $\mathsfbf{b}^\flat\in\mathrm{ker}(\mathrm{curl})$.  In Sec.~\ref{sec:ablation}, we will show how nontrivial solutions of $\mathsfbf{L}_{\mathcal{E}}\mathsfbf{b}^\flat=\mathsfbf{0}$ can arise for a monolayer containing one or more holes.

\subsection{Helmholtz--Hodge decomposition}
\label{sec:hh}

For $\mathsfbf{v}\in \Gamma(T\mathcal{M}_{\mathcal{E}})$, Helmholtz--Hodge decomposition \cite{lim2019} (see (\ref{eq:hh})) suggests that there exists $\phi\in \Omega_1^0(\hat{\mathcal{N}})$, $\mathsf{u}\in \mathcal{C}\times\mathcal{P}$ and a harmonic field $\mathsfbf{x}\in \Gamma(T\mathcal{M}_{\mathcal{E}})$ such that 
\begin{subequations}
    \label{eq:hhp}
    \begin{align}
    \mathsfbf{v}&=\mathrm{grad}\,\phi+\mathrm{rot}\,\mathsf{u}+\mathsfbf{x} = (\mathsf{A}_1^* \phi)^\sharp +(\star_{1,1}^{-1}\mathsf{B}_1^{*\top}\mathsf{u}^\flat)^\sharp+\mathsfbf{x},
\end{align}
where
\begin{align}
    \mathsfbf{L}_{\mathcal{E}}\,\mathsfbf{x}^\flat&=\mathsf{0}, \quad
    -\mathrm{div}\,\mathsfbf{v}=\star_{1,0}^{-1}
\mathsf{A}_1^{*\top}\star_{1,1} \mathsf{A}_1^* \phi, \quad (\mathrm{curl}\,\mathsfbf{v})^\flat=\star_{1,2}\mathsf{B}_1^*\star_{1,1}^{-1}\mathsf{B}_1^{*\top} \mathsf{u}^\flat.
\end{align}
\end{subequations}
The operators in (\ref{eq:hhp}) are defined in (\ref{eq:grad}a), (\ref{eq:rot}a), (\ref{eq:div}a) and (\ref{eq:curl}a).  The Poisson problems in (\ref{eq:hhp}b) decompose into 
\begin{equation}
\label{eq:poip}
    \left\{-\mathrm{div}^v\,\mathsfbf{v} , -\mathrm{codiv}^v\,\mathsfbf{v}\right\}^\top=\mathsf{L}_{\mathcal{V}}\left\{\phi^\parallel,\phi^\perp\right\}^\top,\quad    \left\{\mathrm{cocurl}^c\,\mathsfbf{v} , \mathrm{curl}^c\,\mathsfbf{v}\right\}^\top=\mathsf{L}_{\mathcal{F}}\left\{\mathsf{u}^\parallel,\mathsf{u}^\perp\right\}^\top.
\end{equation}
By using the reduced network $\hat{\mathcal{N}}$, we effectively impose $\phi=\{0,0\}^\top$ at peripheral vertices in (\ref{eq:poip}a); the solvability condition on (\ref{eq:poip}b) is 
\begin{equation}
\label{eq:solvp}
    [\{\mathsf{1}_c,\mathsf{1}_c\}^\top,\mathrm{curl}\,\mathsfbf{v}]_\mathcal{C}=0, 
\end{equation}
using (\ref{eq:inner2}).  The harmonic field in (\ref{eq:hhp}a) has individual harmonic components $\mathsfbf{x}^\flat=\{\mathsf{x}^\parallel, \mathsf{x}^\perp\}^\top$.  These scalar fields both satisfy
\begin{equation}
\label{eq:scled}
    \mathsf{L}_{\mathcal{E}}\mathsf{x}^\parallel=\mathsf{0}, \quad
    \mathsf{L}_{\mathcal{E}}\mathsf{x}^\perp=\mathsf{0}, \quad 
    \mathsf{L}_{\mathcal{E}}\equiv {\textstyle\sum_{j,j'}}\{\hat{\mathsf{A}} \hat{\mathsf{E}}^{-1}\hat{\mathsf{A}}^\top\hat{\mathsf{T}}_e^{-1}+\hat{\mathsf{T}}_e\hat{\mathsf{B}}^\top \mathsf{H}^{-1}\hat{\mathsf{B}}\}_{j,j'}\mathsf{q}_j^*\otimes \mathsf{q}_{j'}^*.
\end{equation}
For a monolayer with $n_h$ holes, there exist $n_h$ eigenmodes $\mathsf{w}^{(m)}$, $m=1,2,\dots,n_h$ satisfying $\mathsf{L}_{\mathcal{E}}\mathsf{w}^{(m)}=\mathsf{0}$, forming (using the interior product (\ref{eq:interior})) the $m$th mode $\mathsfbf{x}^{(m)\flat}=\iota_{\mathsfbf{z}^{(m)}} \mathsf{w}^{(m)}$ for some field $\mathsfbf{z}^{(m)\flat}\equiv \{z^{\parallel(m)}, z^{\perp(m)} \}^\top \mathsf{1}_e$.  $\mathsfbf{z}^{(m)\flat}$ is uniform across all edges, but $\mathsfbf{z}^{(m)}$ is non-uniform in physical space across $\Gamma(T\mathcal{M}_{\mathcal{E}})$, after projection onto the contravariant edge vectors. The corresponding vector field is the linear combination 
\begin{equation}
    \mathsfbf{x}^{(m)}=\left(\iota_{\mathsfbf{z}^{(m)}} \mathsf{w}^{(m)}\right)^\sharp=z^{\parallel(m)} \left({\textstyle{\sum_j}}\mathsf{q}_j w_{j}^{(m)} \mathbf{e}_{j\parallel}\right)+z^{\perp(m)}\left({\textstyle{\sum_j}}\mathsf{q}_j w_{j}^{(m)} \mathbf{e}_{j\perp} \right)
    \label{eq:edgex}
\end{equation}
for $m=1,\dots,n_h$, with overall magnitude   
\begin{equation}
    [\mathsfbf{x}^{(m)},\mathsfbf{x}^{(m)}]_{\hat{\mathcal{E}}}=\left[(z^{\parallel(m)})^2+(z^{\perp(m)})^2\right] {\textstyle \sum_j} F_j \left(w_j^{(m)}/t_j\right)^2.
\end{equation}
From (\ref{eq:ker}), each field has zero divergence and zero curl, \hbox{i.e.}
\begin{equation}
\label{eq:zerodivcurl}
    \hat{\mathsf{E}}^{-1}\hat{\mathsf{A}}^\top \hat{\mathsf{T}}_e^{-1} \mathsf{w}^{(m)}\{z^{\parallel(m)},z^{\perp(m)}\}^\top=\{\mathsf{0},\mathsf{0}\}^\top,\quad  \mathsf{H}^{-1}\hat{\mathsf{B}}\mathsf{w}^{(m)} \{z^{\perp(m)},-z^{\parallel(m)}\}^\top=\{\mathsf{0},\mathsf{0}\}^\top.
\end{equation}
In other words, to find a harmonic vector field $\mathsfbf{x}^\flat\in\Omega_1^1(\hat{\mathcal{N}})$ in the kernel of $\mathsfbf{L}_{\mathcal{E}}$ in (\ref{eq:hhp}b), we solve two scalar problems in (\ref{eq:scled}), for which $\mathsf{L}_{\mathcal{E}}$ acts on the $\parallel$ and $\perp$ components of $\mathsfbf{x}^\flat$.    $z^{\parallel(m)}$ and $z^{\perp(m)}$ are the arbitrary amplitudes assigned to the two instances of the $m$th eigenmode of $\mathsf{L}_\mathcal{E}$.  Once $\mathsfbf{x}^{(m)}$ is reconstructed in (\ref{eq:edgex}), we see how changes in $z^{\parallel(m)}$ and $z^{\perp(m)}$ can reorient the field $\mathsfbf{x}^{(m)}$ uniformly across the whole monolayer.

Likewise for $\mathsfbf{V}\in \Gamma(T\mathcal{M}_{\mathcal{L}})$, there exists $\Phi\in \Omega_1^0(\hat{\mathcal{N}}^\rhd)$, $\mathsf{U}\in \hat{\mathcal{V}}\times\mathcal{P}$ and a harmonic field $\mathsfbf{X}$ such that 
\begin{subequations}
\label{eq:hhd}
\begin{equation}
    \mathsfbf{V}=\mathrm{grad}\,\Phi+\mathrm{rot}\,\mathsf{U}+\mathsfbf{X}
\end{equation}
where
\begin{align}
    \mathsfbf{L}_{\mathcal{L}}\,\mathsfbf{X}^\flat&=\mathsf{0}, \quad
    -\mathrm{div}\,\mathsfbf{V}=(\star_{1,0}^\rhd)^{-1}
\mathsf{B}_1^{*}\star_{1,1}^\rhd \mathsf{B}_1^{*\top} \Phi, \quad (\mathrm{curl}\,\mathsfbf{V})^\flat=\star_{1,2}^{\rhd}\mathsf{A}_1^{*\top}(\star_{1,1}^\rhd)^{-1}\mathsf{A}_1^{*} \mathsf{U}^\flat
\end{align}
\end{subequations}
giving
\begin{equation}
\label{eq:poid}
    \left\{-\mathrm{div}^c\,\mathsfbf{V} , -\mathrm{codiv}^c\,\mathsfbf{V}\right\}^\top=\mathsf{L}_{\mathcal{C}}\left\{\Phi^\parallel,\Phi^\perp\right\}^\top,\quad    \left\{\mathrm{cocurl}^v\,\mathsfbf{V} , \mathrm{curl}^v\,\mathsfbf{V}\right\}^\top=\mathsf{L}_{\mathcal{T}}\left\{\mathsf{U}^\parallel,\mathsf{U}^\perp\right\}^\top.
\end{equation}
The operators in (\ref{eq:hhd}) are defined in (\ref{eq:grad}b), (\ref{eq:rot}b), (\ref{eq:div}b) and (\ref{eq:curl}b).  Here, 
\begin{equation}
    \mathsf{L}_{\mathcal{L}}\mathsf{X}^\parallel=\mathsf{0}, \quad
    \mathsf{L}_{\mathcal{L}}\mathsf{X}^\perp=\mathsf{0}, \quad 
    \mathsf{L}_{\mathcal{L}}\equiv \hat{\mathsf{B}}^\top \mathsf{H}^{-1}\hat{\mathsf{B}}\hat{\mathsf{T}}_l^{-1}+\hat{\mathsf{T}}_l\hat{\mathsf{A}} \hat{\mathsf{E}}^{-1}\hat{\mathsf{A}}^\top.
\end{equation}
We effectively impose $\mathsf{U}=\{0,0\}^\top$ at peripheral vertices in (\ref{eq:poid}b); the solvability condition on (\ref{eq:poid}a) is 
\begin{equation}
\label{eq:solvd}
    [\{\mathsf{1}_c,\mathsf{1}_c\}^\top,(-\mathrm{div}\,\mathsfbf{V})^\sharp]_\mathcal{C}=0.
\end{equation}
For a monolayer with $n_h$ holes, $\mathsfbf{X}^{(m)\flat}=\iota_{\mathsfbf{Z}^{(m)}}\mathsf{W}^{(m)}$ for some uniform field $\mathsfbf{Z}^{(m)\flat}= \{Z^{\parallel(m)}, Z^{\perp(m)} \}^\top$, with $m=1,2,\dots, n_h$, where $\mathsf{W}^{(m)}$ is the $m$th eigenmode satisfying $\mathsf{L}_{\mathcal{E}}\mathsf{W}^{(m)}=\mathsf{0}$.  Thus 
\begin{equation}
    \mathsfbf{X}^{(m)}=Z^{\parallel(m)} \left({\textstyle{\sum_j}}\mathsf{q}_j W_{j}^{(m)} \mathbf{E}_{j\parallel}\right)+Z^{\perp(m)}\left({\textstyle{\sum_j}}\mathsf{q}_j W_{j}^{(m)} \mathbf{E}_{j\perp} \right).
    \label{eq:edgeX}
\end{equation}

\begin{table}[]
    \centering
    \begin{tabular}{|c|c|c|c|c|}
        \hline Potentials & HH operators & Basis & Laplacian & Forcing \\
        \hline 
        $\phi^\parallel$        & $\mathrm{grad}^v$     & $\mathbf{e}_{j\parallel}$ & $\mathsf{L}_\mathcal{V}$ & $-\mathrm{div}^v$ \\
        $\phi^\perp$            & $\mathrm{cograd}^v$      & $\mathbf{e}_{j\perp}$     & $\mathsf{L}_\mathcal{V}$ & $-\mathrm{codiv}^v$ \\
        $\mathsf{u}^\parallel$  & $\mathrm{corot}^c$    & $\mathbf{e}_{j\perp}$     & $\mathsf{L}_\mathcal{F}$ & $\mathrm{cocurl}^c$ \\
        $\mathsf{u}^\perp$      & $\mathrm{rot}^c$      & $\mathbf{e}_{j\parallel}$ & $\mathsf{L}_\mathcal{F}$ & $\mathrm{curl}^c$ \\
        \hline 
        $\Phi^\parallel$        & $\mathrm{grad}^c$     & $\mathbf{E}_{j\parallel}$ & $\mathsf{L}_\mathcal{C}$ & $-\mathrm{div}^c$ \\
        $\Phi^\perp$            & $\mathrm{cograd}^c$      & $\mathbf{E}_{j\perp}$     & $\mathsf{L}_\mathcal{C}$ & $-\mathrm{codiv}^c$ \\
        $\mathsf{U}^\parallel$  & $\mathrm{corot}^v$    & $\mathbf{E}_{j\perp}$     & $\mathsf{L}_\mathcal{T}$ & $\mathrm{cocurl}^v$ \\
        $\mathsf{U}^\perp$      & $\mathrm{rot}^v$      & $\mathbf{E}_{j\parallel}$ & $\mathsf{L}_\mathcal{T}$ & $\mathrm{curl}^v$ \\
        \hline
    \end{tabular}
    \caption{For the potentials listed in column 1, column 2 gives operators used in the Helmholtz--Hodge decomposition (\ref{eq:hhp}a, \ref{eq:hhd}a); column 3 gives the corresponding contravariant basis.  Components of associated Poisson problems (\ref{eq:poip}, \ref{eq:poid}) are indicated by columns 4 and 5.  Operators in columns 2 are adjoint to those in column 5.}
    \label{tab:opsum}
\end{table}

The operators in the Poisson problems (\ref{eq:poip}, \ref{eq:poid}) are summarised in Table~\ref{tab:opsum}.  To recap, for a domain containing a single hole, we expect a vector field defined on edges or links to be represented with up to five scalar fields with respect to the primal network ($\phi^\parallel$, $\phi^\perp$, $\mathsf{u}^\parallel$, $\mathsf{u}^\perp$ and $\mathsf{w}^{(1)}$), and five (similar) scalar fields ($\Phi^\parallel$, $\Phi^\perp$, $\mathsf{U}^\parallel$, $\mathsf{U}^\perp$, $\mathsf{W}^{(1)}$) with respect to the dual network.  Differences between the representations arise because of non-orthogonality of links and edges.  The relationships between the potentials and operators become clearer when considering the special case when edges and links are orthogonal.  From (\ref{eq:cocobases}, \ref{eq:fj}), this leads to $F_j=T_j t_j$ and exact alignment of $\mathbf{e}_{j\perp}$ with $\mathbf{E}_{j\parallel}$ and of $\mathbf{E}_{j\perp}$ with $\mathbf{e}_{j\parallel}$ (Fig.~\ref{fig:Figure1schematic}b), leading in turn to $\star_{1,1}=-(\star_{1,1}^\rhd)^{-1}$, $\hat{\mathsf{T}}_e=\hat{\mathsf{T}}_l^{-1}$ and
\begin{subequations}
    \label{eq:orthog}
\begin{align}
\mathrm{grad}^v&=-\mathrm{corot}^v, & \mathrm{cograd}^v&=\mathrm{rot}^v, & \mathrm{curl}^c&=-\mathrm{codiv}^c, & \mathrm{cocurl}^c&=\mathrm{div}^c, \\
\mathrm{grad}^c&=-\mathrm{corot}^c, & \mathrm{cograd}^c&=\mathrm{rot}^c, & \mathrm{curl}^v&=-\mathrm{codiv}^v, & \mathrm{cocurl}^v&=\mathrm{div}^v, \\
 \mathsf{L}_{\mathcal{V}}&=\mathsf{L}_{\mathcal{T}}, &\phi^\parallel&=-U^\parallel, & \phi^\perp&=U^\perp, & \mathsf{w}^{(1)}&=\mathsf{W}^{(1)},\\
 \mathsf{L}_{\mathcal{C}}&=\mathsf{L}_{\mathcal{F}}, &\Phi^\parallel&=-u^\parallel, & \Phi^\perp&=u^\perp, & \mathsf{L}_{\mathcal{E}}&=\hat{\mathsf{T}}_l^{-1}\mathsf{L}_{\mathcal{L}}\hat{\mathsf{T}}_l.
\end{align}
\end{subequations}
On a network lacking this symmetry, we can anticipate small differences between these operators and potentials.  It is therefore natural to identify four divergence operators (treating $-$cocurl as a form of $-\mathrm{div}$) generating four similar potentials ($\phi^\parallel$, $\Phi^\parallel$, $-u^\parallel$, $-U^\parallel$) and four curl operators (treating $-$codiv as a form of curl) generating four similar potentials ($\phi^\perp$, $\Phi^\perp$, $u^\perp$, $U^\perp$).  These differ in being defined on cells ($\phi$, $u$) or triangles ($\Phi$, $U$), and in being generated by edges and links ($\parallel$) or by rotated edges and rotated links ($\perp$).

\subsection{Application: a vertex model of an ablated monolayer}
\label{sec:vm}

We will apply Helmhotz--Hodge decomposition to a vector field emerging from an implementation of the vertex model that offers some useful biomechanical insight.  We give a statement of the vertex model in Appendix~\ref{app:vm}, using an adjointness relationship resembling (\ref{eq:adj}) to show how osmotic as well as mechanical effects can contribute to cell configurations.  Assuming a free energy of quadratic form, vertex evolution satisfies the force balance
\begin{equation}
\label{eq:vmvan}
    \dot{\mathsfbf{r}}=-\mathrm{grad}_A(\mathsf{A}-\mathsf{1}_c)-\Gamma \,\mathrm{grad}_{L}(\mathsf{L}-\tilde{L}_0\mathsf{1}_c),
\end{equation}
where cell areas $\mathsf{A}\equiv \sum_i A_i \mathsf{q}_i^*$ and perimeters $\mathsf{L}\equiv \sum_i L_i \mathsf{q}_i^*$ are both functions of $\mathsfbf{r}(t)=\sum_k\mathbf{r}_k(t)\mathsf{q}_k^*$ and a dot denotes a time derivative.  $\tilde{L}_0$ is a dimensionless preferred perimeter and the dimensionless parameter $\Gamma$ measures the relative energetic importance of perimeter to bulk effects in cells.  In contrast to standard approaches, (\ref{eq:gamma}) shows how $\Gamma$ can incorporate the energetic influence of two chemical species that occupy the bulk or the perimeter of cells, provided they diffuse between cells more rapidly than the cells change shape.  The operators $\mathrm{grad}_A$ and $\mathrm{grad}_L$ appearing in (\ref{eq:vmvan}), specified in Appendix~\ref{app:vm}, defined in \cite{cowley2024} and present implicitly in standard implementations of the vertex model \cite{farhadifar2007, hernandez2022, ANB2018a}, differ from the gradient operators presented so far because they map scalars defined on cells in $\Omega_0^0(\mathcal{N})$ to vectors on vertices in $\Gamma(T\mathcal{M}_\mathcal{V})$.  

\begin{figure}
    \centering
    \includegraphics[width=0.95\linewidth]{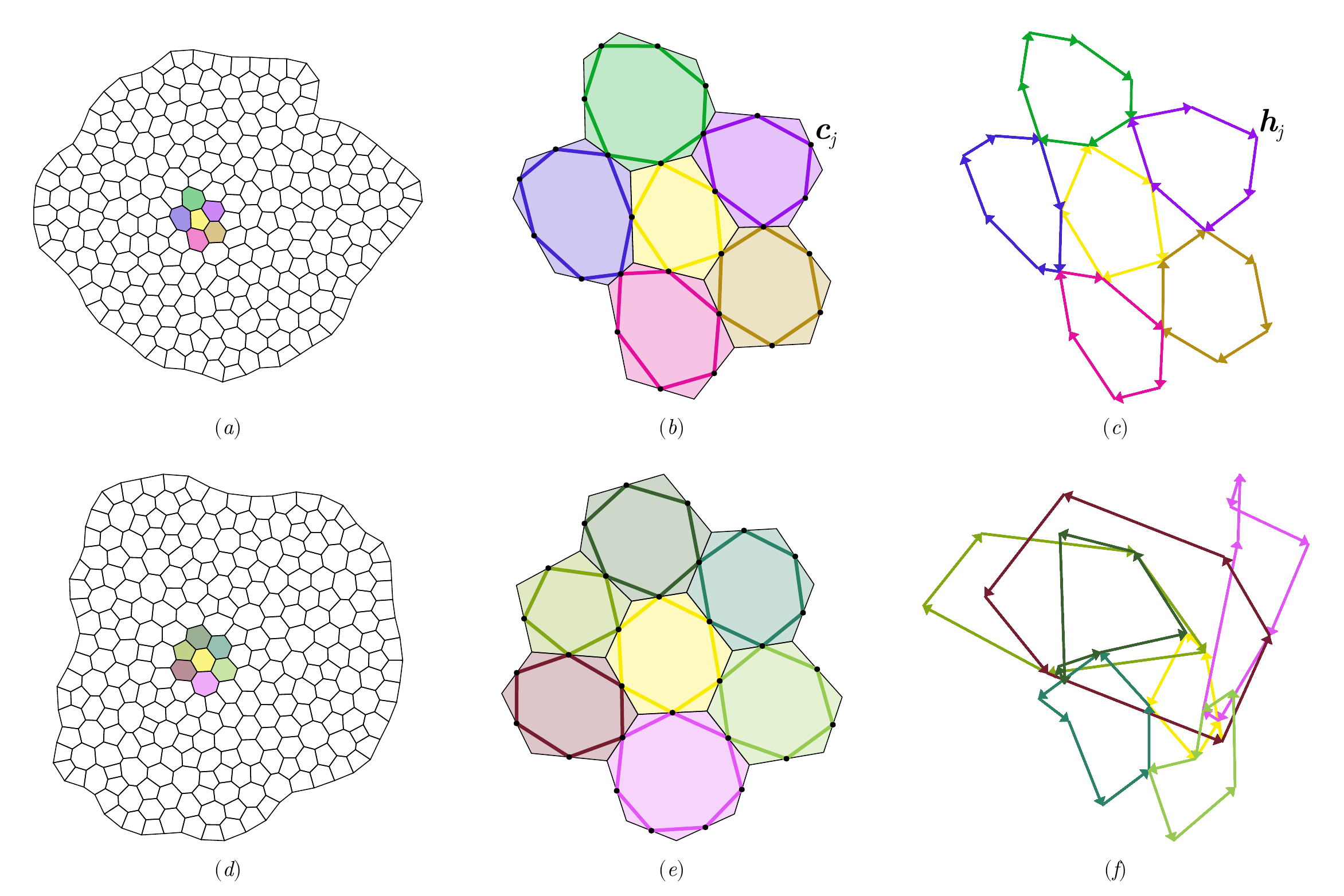}
    \caption{Two illustrations of the rotated force potential.  
    (a,d) show cell monolayers, at mechanical equilibrium, with clusters of central cells highlighted, with cells coloured arbitrarily.  (b,e) show the clusters with the network $N^\Diamond$ connecting adjacent edge centroids $\mathbf{c}_j$ superimposed.  An edge of $N^\Diamond$ in cell $i$ adjacent to vertex $k$ is given by $\mathbf{s}_{ik}$, as defined in  (\ref{eq:sik}).  The corresponding forces $\mathbf{f}_{ik}$ are rotated by $\pi/2$ and assembled to form networks (e,f), with vertices $\mathbf{h}_j$, preserving the colours assigned in (a,d) and (b,e).  The monolayers are in equilibrium, so that rotated forces form closed triangles around vertices ($\sum_i \mathbf{f}_{ik}=\mathbf{0}$) and closed polygons around cells ($\sum_k \mathbf{f}_{ik}=\mathbf{0}$), so sharing the topology of $N^\Diamond$.  The network in (c) is non-overlapping because the monolayer simulated in (a) is subject to isotropic compression at its periphery.  The monolayer in (d) is stress-free at its periphery, showing that the rotated force network is typically a non-planar graph. 
    }
    \label{fig:Figure4ForceNetwork}
\end{figure}

Using (\ref{eq:vmvan}), planar monolayers were simulated using an existing computational implementation of the vertex model \cite{Revell_VertexModel_jl_2026}.  We use $\Gamma=0.2$ and $\tilde{L}_0=0.75$ throughout, ensuring that monolayers remain rigid.  Isolated disordered monolayers were grown using a random division algorithm, imposing a prescribed isotropic stress at the monolayer periphery and allowing T1 transitions. This approach captures prestress within the monolayer that is generated by the growth process \cite{tahaei2025}. After the required number of cells had been created, the system was allowed to relax to equilibrium, resulting in a configuration in which the forces acting on each vertex associated with the three neighbouring cells (the right-hand-side of (\ref{eq:vmvan})) were in equilibrium.  The associated force vectors, after rotating by $\pi/2$ (i.e. the normals to the closed triangle of force vectors around each vertex), form a closed network that matches (topologically) the network $\mathcal{N}^\Diamond$ 
obtained by connecting adjacent edge centroids \cite{jensen2020}, although the graph typically is not planar (Fig.~\ref{fig:Figure4ForceNetwork}).  Nevertheless, the vertices of this rotated-force network, $\mathbf{h}_j$ ($j=1,\dots,N_e$), provide an interpretable vector field defined on edges and links that is suitable for Helmholtz--Hodge decomposition \cite{jensen2022}. (The field $\mathbf{h}_j$ can be described as a potential, because the differencing operator $-\sum_j B_{ij}A_{jk}\mathbf{h}_j$ recovers a rotated force, just as it creates edges of $N^\Diamond$ from edge centroids via (\ref{eq:sik}).)  The monolayers analysed below are all subject to zero peripheral stress. The scalar potentials of $\mathsfbf{h}=\sum_j \mathbf{h}_j \mathsf{q}_j$ for a simply-connected monolayer are analogues of the Airy and Mindlin stress functions of planar elasticity, with the latter function indicating the existence of couple stresses at vertices.  

The stress over cell $i$ can be written \cite{jensen2022}
\begin{equation}
\label{eq:cellstress}
\boldsymbol{\sigma}_i={\textstyle\sum_{j}} A_i^{-1} B_{ij}(\mathbf{t}_j \otimes \mathbf{h}_j)\boldsymbol{\epsilon}_i \quad (i=1,\dots,N_c)
\end{equation}
where the outer product creates a tensor from vectors in $\Gamma(T\mathcal{M}_\mathcal{E})$.  (We do not seek here to formulate (\ref{eq:vmvan}) or (\ref{eq:cellstress}) using exterior calculus, but see \cite{kanso2007} and \cite{rashad2023} for treatment of stress.)  For a monolayer under zero external load, $\sum_i A_i\boldsymbol{\sigma}_i=\mathbf{0}$; correspondingly, $\mathbf{h}_j$ can be set to $\mathbf{0}$ along peripheral edges \cite{jensen2022}.  Given (\ref{eq:cellstress}), the deviatoric cell stress 
$\boldsymbol{\sigma}_i^D=\boldsymbol{\sigma}_i-\tfrac{1}{2}\mathrm{tr}(\boldsymbol{\sigma}_i)\mathsf{I}$, satisfying $\mathrm{tr}(\boldsymbol{\sigma}_i^D)=0$,
is decomposed as $\boldsymbol{\sigma}_i^{Ds}=\tfrac{1}{2}(\boldsymbol{\sigma}_i^D+\boldsymbol{\sigma}_i^{D\top})$ and $\boldsymbol{\sigma}_i^{Da}=\tfrac{1}{2}(\boldsymbol{\sigma}_i^D-\boldsymbol{\sigma}_i^{D\top})$ so that $\boldsymbol{\sigma}_i^{D}=\boldsymbol{\sigma}_i^{Ds}+\boldsymbol{\sigma}_i^{Da}$.  The shear stress is defined as 
\begin{equation}
\zeta_i=\sqrt{-\mathrm{det}\left(\boldsymbol{\sigma}_i^{Ds}\right)}.  
    \label{eq:shearstressexact}
\end{equation}
$\boldsymbol{\sigma}_i^{Da}$ is proportional to $\mathrm{curl}^c\,\mathsfbf{h}$ \cite{jensen2022}.  However, for a monolayer at equilibrium, rotated forces form a closed loop around individual cells (Fig.~\ref{fig:Figure4ForceNetwork}c,f), ensuring that $\mathrm{curl}^c\,\mathsfbf{h}=\mathsf{0}$ and that $\boldsymbol{\sigma}_i$ is symmetric.  Using (\ref{eq:curl}a), the isotropic stress is captured by 
\begin{equation}
\label{eq:hderiv}
P_{\mathrm{eff},i}\equiv\tfrac{1}{2} \mathrm{tr}(\boldsymbol{\sigma}_i)= -\tfrac{1}{2}\mathrm{cocurl}^c\,\mathsfbf{h}.
\end{equation}

Ablation was simulated via removal of one or more internal cells, followed by a further period of relaxation under (\ref{eq:vmvan}).  Code for derivation of scalar potentials and other discrete calculus operations is available via \cite{Revell_DiscreteCalculus_jl}.

\section{Results}
\label{sec:abl}

In equilibrium, the forces at each interior vertex of a planar monolayer at equilibrium are represented by three vectors that sum to zero.  Rotating each force by $\boldsymbol{\epsilon}_i\equiv-\boldsymbol{\epsilon}_k$ builds closed triangles with vertices $\mathsfbf{h}$ (Fig.~\ref{fig:Figure4ForceNetwork}c,f) sitting in a space isomorphic to $\Gamma(T\mathcal{M}_{\mathcal{E}})$.  In Sec.~\ref{sec:ablation} we will apply Helmholtz--Hodge decomposition to this vector force potential, and then in Sec.~\ref{sec:abph} we consider the wider impact of ablation on stress and displacement fields over a monolayer.  We begin by addressing a purely geometric question, namely the nature of harmonic fields in ablated monolayers.

\subsection{Harmonic fields of ablated monolayers}
\label{sec:hm}

\begin{figure}
    \centering
    \includegraphics[width=\textwidth]{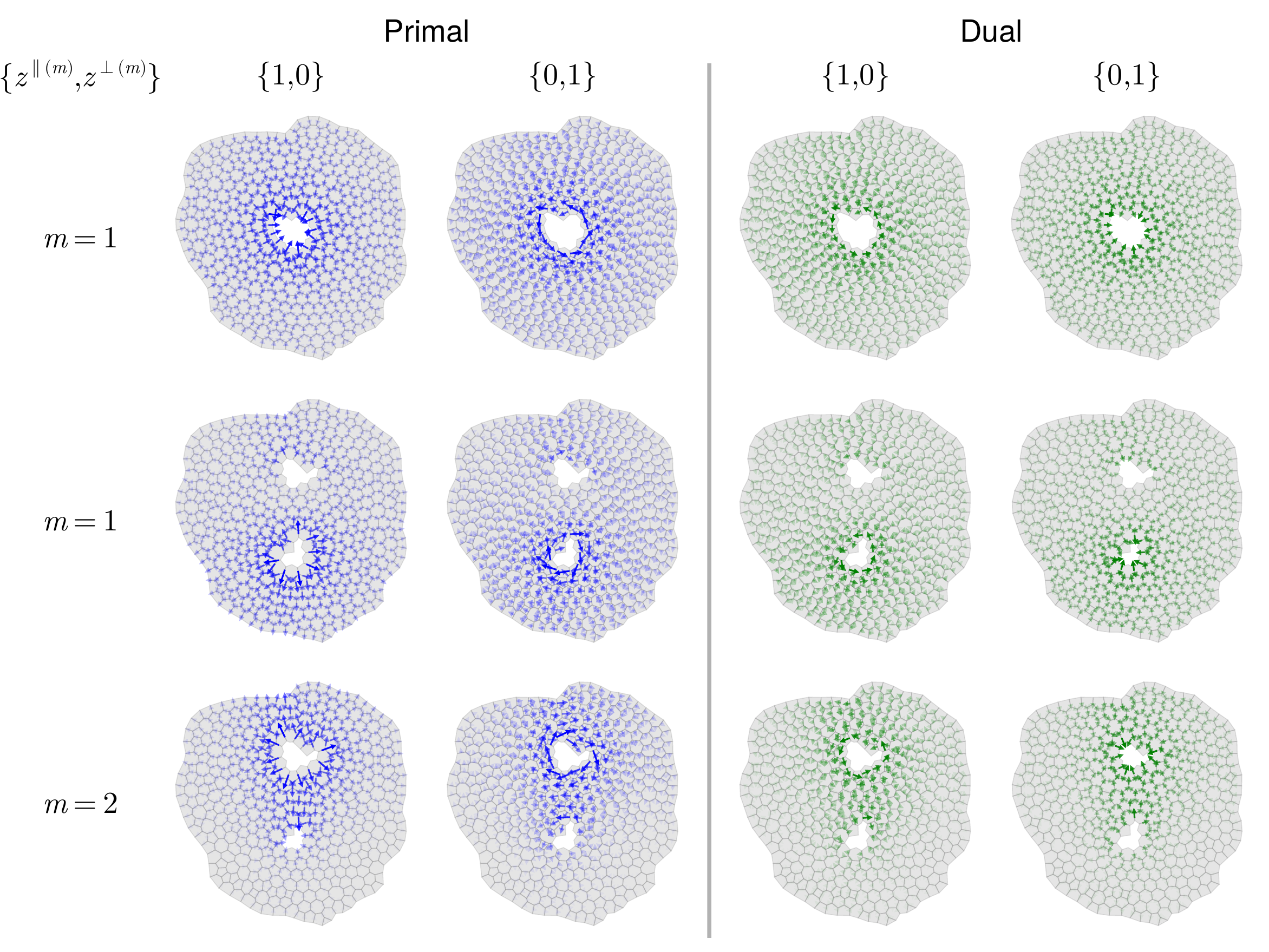}
    \caption{
Harmonic fields on edges $\mathsfbf{x}^{(m)}$ in (\ref{eq:edgex}) are shown in the two left-hand columns; harmonic fields on links $\mathsfbf{X}^{(m)}$ in (\ref{eq:edgeX}) are shown in the two right-hand columns. Top row: the 1st eigenmode ($m=1$) of a system with 1 hole, for $\{z^{\parallel(1)},z^{\perp(1)}\}=\{1,0\}$ or $\{0,1\}$.  Rows 2 and 3: the 1st ($m=1$) and 2nd ($m=2$) eigenmodes of a system with 2 holes. 
    }
    \label{fig:Figure5edgeLaplacianHarmonicFieldVectors}
\end{figure}

Eigenmodes of the edge Laplacian (\ref{eq:veclaps}) having zero eigenvalue (harmonic fields) are represented as vector fields (\ref{eq:edgex}) and (\ref{eq:edgeX}), parametrized by amplitudes $\{z^{\parallel(m)},z^{\perp(m)}\}^\top$.  A monolayer with a single hole has a single harmonic eigenfunction $\mathsf{w}^{(1)}$, generating a vector field $\mathsfbf{x}^{(1)}$ oriented either azimuthally or radially around the hole (Fig.~\ref{fig:Figure5edgeLaplacianHarmonicFieldVectors}, top row).  Closely matching (but rotated) fields appear on the dual network.  A monolayer with two holes has two harmonic modes, each associated with a single hole, on the primal and dual networks (Fig.~\ref{fig:Figure5edgeLaplacianHarmonicFieldVectors}).

\begin{figure}
    \centering
    \includegraphics[width=\textwidth]{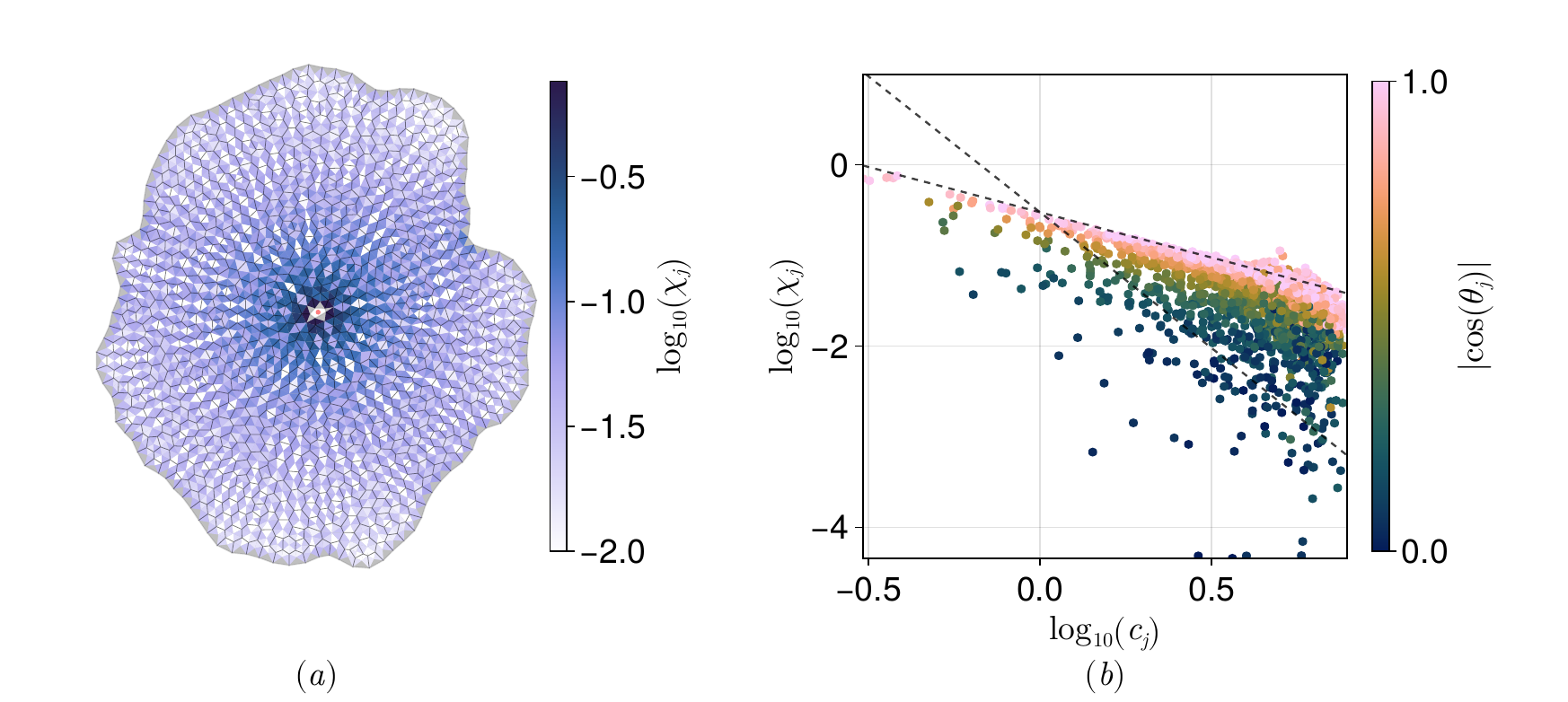}
    \caption{(a) The harmonic field magnitude $\chi\equiv \sum_j \chi_j  \mathsf{q}_j^*$ over edges $j$ associated with the ablation of a single cell at the monolayer centre, where $\chi_j=\vert\mathsf{w}^{(1)}_j\vert/t_j$.  Edge quadrilaterals are coloured by $\log_{10}\chi_j$. The colourmap is truncated at $\chi_j \geq 10^{-2}$.  (b) The full distribution of $\chi_j$ values against the distance, $c_j=\vert\mathbf{c}_j\vert$, of edge $j$ from the centre of the ablated cell, shown with a red dot in (a) and taken to be the spatial origin.  Dashed lines have slope $-1$ and $-3$. Dot colours in (b) show edge orientations relative to the radial direction, $\vert \cos\left(\theta_j\right)\vert$ where $\theta_j$ is the angle between $\mathbf{c}_j$ and $\mathbf{t}_j$, demonstrating how radially-oriented edges carry larger values of $\chi_j$.}
    \label{fig:Figure6harmonicFieldCellRemoved}
\end{figure}

The magnitude of the  of the harmonic field generated by removal of a single cell at the centre of a monolayer (Fig.~\ref{fig:Figure6harmonicFieldCellRemoved}a) reveals dominant contributions from cell edges that are oriented radially with respect to the hole, with a magnitude that has an approximate upper bound that decays proportionally to $1/r$, where $r$ is distance from the hole (Fig.~\ref{fig:Figure6harmonicFieldCellRemoved}b).  This supports an analogy between $\mathsfbf{x}^{(1)}$ and the two-parameter family of smooth harmonic functions in $\mathbb{R}^2\setminus \{0\}$ written in polar coordinates as $(\alpha \hat{\mathbf{r}}+\beta\hat{\boldsymbol{\theta}})/r$ for some constants $\alpha$ and $\beta$, having vanishing divergence and vanishing curl. Fig.~\ref{fig:Figure6harmonicFieldCellRemoved}(b) shows that a very approximate lower bound on the magnitude of the harmonic field is provided by $D/r^3$ for some $D>0$.

\subsection{Scalar stress potentials of ablated monolayers}
\label{sec:ablation}

\begin{figure}
    \centering
    \includegraphics[width=0.95\textwidth]{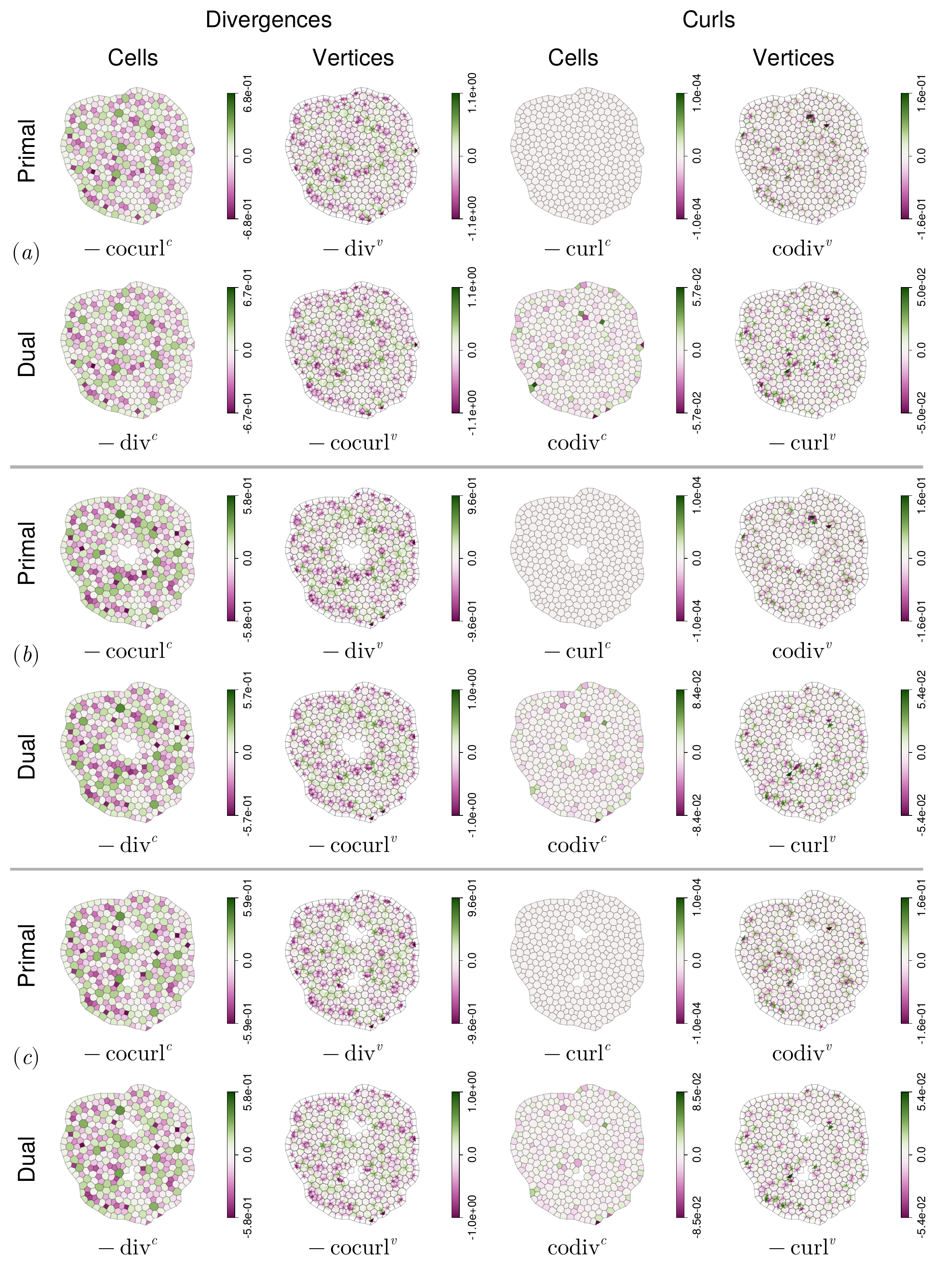}
    \caption{Three monolayers generated from the same initial system, with holes added before a period of relaxation to equilibrium, with (a) no holes, (b) one hole and (c) two holes. Columns 1 and 3 (2 and 4) show derivatives of the rotated force potential $\mathsfbf{h}$ defined over cells (triangles).  Columns 1 and 2 are divergences (including cocurl); columns 3 and 4 are curls (including codiv), as defined in (\ref{eq:curl}, \ref{eq:div}). Rows 1, 3 and 5 (2, 4, and 6) show operators associated with the primal (dual) network.  The monolayers in (b,c) match those shown in Fig.~\ref{fig:Figure5edgeLaplacianHarmonicFieldVectors}.}
    \label{fig:Figure7Derivatives}
\end{figure}

Derivatives of the rotated force potential $\mathsfbf{h}$ are shown in Fig.~\ref{fig:Figure7Derivatives}, for equilibrium monolayers with zero, one and two holes.  Divergences of $\mathsfbf{h}$ (columns 1 and 2, including cocurls) are associated with the isotropic component of the stress field; the divergences show consistent (but heterogeneous) patterns of isotropic stress (over cells and over triangles) across the monolayer. Curls of $\mathsfbf{h}$ (columns 3 and 4, including codivs) capture couple stresses.  Small variations between 1 and 2 in Fig.~\ref{fig:Figure7Derivatives} (and between rows 3 and 4, and 5 and 6) arise primarily from the non-orthogonality of edges and links in the primal and dual networks.  Variations between columns 1 and 2 (and between columns 3 and 4) arise primarily because derivatives of a common underlying field are mapped onto either cells or the triangles associated with vertices.

\begin{table}
    \centering
   \setlength{\tabcolsep}{1mm}
    \begin{tabular}{|c|c|c|c|}
        \hline 
        & No hole                 & 1 hole   & 2 holes \\
        \hline
        $\langle\sum_i A_i \mathsf{q}_i^*\vert\mathrm{cocurl}^c\,\mathsfbf{h}\rangle$   & $-4.36\times 10^{-7}$  & $-1.82\times 10^{-8}$  & $1.49\times 10^{-10}$\\
        $\langle\sum_i A_i \mathsf{q}_i^*\vert\mathrm{curl}^c\,\mathsfbf{h}\rangle$     & $8.70\times 10^{-8} $& $1.42\times 10^{-8}$ & $6.86\times 10^{-11}$\\
        $\langle-\mathrm{div}^c\,\mathsfbf{h}\vert \sum_i A_i \mathsf{q}_i\rangle$      & $0.00515$             & $0.0115$              & $0.0101$              \\
        $\langle -\mathrm{codiv}^c\,\mathsfbf{h}\vert \sum_i A_i \mathsf{q}_i\rangle$   & $0.00244$             & $0.00905$             & $0.00734$             \\
        \hline
    \end{tabular}
    \caption{The components of the solvability conditions (\ref{eq:solvp}) and (\ref{eq:solvd}), evaluated using the derivatives shown in Fig.~\ref{fig:Figure7Derivatives}; integrals are expressed using the natural pairing.}
    \label{tab:integrals}
\end{table}

The condition for individual cells to experience zero net force is $\mathrm{curl}^c\,\mathsfbf{h}=\mathsf{0}$.  The condition for the monolayer to experience zero net isotropic stress (because it is under zero external load) is for the integral of $\mathrm{cocurl}^c\,\mathsfbf{h}$ to vanish (by (\ref{eq:hderiv}), this is $2\sum_i A_i P_{\mathrm{eff},i}=0$).  Both conditions are comfortably satisfied in computations (Fig.~\ref{fig:Figure7Derivatives}; Table~\ref{tab:integrals}), showing that monolayers are equilibrated. Each derivative has an analogue on the dual network: integrals of $-\mathrm{div}^c\,\mathsfbf{h}$ and $-\mathrm{codiv}^c\,\mathsfbf{h}$ deviate slightly from zero, which we attribute to non-orthogonality.  $\mathrm{curl}^v\,\mathsfbf{h}$ in Fig.~\ref{fig:Figure7Derivatives} reveals weak couple stresses at internal vertices; its representation on the primal network ($\mathrm{codiv}^v\,\mathsfbf{h}$) is also non-zero.  

\begin{figure}
    \centering
    \includegraphics[width=0.95\textwidth]{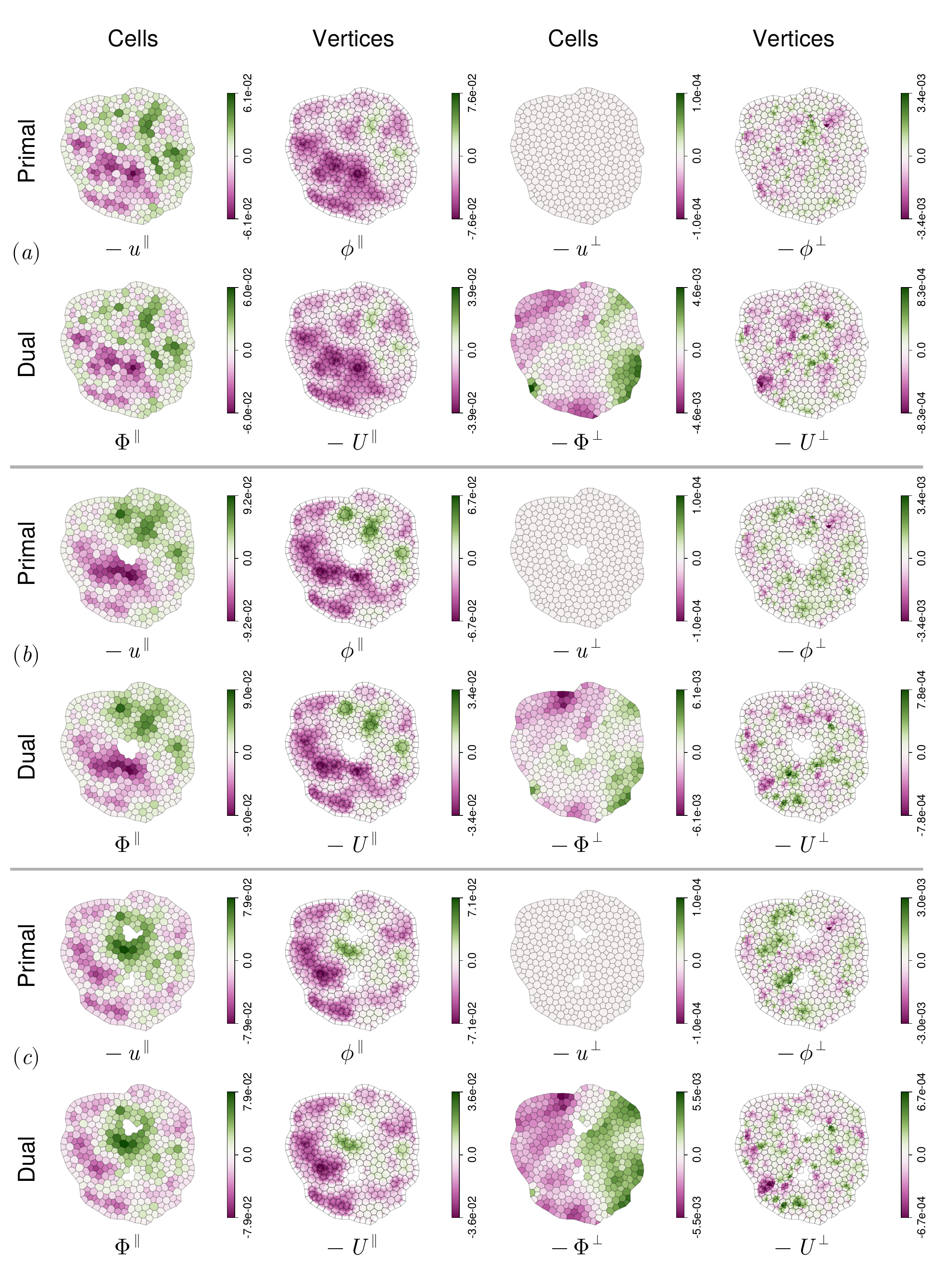}
    \caption{The potentials associated with the derivatives of $\mathsfbf{h}$ over the primary and dual network for the same three monolayers illustrated in Fig.~\ref{fig:Figure7Derivatives}. Mirroring the layout of Fig.~\ref{fig:Figure7Derivatives}, rows 1, 3 and 5 (2, 4 and 6) show potentials defined over the primal (dual) networks. Columns 1 and 2 show representations of the Airy stress function; column 4 shows representations of the Mindlin stress function.}
    \label{fig:Figure8Potentials}
\end{figure}

The Poisson problems (\ref{eq:hhp}b) and (\ref{eq:hhd}b) require the forcing to have zero integral for a solution to exist, as specified in (\ref{eq:solvp}, \ref{eq:solvd}).  We enforced zero mean of the forcing before implementing Moore--Penrose inversion; given the data in Table~\ref{tab:integrals}, a small correction must be introduced to accommodate non-zero forcing in (\ref{eq:hhd}b), as explained in Appendix~\ref{app:valid}.  The scalar potentials of $\mathsfbf{h}$, obtained by inverting the Poisson problems summarised in Table~\ref{tab:opsum}, are shown in Fig.~\ref{fig:Figure8Potentials}. Columns 1 and 2 show consistent representations of the Airy stress function.  Because $\mathrm{curl}^c\,\mathsfbf{h}=\mathsf{0}$ for a monolayer strictly in equilibrium, the associated potential satisfies $\mathsf{u}^\perp=0$ (column 3); its representation over the dual network, $\Phi^\perp$, is correspondingly small in magnitude. Column 4 reflects the Mindlin stress function, illustrating couple stress effects.  These are defined over vertices via $-\mathsf{U}^\perp$, and represented over cells in $-\phi^\perp$.  

\begin{figure}
    \centering
    \includegraphics[width=\textwidth]{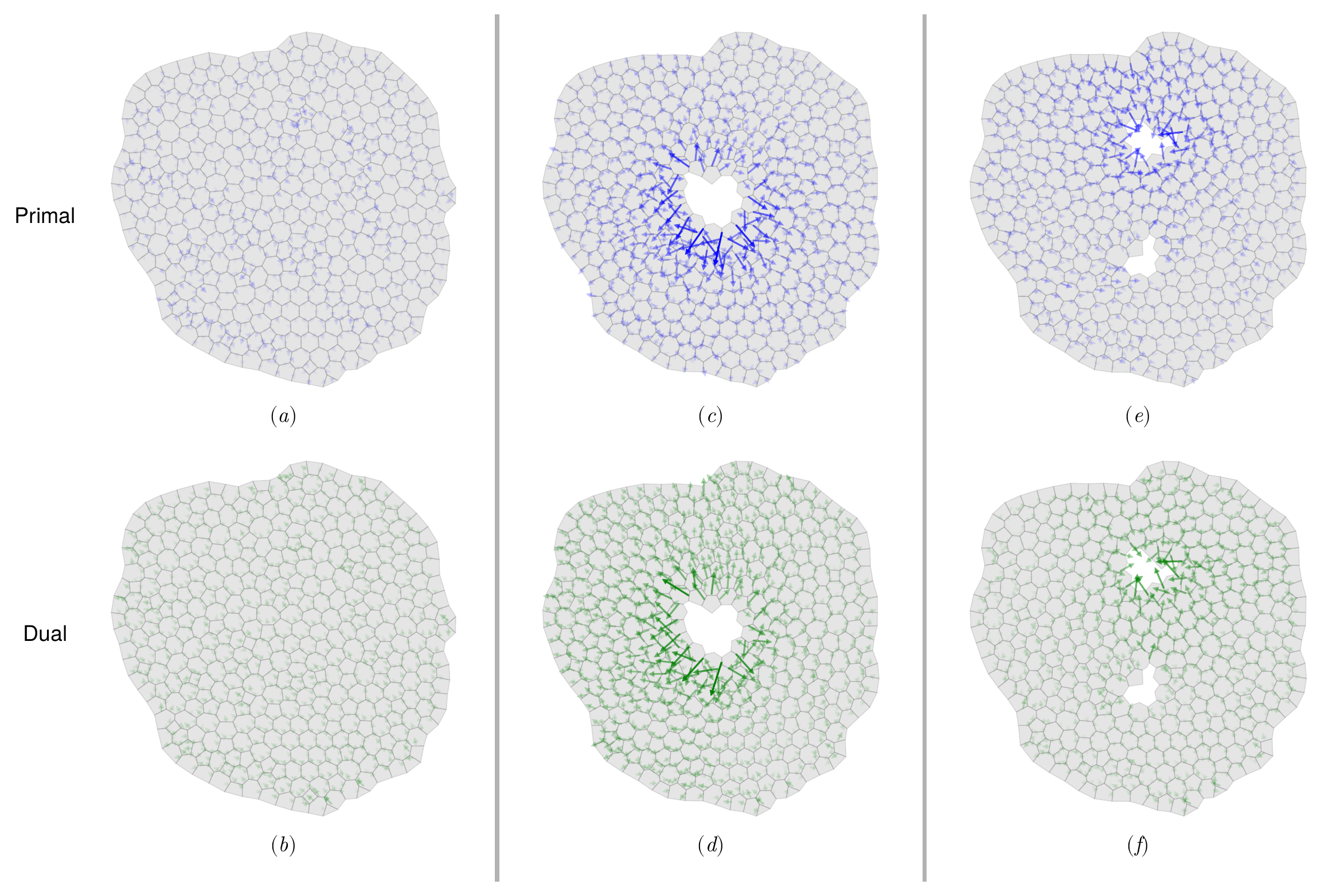}
    \caption{
    Differences $\breve{\mathsfbf{x}}$ and $\breve{\mathsfbf{X}}$ between the original $\mathsfbf{h}$ field and that reconstructed from potentials in Fig.~\ref{fig:Figure8Potentials} using (\ref{eq:hhp}a) and (\ref{eq:hhd}a), assuming no harmonic component, over (a, c, e) primal and (b, d, f) dual networks for monolayers with (a, b) zero, (c, d) one and (e, f) two holes. 
    Vectors $\breve{\mathbf{x}}_j$ (a, c, e) or $\breve{\mathbf{X}}_j$ (b, d, f) are mapped onto corresponding edge centroids $\mathbf{c}_j$ or link midpoints $\mathbf{C}_j$. Vector opacity is set by the length of that vector relative to the maximum vector length across all 6 panels. 
}
\label{fig:Figure9HelmholtzHodgeHoles}
\end{figure}

\begin{figure}
    \centering
    \includegraphics[width=\textwidth]{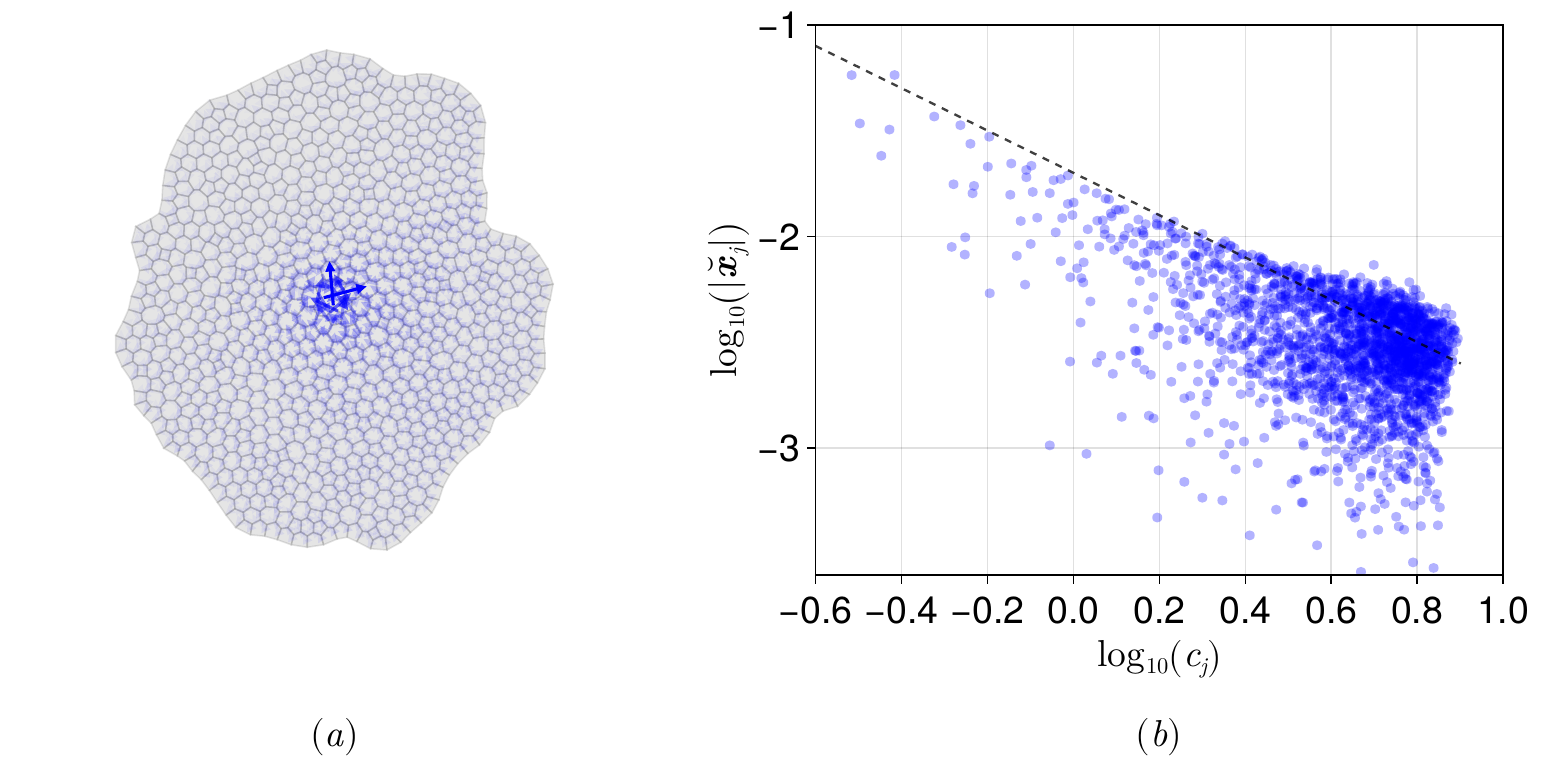}
    \caption{
    (a) Vectors $\breve{\mathbf{x}}_j$ (where $\breve{\mathsfbf{x}}=\mathsfbf{h}-\mathrm{grad}\,\phi-\mathrm{rot}\,\mathsf{u}$) plotted at edge centroids $\mathbf{c}_j$ for a large equilibrated monolayer following ablation of a single central cell.  (b) $\vert\breve{\mathbf{x}}_j\vert$ plotted against distance of the edge centroid from the hole, $c_j=\vert\mathbf{c}_j\vert$, taking the origin to be the centre of the removed cell, on a log scale. The dashed line has slope $-1$.  
    }
\label{fig:Figure10xFieldCellRemoved}
\end{figure}

For the simply-connected monolayer, we confirmed that the four scalar potential fields are sufficient to recover $\mathsfbf{h}$, to within reasonable accuracy.  Considering (\ref{eq:hhp}) and (\ref{eq:hhd}), Fig.~\ref{fig:Figure9HelmholtzHodgeHoles}(a,b) plots the differences $\breve{\mathsfbf{x}}\equiv \mathsfbf{h}-\mathrm{grad}\,\phi -\mathrm{rot}\,\mathsf{u}$ and $\breve{\mathsfbf{X}}\equiv \mathsfbf{h}-\mathrm{grad}\,\Phi -\mathrm{rot}\,\mathsf{U}$ over the primal and dual networks respectively.  The maximum value of $\vert\breve{\mathbf{x}}_j\vert$ and $\vert\breve{\mathbf{X}}_j\vert$ is bounded by $0.012$,
which compares to the maximum value of $\vert\mathbf{h}_j\vert$, $0.18$.  Some of this error can be attributed to non-orthogonality: the correction introduced to accommodate solvability conditions (Table~\ref{tab:integrals}) requires adjustment of Laplacian operators at the monolayer periphery (Appendix~\ref{app:valid}), leading to imperfections in the representation of $\mathsfbf{h}$.  Nevertheless, introduction of one hole (Fig.~\ref{fig:Figure9HelmholtzHodgeHoles}c,d) reveals numerical predictions $\breve{\mathsfbf{x}}$ of the harmonic field $\mathsfbf{x}^{(1)}$, directed radially to the hole with an amplitude near the hole that is elevated above the background numerical error, consistent with Fig.~\ref{fig:Figure5edgeLaplacianHarmonicFieldVectors}.  Introduction of two holes (Fig.~\ref{fig:Figure9HelmholtzHodgeHoles}e,f) reveals a pattern reminiscent of the harmonic eigenmode $\mathsfbf{x}^{(1)}$ focused around the upper hole, shown in Fig.~\ref{fig:Figure5edgeLaplacianHarmonicFieldVectors}.   The amplitude of the reconstructed field $\breve{\mathsfbf{x}}$ for a monolayer in which a single cell has been removed (Fig.~\ref{fig:Figure10xFieldCellRemoved}) demonstrates a $1/r$ decay in maximum amplitude, consistent with Fig.~\ref{fig:Figure6harmonicFieldCellRemoved}, although this scaling is obscured by numerical error further from the hole.  In summary, despite some imperfections, this data provides evidence that the force potential $\mathsfbf{h}$ gains a contribution from the harmonic field after ablation.

Because it has zero divergence, the harmonic component of $\mathsfbf{h}$ makes no contribution to the isotropic stress.  The shear-stress component $\zeta^{(1)}$ associated with the harmonic field of an ablated monolayer shown in Fig.~\ref{fig:Figure6harmonicFieldCellRemoved} is evaluated in Appendix~\ref{app:harmonicstress}.  As shown in Fig.~\ref{fig:FigureG1EdgeLaplacianShearStressCellRemoved} below, it shares the approximate $1/r$ decay of the harmonic field.  We now investigate its possible contribution to the full stress field in an ablated monolayer.


\subsection{Stress and displacement in an ablated monolayer}
\label{sec:abph}

\begin{figure}
    \centering
    \includegraphics[width=0.99\textwidth]{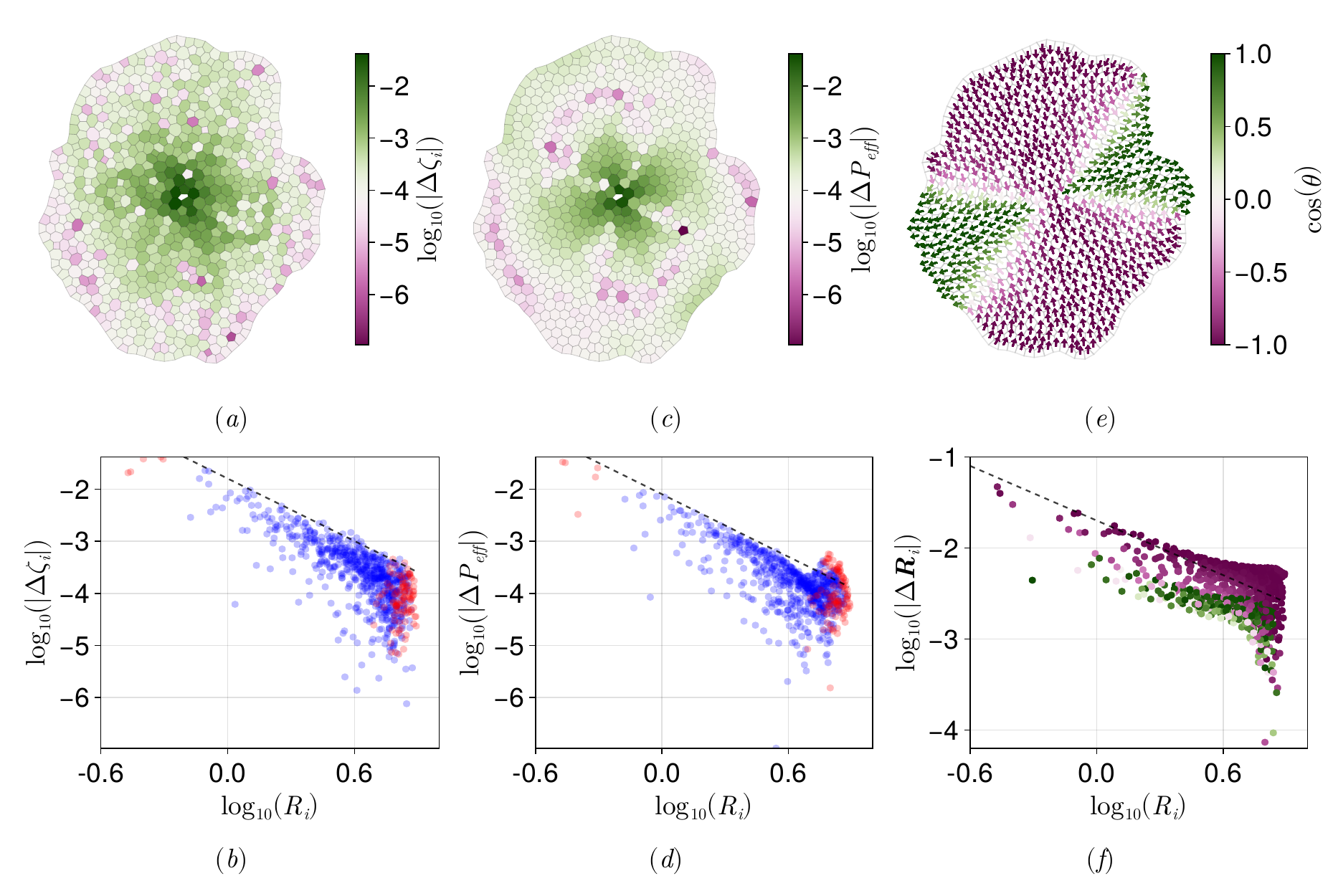}
    \caption{
    (a) Difference in cell shear-stress magnitude $\vert \zeta_i\vert$, as defined by (\ref{eq:shearstressexact}), before and after ablation of one cell in a monolayer. (b) Scatterplot of values in (a) against distance $R_i=\vert\mathbf{R}_i\vert$ of cell $i$ ($i=1,\dots, N_c$) from the centre of the ablated cell (taken to lie at the origin); peripheral (internal) cells are shown with red (blue) dots.  (c, d) show the corresponding difference in cell isotropic-stress magnitude ($\vert P_{\mathrm{eff},i}\vert$), as defined by (\ref{eq:hderiv}).  Dashed lines in (b,d) have slope $-2$. (e) Arrows show cell centre displacements $\Delta\mathbf{R}_i$ following ablation; colours show normalised radial component $\Delta\hat{\mathbf{R}}_i\cdot \hat{\mathbf{R}}_i$.   (f) Scatterplot of $\log_{10}\vert \Delta \mathbf{R}_i\vert $ against $\log_{10} {R}_i$; dashed line has slope $-1$.  Points for each cell are coloured to show $\Delta\hat{\mathbf{R}}_i\cdot \hat{\mathbf{R}}_i$, as in (e).  
    }
    \label{fig:Figure11AblationStressComparison}
\end{figure}


Figure~\ref{fig:Figure11AblationStressComparison}(a-d) illustrates the change in the magnitude of the shear and isotropic cell-stress resulting from ablation of a single cell at the centre of a monolayer.  Both fields decay in magnitude at a rate bounded approximately by $\alpha/r^2$ for some $\alpha>0$.  (Cells at the periphery, which are elongated because they have only one peripheral edge, behave slightly differently.)  While the harmonic field shown in Fig.~\ref{fig:FigureG1EdgeLaplacianShearStressCellRemoved} may be present, its amplitude is likely too small to reveal a clear $1/r$ scaling near the hole in Fig.~\ref{fig:Figure11AblationStressComparison}(b).   The $1/r^2$ decay rate in shear-stress magnitude is consistent with the behaviour of a punctured linearly elastic disc (see (\ref{eq:lindiscstress})).  However the simple elastic problem lacks the heterogeneous prestress illustrated by $\mathrm{cocurl}^c\,\mathsfbf{h}$ in Fig.~\ref{fig:Figure7Derivatives}.
This may explain why (\ref{eq:lindiscstress}) does not predict $1/r^2$ component of isotropic perturbation stress.  

The isotropic perturbation stress field (Fig.~\ref{fig:Figure11AblationStressComparison}c) shows evidence of a long-range quadrupolar structure in this example.  Its origin is revealed by examination of the displacement of cell centres arising as a result of ablation (Fig.~\ref{fig:Figure11AblationStressComparison}e,f), which also has strongly quadrupolar features.  The displacement field partitions into two wedge-shaped regions in which cells move away from the ablation (green), and two regions in which they move towards it (purple).  Both inward and outward moving fields exhibit 
a $1/r$ scaling near the hole, consistent with a partial contribution from the harmonic field.  The coherence of the motion supports an approximate continuum description, in which radial displacements have the approximate form $f(r)[a+b\cos(2(\theta-\theta_0)]$ in polar coordinates, for some $f(r)$ and some $\theta_0$.  The monopolar term ($a$) and the quadrupolar term ($b$) share the same radial dependence in this approximation, allowing lines of zero radial displacement to be straight.  For $-b<a<0<b$, for example, the wedge of inward-moving cells is wider than that of the outward moving cells.  


\begin{figure}
    \centering
    \includegraphics[width=0.99\textwidth]{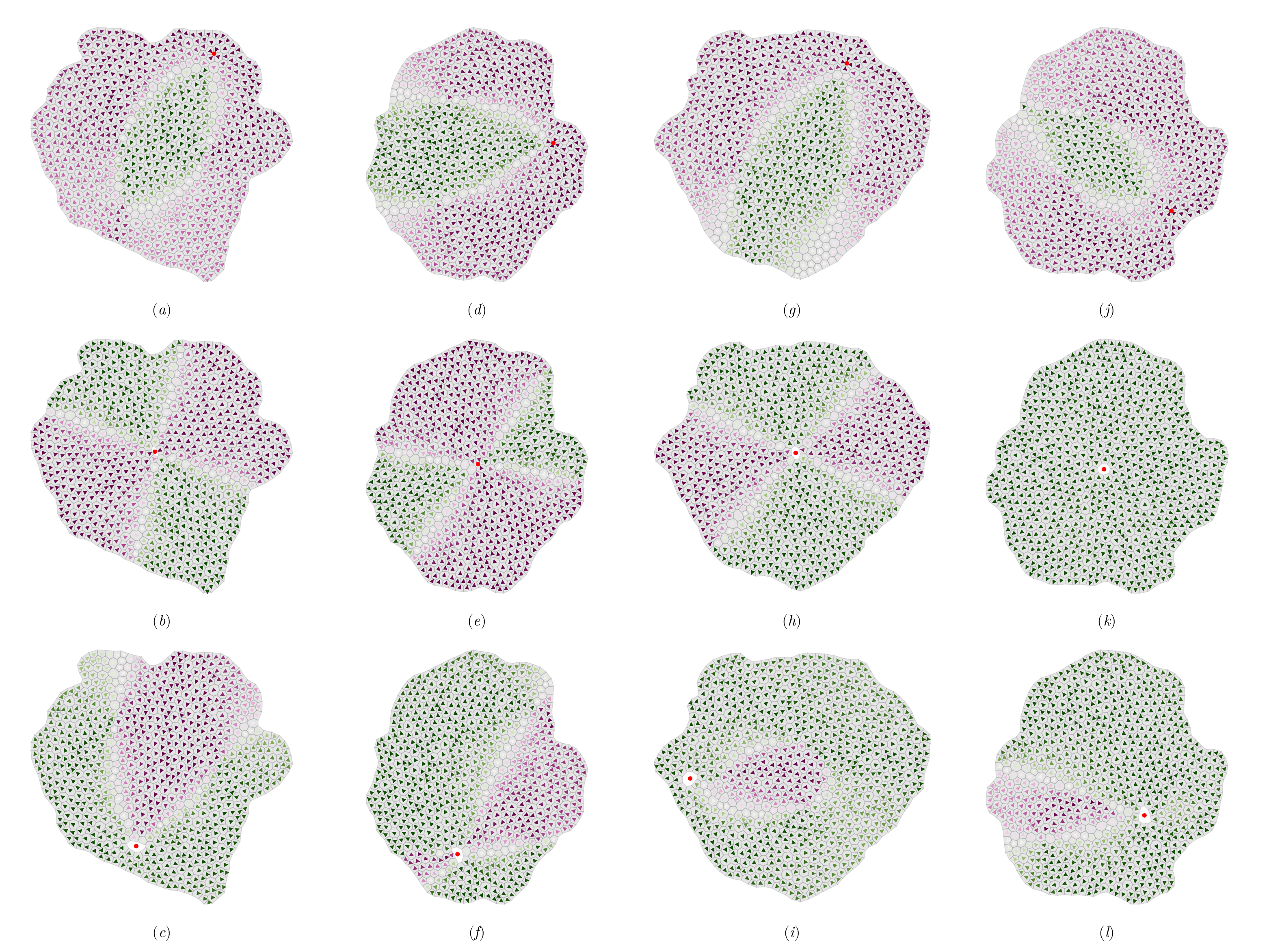}
    \caption{Displacement fields following ablation of a single cell (marked with a red dot) in four monolayers.  Cell displacements are shown using the colour scheme of Fig.~\ref{fig:Figure11AblationStressComparison}(e).  For each monolayer, the ablated cell is either that with smallest isotropic stress (a,d,g,j), the central cell (b,e,h,k), or that with largest isotropic stress (c,f,i,l).  Values of shear and isotropic stress of the targetted cells, immediately prior to ablation, are shown in Table~\ref{tab:stressvals}.}
    \label{fig:Figure12DisplacementGrid}
\end{figure}

\begin{table}[]
    \centering
    \setlength{\tabcolsep}{1mm}
    \begin{tabular}{|c|c|c|c|c|c|}
        \hline 
        Panel & 
        $\zeta_i$ & $2P_{\mathrm{eff},i}$ 
        & Panel & $\zeta_i$ & $2P_{\mathrm{eff},i}$ 
        \\
        \hline 
        (a) & 
        0.324 & -0.605 &
               (g) & 
        0.351 & -0.670 \\
        (b) & 
        0.119 & -0.0300 &
        (h) & 
        0.128 & 0.150 \\
        (c) & 
        0.309 & 0.435 & 
        (i) & 
        0.153 & 0.401 \\ \hline 
        (d) & 
        0.325 & -0.695 &
        (j) & 
        0.370 & -0.736 \\
        (e) & 
        0.179 & -0.226 & 
        (k) & 
        0.109 & 0.215 \\
        (f) & 
        0.306 & 0.401 &
        (l) & 
        0.290 & 0.480 \\
        \hline
    \end{tabular}
    \caption{Values of shear and isotropic stress of the cells that are ablated in the examples shown in Fig.~\ref{fig:Figure12DisplacementGrid}.}
    \label{tab:stressvals}
\end{table}

Further examples of displacement fields following ablation are given in Fig.~\ref{fig:Figure12DisplacementGrid}.  We created four monolayers and in each ablated one of three cells.  When the cell with lowest $P_{\mathrm{eff},i}$ in the monolayer is ablated, the motion is predominantly directed towards the ablation (Fig.~\ref{fig:Figure12DisplacementGrid}(a,d,g,j)).  These small, strongly compressed cells lie near the edge of the monolayer in these examples and the wedge pattern is replaced by a tear-drop-shaped region of cells moving away from the ablation.  Removal of these strongly compressed cells leads, as expected, to shrinkage of the hole.  In contrast, ablation of the cell with the highest $P_{\mathrm{eff},i}$ generates predominantly outward motion (Fig.~\ref{fig:Figure12DisplacementGrid}(c,f,i,l).  Again, these initially large cells arise near the monolayer periphery in these examples; their removal  leads to expansion of the hole.  Ablation of the cells at the centre of the monolayer typically recovers a quadrupolar field: examples (b) and (e) in Fig.~\ref{fig:Figure12DisplacementGrid} have $P_{\mathrm{eff},i}<0$ (Table~\ref{tab:stressvals}), consistent with a wider contractile wedge; example (h) has $P_{\mathrm{eff},i}>0$, widening the dilational wedge.  Curiously, it is also possible for fully contractile motion to arise (Fig.~\ref{fig:Figure12DisplacementGrid}(k)).  The shear-stress value $\zeta_i$ of this cell prior to ablation is lower than other examples (Table~\ref{tab:stressvals}), but not sufficiently to suggest that $\zeta_i$ is predictive of this outcome.

\section{Discussion}

The present study makes three primary contributions.  First, we have identified a framework of spaces and maps ((\ref{eq:dr}), Fig.~\ref{fig:Figure7Derivatives}) allowing an existing set of differential operators that exploit the intrinsic polygonal structure of confluent cells \cite{jensen2022} to be expressed in the language of DEC.  To accommodate the natural irregularities of epithelia, the framework avoids imposing orthogonality of links and edges and accommodates boundary conditions arising at monolayer boundaries, including those around an ablation. A non-standard wedge product acting on the value-legs of cochains (\ref{eq:wedge1}, \ref{eq:wedge7}) has been deployed in order to construct appropriate inner products.  Second, we have used this DEC-inspired framework to evaluate global harmonic fields induced by holes in monolayers (Fig.~\ref{fig:Figure5edgeLaplacianHarmonicFieldVectors}) and demonstrated an approximate $1/r$ decay of amplitude with distance $r$ from a hole (Fig.~\ref{fig:Figure6harmonicFieldCellRemoved}).  Third, using a version of the vertex model that predicts equilibrium forces across a disordered monolayer incorporating osmotic effects (Appendix~\ref{app:vm}), we have investigated the mechanical impact of ablation, demonstrating striking long-range coherence in perturbation displacement fields (Fig.~\ref{fig:Figure12DisplacementGrid}).  For a simply-connected monolayer, we reconstructed the vector force potential of an ablated monolayer with scalar stress functions; for an ablated monolayer, we demonstrated excitation of the harmonic field (Figs~\ref{fig:Figure9HelmholtzHodgeHoles}, \ref{fig:Figure10xFieldCellRemoved}).  We showed that the perturbation stress fields (Fig.~\ref{fig:Figure11AblationStressComparison}) have a $1/r^2$ decay rate in upper bound, a feature relevant to the mechanoresponse of a monolayer to ablation. This scaling, and the associated quadrupolar pattern, is consistent with the long-range stress induced in a planar elastic sheet by a force dipole or Eshelby inclusion \cite{tahaei2025}. The cellular shear-stress contribution induced by the harmonic field (Fig.~\ref{fig:FigureG1EdgeLaplacianShearStressCellRemoved}) has an approximate $1/r$ decay of amplitude with distance, not directly evident in simulation data.  However the displacement field shows evidence of $1/r$ decay,  suggesting involvement of a harmonic component.  The quadrupolar component of discrete displacement fields (Fig.~\ref{fig:Figure12DisplacementGrid}) awaits further analysis; however it mirrors reported features of fields generated by cell divisions \cite{tahaei2025}, ablation \cite{tahaei2025} and neighbour exchanges \cite{lemaitre2021}.

In a rigid monolayer, our results illustrate a long-range response of the stress field in a monolayer to ablation, which decays algebraically rather than exponentially with distance.  A far-field shear-stress distribution requires cells to be in a jammed state, because the shear stress predicted by the vertex model is determined by tensions in cell edges \cite{ANB2018a}.  Fluidization of cells in the tissue surrounding a wound, as reported in \textit{Drosophila} wing imaginal disc \cite{tetley2019}, would therefore suppress the spatial extent of this mechanical signal.  We have not sought here to incorporate the inflammatory response, re-epithelialization, matrix deposition and other processes that lead to resolution of a wound \cite{pena2024}.  However, loosely motivated by the action of mechanoregulatory factors such as YAP/TAZ \cite{perez2018} and ERK \cite{boocock2021}, we have shown how the geometric operators that appear naturally in the vertex model (\ref{eq:vmvan}) can be used to model diffusion of mobile chemical signals between cells, enabling osmotic effects to be incorporated into the vertex model via modification of the parameter $\Gamma$ (\ref{eq:fullvm1}) that measures the relative importance of peripheral to bulk free energy.  Shear stress in a cell is proportional to $\Gamma$.  Thus a chemical that spreads rapidly between cells and which promotes cell swelling (which, in the present 2D formulation, is equivalent to expansion of the cell's apical face) lowers the parameter $\Gamma$; likewise rapid spreading of a chemical that occupies the perimeter of the apical face and promotes its elongation leads to an increase in $\Gamma$, and hence shear stress.  We leave investigation of the coupled mechanochemical system (\ref{eq:fullvm1}) for a future study.

To put these results in a broader mathematical context, it is helpful to consider the different operators that arise when representing fields over polygonal networks, those underpinning the cell vertex model (involving particular constitutive assumptions), and their representations using DEC.  Starting from a weak representation of $\nabla$, DeGoes et \hbox{al.} \cite{degoes2020} defined $\mathrm{d}$, $\sharp$ and $\flat$ operators appropriate for a polygonal network on a curved manifold (that may have non-planar faces), having scalar fields defined on vertices and vector fields on cells. When restricted to a flat manifold, a set of dual operators can be defined that are appropriate for vector fields defined on vertices and scalar fields on cells, and which emerge naturally when considering operators associated with cell area changes \cite{cowley2024}.  These exploit the extended network $\mathcal{N}^\Diamond$ in which links are added between adjacent edge centroids (Fig.~\ref{fig:Figure4ForceNetwork}c), forming closed loops around vertices and cells.  $\mathcal{N}^\Diamond$ serves an additional purpose as the template for an equilibrium force balance in a monolayer \cite{jensen2020}.  The approach taken in this study is complementary to that of \cite{degoes2020}, by defining vectors on cell edges and scalars on cells, with edges between cell vertices and links between cell centres providing bases for expression of discrete vector fields.  We have shown that it is convenient to formulate operators that act on covector-valued cochains defined over vertices and faces, holding $\parallel$ and $\perp$ components associated with projections of vectors onto (or orthogonally to) edges of the primal network, or links of the dual network.  Many of the operators have a clear interpretation as a discretization of an integral representation of a standard operator.  As we have demonstrated here and in \cite{jensen2022}, the resulting structure supports use of Helmholtz--Hodge decomposition.  
The mathematical framework proposed here provides a foundation for future studies addressing a wider set of mechanical and transport processes that may require more exotic differential operators, including covariant and Lie derivatives.  It is also generalisable, in principle, to more complex geometries, such as cells on a curved substrate and tissues formed from polyhedral, rather than  polygonal, cells.

Non-orthogonality of links and edges is a generic feature of epithelia and of many implementations of the vertex model, but it leads to a degree of complexity in the present formulation.  Alternative models that impose a Voronoi structure (for example \cite{bi2016}), for which cell centres are degrees of freedom, benefit from simpler differential operators and successfully capture features such as jamming transitions.  However, by discarding the many additional degrees of freedom associated with vertex displacements, some dynamical features can be lost, such as the numerous zero modes of the Hessian of a monolayer at equilibrium which underpin its geometric stiffness \cite{damavandi2022, cowley2024}.

For monolayers that are not simply connected, a family of harmonic fields (with zero divergence and zero curl) is needed to provide a full description of vector fields defined over the monolayer.  The harmonic fields are found by evaluating the eigenmodes lying in the kernel of a Laplace--de Rahm operator defined on edges.  Each hole generates a one-parameter harmonic field of arbitrary amplitude (illustrated in Fig.~\ref{fig:Figure5edgeLaplacianHarmonicFieldVectors}) which can have radial or azimuthal form; the field for a single hole decays approximately like $1/r$ with distance $r$ from the hole (Fig.~\ref{fig:Figure6harmonicFieldCellRemoved}).  An anology with the smooth harmonic fields $\hat{\mathbf{r}}/r$ and $\hat{\boldsymbol{\theta}}/r$ (in polar coordinates) is evident, however the present fields accommodate boundaries and irregularities in the pattern of cells. We found that the vector force potential of an ablated monolayer could not be fully described in terms of scalar potentials (Fig.~\ref{fig:Figure8Potentials}), requiring a contribution from the harmonic field.  However this contribution does not explain the observed $1/r^2$ scaling in perturbation stress magnitudes.  Near-hole perturbation displacements show a $1/r$ scaling in their upper bound (Fig.~\ref{fig:Figure11AblationStressComparison}f), consistent with involvement of the harmonic field.  A continuum quadrupolar field can be obtained via two covariant derivatives of a harmonic field.  It remains to be seen if such an approach in the present problem, developing the proposed DEC formalism, might explain the quadrupolar features that are evident in displacement fields arising as a result of ablation in Fig.~\ref{fig:Figure12DisplacementGrid}, and as revealed in experimental measures of post-ablation strain fields \cite{tahaei2025}.  

\backmatter

\section*{Declarations}

\begin{itemize}
\item Funding: This work was supported by The Leverhulme Trust (RPG-2021-394) and the Biotechnology and Biological Sciences Research Council (BB/T001984/1).
\item Conflict of interest: None
\item Ethics approval and consent to participate: Not applicable
\item Consent for publication: For the purpose of open access, the authors have applied a Creative Commons Attribution (CCBY) licence to any Author Accepted Manuscript version arising.
\item Data availability: All data presented were generated using scripts in the GitHub repository~\cite{Revell_Ablation_jl} .
\item Materials availability: Not applicable
\item Code availability: In addition to code used to generate results for this paper, found at ~\cite{Revell_Ablation_jl}, the \texttt{VertexModel.jl} code is available on GitHub, specifically in the \texttt{ablation} branch~\cite{Revell_VertexModel_jl_2026}, as is the new \texttt{DiscreteCalculus.jl} package~\cite{Revell_DiscreteCalculus_jl}, which contains all methods and operators described in this paper. All code was written in the Julia language~\cite{bezansonJuliaFreshApproach2017}, making use of the \texttt{DifferentialEquations.jl} library~\cite{rackauckasDifferentialEquationsjlPerformantFeaturerich2017}, and all results visualisations were created with the \texttt{Makie.jl} plotting package~\cite{danischMakiejlFlexibleHighperformance2021} and Scientific Colormaps~\cite{crameri_2023_8409685}.
\item Author contribution: Conceptualization: OEJ; Methodology: OEJ, CKR; Formal analysis and investigation: OEJ, CKR; Writing - original draft preparation: OEJ; Writing - review and editing: OEJ, CKR; Funding acquisition: OEJ.

\end{itemize}

\begin{appendices}

\section{Ablation of an elastic disc}
\label{app:disc}

We recall the behaviour of a linearly-elastic disc of radius $a$ in the plane-strain approximation.  The stress $\boldsymbol{\sigma}$ is related to the displacement field $\mathbf{u}$ via $\boldsymbol{\sigma}=\lambda \mathsf{I}\nabla \cdot\mathbf{u}+\mu(\nabla \mathbf{u}+\nabla\mathbf{u}^\top)$, where $\lambda$ and $\mu$ are Lam\'e constants.  The equilibrium condition $\nabla\cdot \boldsymbol{\sigma}=0$ requires
\begin{equation}
    0=(\lambda+\mu)\nabla(\nabla\cdot \mathbf{u})+\mu \nabla^2 \mathbf{u}.
\end{equation}
With $\mathbf{u}$ dependent only on coordinates in the plane of the disc, the Laplace--de Rahm operator can be written $\nabla^2\mathbf{u}\equiv -(-\mathrm{grad}\,\circ\, \mathrm{div}+\mathrm{rot}\,\circ\, \mathrm{curl})\mathbf{u}$.  Under an external pressure $P$ at $r=a$ ($r$ is the radial cylindrical polar coordinate, with unit vector $\hat{\mathbf{r}}$), the disc has uniform isotropic in-plane stress $\boldsymbol{\sigma}^{2D}=-P\mathsf{I}_2$, zero shear stress and radial displacement $\mathbf{u}=-Pr\hat{\mathbf{r}}/[2(\lambda+\mu)]$.  Thus when $P<0$, a small hole introduced in the centre of the disc is expected to grow before equilibrating.   An annular disc occupying $b<r<a$, under zero stress on $r=b$ and under pressure $P$ at $r=a$, has radial displacement \cite{howell2009}
\begin{equation}
\label{eq:uaxi}
    \mathbf{u}=-\hat{\mathbf{r}} P \left[ \frac{1}{2(\lambda+\mu)}\frac{a^2 r}{a^2-b^2}+\frac{1}{2\mu r}\frac{a^2 b^2}{a^2-b^2}\right],\quad (b<r<a).
\end{equation}
The displacement component $\hat{\mathbf{r}}/r$ is a harmonic field: it has vanishing div and curl, so that $\nabla^2(\hat{\mathbf{r}}/r)=\mathbf{0}$.  The corresponding in-plane stress can be written 
\begin{equation}
\label{eq:lindiscstress}
    \boldsymbol{\sigma}^{2D}=\frac{a^2 P}{a^2-b^2}\left[-\mathsf{I}_2+ \frac{b^2}{r^2}(\hat{\mathbf{r}}\hat{\mathbf{r}}-\hat{\boldsymbol{\theta}}\hat{\boldsymbol{\theta}})\right].
\end{equation}
The hole induces an inhomogeneous shear stress proportional to $b^2/r^2$, arising from a derivative of the harmonic displacement field in (\ref{eq:uaxi}).  
The perturbation displacement, subtracting (\ref{eq:uaxi}) from $-Pr\hat{\mathbf{r}}/[2(\lambda+\mu)]$, gives 
\begin{equation}
    \Delta \mathbf{u}=-\frac{\hat{\mathbf{r}} P b^2}{2(a^2-b^2)}\left[\frac{r}{\lambda+\mu}+\frac{a^2}{\mu r}\right],\quad (b<a<r)
    \label{eq:lindiscdisp}
\end{equation}
This is monotonic across the disc, with the $1/r$ component becoming prominent closer to the hole. 

\section{Boundary conditions}
\label{app:bc1}

Fig.~\ref{fig:Figure2constructors} illustrates the geometric construction of a monolayer.  Peripheral links terminate at edge centroids (Fig.~\ref{fig:Figure2constructors}a); in the present formulation, cells at the outer monolayer boundary have single peripheral edges (Fig.~\ref{fig:Figure2constructors}b); areas $E_k$ associated with internal vertices are triangular (Fig.~\ref{fig:Figure2constructors}c); areas $\tfrac{1}{2}F_j$ associated with internal edges are quadrilateral (Fig.~\ref{fig:Figure2constructors}d).  
Below, we focus on the topological approach to implementing boundary conditions; geometric information is introduced when defining differential operators in Sec.~\ref{sec:do}.

For an isolated monolayer, we partition vertices into $N_{vp}$ peripheral and $N_{vi}$ interior vertices, and edges into $N_{ep}$ peripheral, $N_{en}$ normal and $N_{ei}$ interior edges.  Normal edges are those that connect an interior vertex to a peripheral vertex; they are approximately normal to peripheral edges.  Thus $N_v=N_{vp}+N_{vi}$ and $N_e=N_{ep}+N_{en}+N_{ei}$.  The incidence matrices then take the form 
\begin{equation}
\label{eq:bctop}
\mathsf{A}=\left(    \begin{matrix}
        \mathsf{A}^{pp} & \mathsf{0} \\ 
        \mathsf{A}^{np} & \mathsf{A}^{ni} \\
        \mathsf{0} & \mathsf{A}^{ii}
    \end{matrix}
    \right),\quad
    \mathsf{B}=
    \left(    \begin{matrix}
        \mathsf{B}^{p} & \mathsf{B}^{n} & \mathsf{B}^i 
    \end{matrix}
    \right)
\end{equation}
(suppressing $*$'s on incidence matrices; bases are implicit), with
\begin{equation}
\mathsf{B}\mathsf{A}=\left(
\begin{matrix}
    \mathsf{B}^p\mathsf{A}^{pp}+\mathsf{B}^n\mathsf{A}^{np}~ &
    ~\mathsf{B}^n\mathsf{A}^{ni}+\mathsf{B}^i\mathsf{A}^{ii}
\end{matrix}\right)=(\begin{matrix}
    \mathsf{0} & \mathsf{0}
\end{matrix}).
\end{equation}
Laplacians become
\begin{subequations}
\label{eq:topls}
    \begin{align}
    \mathsf{A}^\top\mathsf{A}&=\left(\
\begin{matrix}
\mathsf{A}^{pp\top}\mathsf{A}^{pp}+\mathsf{A}^{np\top}\mathsf{A}^{np} &
\mathsf{A}^{np\top}\mathsf{A}^{ni} \\
\mathsf{A}^{ni\top}\mathsf{A}^{np} & \mathsf{A}^{ni\top}\mathsf{A}^{ni}+ \mathsf{A}^{ii\top}\mathsf{A}^{ii}
\end{matrix}
    \right), \\
    \mathsf{B}\mathsf{B}^\top&=\mathsf{B}^{p}\mathsf{B}^{p\top}+\mathsf{B}^{n}\mathsf{B}^{n\top}+\mathsf{B}^{i}\mathsf{B}^{i\top} \\
    \mathsf{A}\mathsf{A}^\top+\mathsf{B}^\top\mathsf{B}
    &=\left(
\begin{matrix}
\mathsf{A}^{pp}\mathsf{A}^{pp\top} + \mathsf{B}^{p\top}\mathsf{B}^{p}& \mathsf{A}^{pp}\mathsf{A}^{np\top} + \mathsf{B}^{p\top}\mathsf{B}^{n}& \mathsf{B}^{p\top}\mathsf{B}^{i} \\
\mathsf{A}^{np}\mathsf{A}^{pp\top} +\mathsf{B}^{n\top}\mathsf{B}^{p}& \mathsf{A}^{np}\mathsf{A}^{np\top}+ \mathsf{A}^{ni}\mathsf{A}^{ni\top}+\mathsf{B}^{n\top}\mathsf{B}^{n}& \mathsf{A}^{ni}\mathsf{A}^{ii\top} + \mathsf{B}^{n\top}\mathsf{B}^{i} \\
\mathsf{B}^{i\top}\mathsf{B}^{p} & \mathsf{A}^{ii}\mathsf{A}^{ni\top}+ \mathsf{B}^{i\top}\mathsf{B}^{n}& \mathsf{A}^{ii}\mathsf{A}^{ii\top} + \mathsf{B}^{i\top}\mathsf{B}^{i}
\end{matrix}
   \right) 
\end{align}
\end{subequations}

To solve $\mathsf{B}\mathsf{B}^\top \theta=\mathsf{S}$ with a Neumann-type boundary condition, for some $\mathsf{S}$, we modify $\mathsf{B}\mathsf{B}^\top$ in (\ref{eq:topls}b) by suppressing $\mathsf{B}^{p}\mathsf{B}^{p\top}$, \hbox{i.e.} suppressing contributions from links connecting peripheral cells to peripheral edge centroids.  Thus we solve
\begin{equation}
(\mathsf{B}^{n}\mathsf{B}^{n\top}+\mathsf{B}^{i}\mathsf{B}^{i\top})\theta=\mathsf{S} 
\end{equation}
subject to the solvability condition $\mathsf{1}_c^\top \mathsf{S}={0}$, where $\mathsf{1}_c$ is the chain identifying all cells.  

To solve $\mathsf{A}^\top\mathsf{A} \phi = \mathsf{s}$, we write $\phi=(\phi^p,\phi^i)^\top$ and $\mathsf{s}=(\mathsf{s}^p,\mathsf{s}^i)^\top$.  Imposing a Dirichlet boundary condition $\phi^p=\mathsf{0}$, then we do not need to evaluate $\mathsf{s}^p$ and using (\ref{eq:topls}a) we solve
\begin{equation}
(\mathsf{A}^{ni\top}\mathsf{A}^{ni}+ \mathsf{A}^{ii\top}\mathsf{A}^{ii}){\phi^i}=\mathsf{s}^i.
\end{equation}
This Laplacian is expected to be non-singular for a simply-connected monolayer and can be inverted directly.  

Helmholtz-Hodge decomposition $\psi=\mathsf{A}\phi+\mathsf{B}^\top\theta+\mathsf{x}$ (as in (\ref{eq:hh})) can be decomposed similarly.  We ignore $\psi^p$, impose $\phi^p=0$ and restrict attention to the reduced incidence matrices
\begin{equation}
\label{eq:abred}
    \hat{\mathsf{A}}=\left(    \begin{matrix}
        \mathsf{A}^{ni} \\
        \mathsf{A}^{ii}
    \end{matrix}
    \right),\quad
    \hat{\mathsf{B}}=
    \left(    \begin{matrix}
    \mathsf{B}^{n} & \mathsf{B}^i 
    \end{matrix}
    \right).
\end{equation}
In addition to suppressing $\mathsf{B}^p$, we also suppress $\mathsf{A}^{np}\mathsf{A}^{np\top}$ to decouple the harmonic problem from peripheral edges.  We solve for $\phi^i$ (defined over internal vertices), $\theta$ (over all cells) and $\mathsf{x}^n$ and $\mathsf{x}^i$ (over all but peripheral edges), using $\hat{\psi}\equiv(\psi^n,\psi^i)^\top$, $\hat{x}\equiv(x^n,x^i)^\top$ with
\begin{subequations}
    \begin{align}
\hat{\psi}&=\hat{\mathsf{A}}\phi^i+\hat{\mathsf{B}}^\top\theta+\hat{\mathsf{x}}, & 
\hat{\mathsf{A}}^\top\hat{\mathsf{A}}\phi^i&=\hat{\mathsf{A}}^{\top}\hat{\psi},\\
    \left(
 \hat{\mathsf{A}}\hat{\mathsf{A}}^{\top}+\hat{\mathsf{B}}^{\top}\hat{\mathsf{B}}
   \right) \hat{\mathsf{x}}&=\mathsf{0}, &
\hat{\mathsf{B}}\hat{\mathsf{B}}^{\top}\theta&=\hat{\mathsf{B}}\hat{\psi}.
\end{align}
    \end{subequations}
This relies on $\hat{\mathsf{B}}\hat{\mathsf{A}}=\mathsf{0}$
and ensures $\hat{\mathsf{A}}^{\top} \hat{\mathsf{x}}=\mathsf{0}$ and $\hat{\mathsf{B}} \hat{\mathsf{x}}=\mathsf{0}$.  
The reduced incidence matrices $\hat{\mathsf{A}}$ and $\hat{\mathsf{B}}$ operate over reduced networks $\hat{\mathcal{N}}$ (without periperal edges and peripheral vertices) and $\hat{\mathcal{N}}^\rhd$ (without peripheral links), illustrated in Fig.~\ref{fig:Figure2constructors}(d,c) respectively.

\section{Interior product}
\label{app:interior}

The interior product pairing scalar-valued and $\mathcal{P}$-valued cochains defined on edges, ${\phi}\in\Omega_0^1(\hat{\mathcal{N}})$ and $\mathsf{w}\in\Omega_1^1(\hat{\mathcal{N}})$ respectively, with a vector field $\mathsfbf{v}\in\Gamma(T\mathcal{M}_{\mathcal{E}})$, is defined as
\begin{subequations}
\label{eq:interior}
    \begin{align}
\iota_{\mathsfbf{v}}\phi& \defeq{\textstyle \sum_j}^\top \phi_j\{v_j^\parallel, v_j^\perp\}^\top\mathsf{q}_j^*\in \Omega_1^1(\hat{\mathcal{N}}),\\
    \iota_{\mathsfbf{v}}\mathsf{w}
    &\defeq{\textstyle{\sum_j}} (w_j^\parallel v_j^\parallel +w_j^\perp v_j^\perp)\mathsf{q}^*_j\in \Omega_0^1(\hat{\mathcal{N}}).
\end{align}
\end{subequations}
Consistent with other DEC formulations \cite{wang2023}, the interior product can be connected to the wedge product via
\begin{equation}    
\label{eq:drint}
\begin{tikzcd}
\Omega_2^1(\hat{\mathcal{N}}) \arrow[r, "\star_{2,1}"]
&  \Omega_0^1(\hat{\mathcal{N}}^\rhd)  \\
\Omega_1^1(\hat{\mathcal{N}})   \arrow[u, "\mathsfbf{v}^\flat\wedge"] 
\arrow[r,"\star_{1,1}"] 
& 
\Omega_1^1(\hat{\mathcal{N}}^\rhd) \arrow[u, "\iota_{\mathsfbf{v}}"] \\ 
\Omega_0^1(\hat{\mathcal{N}}) \arrow[r, "\star_{0,1}"]  \arrow[u, "\mathsfbf{v}^\flat\wedge"]
& \Omega_2^1(\hat{\mathcal{N}}^\rhd) \arrow[u, "\iota_{\mathsfbf{v}}"]
\end{tikzcd}, \qquad
\begin{tikzcd}
\Omega_2^1(\hat{\mathcal{N}}) \arrow[d, "\iota_{\mathsfbf{V}}"]
&  \arrow[l, "\star_{2,1}^\rhd"] \Omega_0^1(\hat{\mathcal{N}}^\rhd) \arrow[d, "\mathsfbf{V}^\flat\wedge"] \\
\Omega_1^1(\hat{\mathcal{N}})    \arrow[d, "\iota_{\mathsfbf{V}}"]
& 
\arrow[l,"\star_{1,1}^\rhd"] \Omega_1^1(\hat{\mathcal{N}}^\rhd)  \arrow[d, "\mathsfbf{V}^\flat\wedge"]\\ 
\Omega_0^1(\hat{\mathcal{N}})   
& \arrow[l, "\star_{0,1}^\rhd"]\Omega_2^1(\hat{\mathcal{N}}^\rhd) 
\end{tikzcd}.
\end{equation}
The three sequences in (\ref{eq:spacen}), for $n=0,1,2$, which incorporate Hodge stars mapping between $\hat{\mathcal{N}}$ and $\hat{\mathcal{N}}^\rhd$ for fixed $n$, are here connected by wedge (interior) products that raise (lower) the value $n$ of cochains.  More precisely, for $\phi\in\Omega_0^1(\hat{\mathcal{N}})$, $\mathsf{w}\in \Omega_1^1(\mathsf{N})$, $\mathsfbf{v}\in\Gamma(T\mathcal{M}_\mathcal{E})$, $\Phi\in\Omega_0^1(\hat{\mathcal{N}}^\rhd)$, $\mathsf{W}\in\Omega_1^1(\hat{\mathcal{N}}^\rhd)$, $\mathsfbf{V}\in\Gamma(T\mathcal{M}_\mathcal{L})$, 
\begin{subequations}
\begin{align}
\star_{1,1}(\mathsfbf{v}^\flat \wedge \phi) &=\iota_{\mathsfbf{v}}\star_{0,1}\phi={\textstyle \sum_j}\phi_j (F_j/t_j^2)\{v_j^\parallel,v_j^\perp\}^\top\mathsf{q}_j^*\in\Omega_1^1(\hat{\mathcal{N}}^\rhd), \\
\star_{2,1} (\mathsfbf{v}^\flat \wedge \mathsf{w})&=-\iota_{\mathsfbf{v}}\star_{1,1}\mathsf{w}={\textstyle \sum_j} (F_j/t_j^2)(v_j^\parallel w_j^\perp-v_j^\perp w_j^\parallel)\mathsf{q}_j^* \in\Omega_0^1(\hat{\mathcal{N}}^\rhd),\\
    \star_{1,1}^\rhd(\mathsfbf{V}^\flat \wedge \Phi) &=\iota_{\mathsfbf{v}}\star_{2,1}^\rhd\Phi={\textstyle \sum_j}\Phi_j (F_j/T_j^2)\{V_j^\parallel,V_j^\perp\}^\top\mathsf{q}_j^* \in\Omega_1^1(\hat{\mathcal{N}}), \\
        \star_{0,1}^\rhd (\mathsfbf{V}^\flat \wedge \mathsf{W}) &=-\iota_{\mathsfbf{v}}\star_{1,1}^\rhd\mathsf{W}={\textstyle \sum_j} (F_j/T_j^2)(V_j^\parallel W_j^\perp-V_j^\perp W_j^\parallel)\mathsf{q}_j^*\in\Omega_0^1(\hat{\mathcal{N}}),
\end{align}
\end{subequations}
using the definitions of $\star_{2,1}$, $\star_{0,1}$, $\star_{2,1}^\rhd$ and $\star_{0,1}^\rhd$ shown in Table~\ref{tab:maps}.  Now
\begin{align}
   {v}_j^\parallel {w}_j^\perp-{v}_j^\perp {w}_j^\parallel&=
(\mathbf{v}_j\cdot \mathbf{t}_j) (\mathbf{w}_j\cdot \boldsymbol{\epsilon}_i \mathbf{t}_j)-(\mathbf{v}_j\cdot \boldsymbol{\epsilon}_i \mathbf{t}_j) (\mathbf{w}_j\cdot  \mathbf{t}_j) \nonumber \\
&={t_j^2}\left[ (\mathbf{v}_j\cdot \hat{\mathbf{t}}_j) (\mathbf{w}_j\cdot \boldsymbol{\epsilon}_i \hat{\mathbf{t}}_j)-(\mathbf{v}_j\cdot \boldsymbol{\epsilon}_i \hat{\mathbf{t}}_j) (\mathbf{w}_j\cdot  \hat{\mathbf{t}}_j)\right] \nonumber \\
&=t_j^2 \mathbf{v}_j^T\left[
 \hat{\mathbf{t}}_j (\boldsymbol{\epsilon}_i \hat{\mathbf{t}}_j)^T-(\boldsymbol{\epsilon}_i \hat{\mathbf{t}}_j)   \hat{\mathbf{t}}_j^T \right]\mathbf{w}_j
 =t_j^2 \mathbf{v}_j^T \left[
 -\boldsymbol{\epsilon}_i \right]\mathbf{w}_j.
 \label{eq:vxw}
\end{align}
Thus (with a suitable choice of orientiation $\boldsymbol{\epsilon}_i$) the area-weighted scalar and vector products of $\mathsfbf{v}$ and $\mathsfbf{w}$ are recovered in the form
\begin{equation}
\star_{2,1}\iota_{\mathsfbf{v}}\mathsfbf{w}^\flat ={\textstyle\sum_j}F_j (\mathbf{v}_j\cdot\mathbf{w}_j)\mathsf{q}_j^*, \quad
    \star_{2,1}(\mathsfbf{v}^\flat\wedge \mathsfbf{w}^\flat)=-\iota_{\mathsfbf{v}}\star_{1,1}\mathsfbf{w}^\flat={\textstyle\sum_j}F_j (\mathbf{v}_j\times \mathbf{w}_j)\mathsf{q}_j^*.
\end{equation}

\section{Evaluation of operators}
\label{app:evalop}

\subsection{Primary operators}
\label{app:primary}

For $\phi\in \Omega_1^0(\hat{\mathcal{N}})$ and $\Phi\in \Omega_1^0(\hat{\mathcal{N}}^\rhd)$, we define
\begin{subequations}
    \label{eq:grad}
\begin{align}
    \mathrm{grad}\,\phi
    &=(\mathrm{d}\phi)^\sharp 
    =(\mathsf{A}_1^*\phi)^\sharp
    =(\mathsf{A}_1^*\{\phi^\parallel,\phi^\perp\}^\top)^\sharp 
    =(\mathsf{A}_1^*{\textstyle{\sum_k}}\{\phi_k^\parallel,\phi_k^\perp\}^\top\mathsf{q}_k^*)^\sharp  \nonumber \\
   & =({\textstyle{\sum_{j,k}}}\mathsf{q}_j^* \hat{A}_{jk}\{\phi_k^\parallel,\phi_k^\perp\}^\top)^\sharp  
={\textstyle{\sum_{j,k}}}\mathsf{q}_j (\mathbf{e}_{j\parallel} \hat{A}_{jk}\phi_k^\parallel+\mathbf{e}_{j\perp}\hat{A}_{jk}\phi_k^\perp)  \nonumber \\
 & \defeq \mathrm{grad}^v\,\phi^\parallel+\mathrm{cograd}^v\,\phi^\perp\in \Gamma(T\mathcal{M}_{\mathcal{E}}),  \\
    \mathrm{grad}\,\Phi
    &=(\mathrm{d}\Phi)^\sharp 
    =(\mathsf{B}_1^{*\top}\Phi)^\sharp
    =(\mathsf{B}_1^{*\top}\{\Phi^\parallel,\Phi^\perp\}^\top)^\sharp 
    =(\mathsf{B}_1^{*\top}{\textstyle{\sum_i}}\{\Phi_i^\parallel,\Phi_i^\perp\}^\top\mathsf{q}_i^*)^\sharp \nonumber  \\
  &  =({\textstyle{\sum_{i,j}}} \{\Phi_i^\parallel,\Phi_i^\perp\}^\top \hat{B}_{ij}\mathsf{q}_j^*)^\sharp  
={\textstyle{\sum_{i,j}}}(\Phi_i^\parallel \hat{B}_{ij}\mathbf{E}_{j\parallel} +\Phi_i^\perp \hat{B}_{ij}\mathbf{E}_{j\perp})\mathsf{q}_j  \nonumber  \\
& \defeq \mathrm{grad}^c\,\Phi^{\parallel}+\mathrm{cograd}^c\,\Phi^{\perp}\in \Gamma(T\mathcal{M}_{\mathcal{L}}).  \end{align}
\end{subequations}
The superscript $v$ [$c$] denotes operators involving $\hat{\mathsf{A}}$ [$\hat{\mathsf{B}}$].  The operator cograd$^\bullet$ is orthogonal (with respect to the scalar product defined by the metric $\mathsfbf{g}$ or $\mathsfbf{G}$ in (\ref{eq:loceu})) to grad$^\bullet$ (with $\bullet=v$ or $c$).  Here, $\hat{\mathsf{A}}$ and $\hat{\mathsf{B}}$ act as difference operators, with the contravariant bases providing orientation and introducing an inverse length dependence to the gradient operators.  The magnitude of a grad field is measured by
\begin{subequations}
    \begin{align}
    [ \mathrm{grad}\,\phi, \mathrm{grad}\,\phi]_{\hat{\mathcal{E}}}&=
    \langle (\mathrm{d}\phi)\wedge \star \mathrm{d}\phi \vert\mathsf{1}_e\rangle=
    \langle [\mathsf{A}_1^* \phi]^\top (-\boldsymbol{\epsilon}_\mathcal{P})(\hat{\mathsf{T}}_e^{-1}\otimes \boldsymbol{\epsilon}_{\mathcal{P}})\mathsf{A}_1^*\phi\vert \mathsf{1}_e\rangle \nonumber \\ 
    &= \langle \phi^\top [(\hat{\mathsf{A}}^\top \hat{\mathsf{T}}_e^{-1}\hat{\mathsf{A}})\otimes\mathsf{I}_{\mathcal{P}}]\phi\vert\mathsf{1}_e\rangle \nonumber \\
    &={\textstyle{\sum_{k,k'}}}\left[\phi^\parallel_k\{\hat{\mathsf{A}}^\top \hat{\mathsf{T}}_e^{-1}\hat{\mathsf{A}}\}_{k,k'}\phi^\parallel_{k'}+ \phi^\perp_k\{\hat{\mathsf{A}}^\top \hat{\mathsf{T}}_e^{-1}\hat{\mathsf{A}}\}_{k,k'}\phi^\perp_{k'}\right],
\end{align}
making use of (\ref{eq:mops}b). Likewise 
\begin{equation}
\label{eq:gradmag}
[\mathrm{grad}\,\Phi,\mathrm{grad}\,\Phi]_{\hat{\mathcal{L}}}={\textstyle{\sum_{i,i'}}}\left[\Phi^\parallel_i\{\hat{\mathsf{B}} \hat{\mathsf{T}}_l^{-1}\hat{\mathsf{B}}^\top\}_{i,i'}\Phi^\parallel_{i'}+ \Phi^\perp_i\{\hat{\mathsf{B}} \hat{\mathsf{T}}_l^{-1}\hat{\mathsf{B}}^\top\}_{i,i'}\Phi^\perp_{i'}\right].
\end{equation}
\end{subequations}
The directional derivative of a $\mathcal{P}$-valued 0-cochain $\phi\in \Omega_1^0(\hat{\mathcal{N}})$ is captured using (\ref{eq:interior}) by 
\begin{equation}
\iota_{\mathsfbf{v}}(\mathrm{grad}\,\phi)^\flat=
\iota_{\mathsfbf{v}}\mathrm{d}\phi=
\langle \mathrm{d}\phi\vert \mathsfbf{v} \rangle_{\mathcal{P}} =
{\textstyle{\sum_j} }\hat{A}_{jk}(v_j^\parallel \phi_k^\parallel +v_j^\perp \phi_k^\perp)\mathsf{q}_j^*,
\end{equation}
so that $(\iota_{\mathsfbf{v}}\mathrm{d}\phi)^\sharp={\textstyle{\sum_j} }\hat{A}_{jk}[(\mathbf{v}_j\cdot \hat{\mathbf{e}}_j^\parallel) \phi_k^\parallel +(\mathbf{v}_j\cdot\hat{\mathbf{e}}_j^\perp) \phi_k^\perp]\mathsf{q}_j^*$, where hats denote unit vectors.  In addition, $\mathrm{grad}\,\{\mathsf{1}_c,\mathsf{1}_c\}^\top=\mathsfbf{0}$, reflecting the Neumann conditions implicit in $\hat{\mathsf{B}}$; however $\mathrm{grad}\,\{\mathsf{1}_v,\mathsf{1}_v\}^\top$ is non-zero at the monolayer periphery, as $\hat{\mathsf{A}}\mathsf{1}_v$ identifies the "spiky" edges at the periphery of $\hat{\mathcal{N}}$.

For $\mathsfbf{v}\in \Gamma(T\mathcal{M}_{\mathcal{E}})$ and $\mathsfbf{V}\in \Gamma(T\mathcal{M}_{\mathcal{L}})$, we define
\begin{subequations}
    \label{eq:curl}
\begin{align}
\mathrm{curl}\,\mathsfbf{v}&=(\star \mathrm{d}\mathsfbf{v}^\flat)^\sharp 
=(\star_{1,2} \mathsf{B}_1^*\mathsfbf{v}^\flat)^\sharp=(\star_{1,2}\mathsf{B}_1^* {\textstyle{\sum_{{j}}}} (\mathsf{q}_j \mathbf{v}_j)^\flat )^\sharp  \nonumber\\ 
& =(\star_{1,2} \mathsf{B}_1^*{\textstyle{\sum_{{j}}}}  \mathsf{q}_j^* \{\mathbf{v}_j\cdot \mathbf{e}^{\parallel}_j,\mathbf{v}_j\cdot \mathbf{e}^{\perp}_j\}^\top)^\sharp
= (\star_{1,2} {\textstyle{\sum_{{i,j}}}} \mathsf{q}_i^* \hat{B}_{ij} \{\mathbf{v}_j\cdot \mathbf{e}^{\parallel}_j,\mathbf{v}_j\cdot \mathbf{e}^{\perp}_j\}^\top)^\sharp \nonumber \\
& =  {\textstyle{\sum_{{i,j}}}} (\mathsf{q}_i/{A_i}) \hat{B}_{ij} \{- \mathbf{v}_j\cdot \mathbf{e}^{\perp}_j, \mathbf{v}_j\cdot \mathbf{e}^{\parallel}_j\}^\top  \defeq \left\{\mathrm{cocurl}^c \,\mathsfbf{v}, \mathrm{curl}^c\,\mathsfbf{v}\right\}^\top\in \mathcal{C}\times\mathcal{P}, \\
\mathrm{curl}\,\mathsfbf{V}&=(\star \mathrm{d}\mathsfbf{V}^\flat)^\sharp 
=(\star_{1,2}^\rhd \mathsf{A}_1^{*\top}\mathsfbf{V}^\flat)^\sharp=(\star_{1,2}^\rhd\mathsf{A}_1^{*\top} {\textstyle{\sum_{{j}}}} (\mathsf{q}_j \mathbf{V}_j)^\flat )^\sharp  \nonumber\\
 =&(\star_{1,2}^\rhd \mathsf{A}_1^{*\top}{\textstyle{\sum_{{j}}}}  \mathsf{q}_j^* \{\mathbf{V}_j\cdot \mathbf{E}^{\parallel}_j,\mathbf{V}_j\cdot \mathbf{E}^{\perp}_j\}^\top)^\sharp 
= (\star_{1,2}^\rhd {\textstyle{\sum_{{j,k}}}}  \{\mathbf{V}_j\cdot \mathbf{E}^{\parallel}_j,\mathbf{V}_j\cdot \mathbf{E}^{\perp}_j\}^\top \hat{A}_{jk} \mathsf{q}_k^*)^\sharp \nonumber \\
 = & {\textstyle{\sum_{{j,k}}}} \{-\mathbf{V}_j\cdot \mathbf{E}^{\perp}_j, \mathbf{V}_j\cdot \mathbf{E}^{\parallel}_j \}^\top \hat{A}_{jk} (\mathsf{q}_k/{E_k}) 
 \defeq \left\{\mathrm{cocurl}^v \,\mathsfbf{V}, \mathrm{curl}^v\,\mathsfbf{V}\right\}^\top\in \hat{\mathcal{V}}\times\mathcal{P}.
\end{align}
\end{subequations}
It follows from the identity $A_i \mathsf{I}_2 =\sum_j \hat{B}_{ij}(-\boldsymbol{\epsilon}_i \mathbf{t}_j)\otimes \mathbf{c}_j$ \cite{jensen2020} that $\mathrm{curl}\,\mathsfbf{c}=\{2,0\}^\top$, when $\mathsfbf{c}$ are edge centroids. 
The operator curl integrates a vector field around a cell or a triangle, along the closed paths 
\begin{equation}
\label{eq:closedpaths}
{\textstyle{\sum_j}} \hat{B}_{ij}\mathbf{e}_j^\parallel=\mathbf{0}, \quad {\textstyle{\sum_j}} \hat{B}_{ij}\mathbf{e}_j^\perp=\mathbf{0}, \quad
{\textstyle{\sum_j}} \hat{A}_{jk}\mathbf{e}_j^\parallel=\mathbf{0}, \quad
{\textstyle{\sum_j}}\hat{A}_{jk}\mathbf{E}_j^\perp=\mathbf{0},    
\end{equation}
for all cells $i=1,\dots,N_c$ and for all internal triangles $k$.  The rotated operator cocurl can therefore be interpreted as a divergence.  
The magnitude of a curl field is given by
\begin{align}
\label{eq:magcurl}
    [ \mathrm{curl}\,\mathsfbf{v}, \mathrm{curl}\,\mathsfbf{v}]_{\mathcal{C}}&=
    \langle (\star \mathrm{d}\mathsfbf{v}^\flat)\wedge \star (\star \mathrm{d}\mathsfbf{v}^\flat) \vert\mathsf{1}_c\rangle=
    \langle (\star_{1,2} \mathsf{B}_1^* \mathsfbf{v}^\flat)\wedge (\mathsf{B}_1^*\mathsfbf{v}^\flat )\vert \mathsf{1}_c\rangle \nonumber \\
    & = \langle \mathsf{v}^\top [(\hat{\mathsf{B}}^\top \mathsf{H}^{-1}\hat{\mathsf{B}})\otimes\mathsf{I}_{\mathcal{P}}]\mathsf{v}\vert\mathsf{1}_c\rangle  \nonumber \\
    &={\textstyle{\sum_{j,j'}}}\left[v^\parallel_j\{\hat{\mathsf{B}}^\top \mathsf{H}^{-1}\hat{\mathsf{B}}\}_{j,j'}v^\parallel_{j'}+ v^\perp_j\{\hat{\mathsf{B}}^\top \mathsf{H}^{-1}\hat{\mathsf{B}}\}_{j,j'}v^\perp_{j'}\right].
\end{align}
Likewise, 
\begin{equation}
    [ \mathrm{curl}\,\mathsfbf{V}, \mathrm{curl}\,\mathsfbf{V}]_{\hat{\mathcal{V}}}={\textstyle{\sum_{j,j'}}}\left[V^\parallel_j\{\hat{\mathsf{A}} \hat{\mathsf{E}}^{-1}\hat{\mathsf{A}}^\top\}_{j,j'}V^\parallel_{j'}+ V^\perp_j\{\hat{\mathsf{A}} \hat{\mathsf{E}}^{-1}\hat{\mathsf{A}}^\top\}_{j,j'}V^\perp_{j'}\right].
\end{equation}   
Definitions (\ref{eq:grad}) and (\ref{eq:curl}) ensure that $\mathrm{curl}\,\circ\,\mathrm{grad}\,\phi=0$ and $\mathrm{curl}\,\circ\,\mathrm{grad}\,\Phi=0$, because $\mathrm{d}\,\circ\,\mathrm{d}=0$.

\subsection{Derived operators}
\label{app:derived}

We now define the operators that are adjoint to grad and curl under the inner products (\ref{eq:inner1}, \ref{eq:inner0}, \ref{eq:inner2}), satisfying (\ref{eq:adj}).

For $\mathbf{v}\in\Gamma(T\mathcal{M}_{\mathcal{E}})$ and $\mathbf{V}\in\Gamma(T\mathcal{M}_{\mathcal{L}})$, using the standard definition of $-\mathrm{div}$, 
\begin{subequations}
    \label{eq:div}
\begin{align}
    -\mathrm{div}\,\mathsfbf{v}&=\star\,\mathrm{d}\star \mathsfbf{v}^\flat=\star_{1,0}^{-1} \,\mathsf{A}_1^{*\top} \star_{1,1}  ({\textstyle{\sum}_j}\mathbf{v}_j \mathsf{q}_j)^\flat = \star_{1,0}^{-1} \,\mathsf{A}_1^{*\top}\star_{1,1} {\textstyle{\sum}_j}\mathsf{q}_j^* \{\mathbf{v}_j\cdot{\mathbf{e}_j^\parallel}, \mathbf{v}_j\cdot{\mathbf{e}_j^\perp}\}^\top\nonumber\\
    &
= \star_{1,0}^{-1} \,\mathsf{A}_1^{*\top} {\textstyle{\sum}_j} ({F_j}/{t_j^2}) \mathsf{q}_j^* \{-\mathbf{v}_j\cdot{\mathbf{e}_j^\perp,\mathbf{v}_j\cdot{\mathbf{e}_j^\parallel}}\}^\top \nonumber \\
& = \star_{1,0}^{-1}  {\textstyle{\sum}_{j,k}} \{-\mathbf{v}_j\cdot{\mathbf{e}_j^\perp}, \mathbf{v}_j\cdot{\mathbf{e}_j^\parallel}\}^\top ({F_j}/{t_j^2})  \hat{A}_{jk}  \mathsf{q}_k^* \nonumber \\
&=   {\textstyle{\sum}_{j,k}} \{\mathbf{v}_j\cdot{\mathbf{e}_j^\parallel},\mathbf{v}_j\cdot{\mathbf{e}_j^\perp}\}^\top ({F_j}/{t_j^2}) \hat{A}_{jk} \mathsf{q}_k^*/E_k \nonumber \\ &
\defeq\{-\mathrm{div}^v\,\mathsfbf{v},-\mathrm{codiv}^v\,\mathsfbf{v}\}^\top\in \Omega_1^0(\hat{\mathcal{N}}), \\
    -\mathrm{div}\,\mathsfbf{V}&=\star\,\mathrm{d}\star \mathsfbf{V}^\flat=(\star_{1,0}^\rhd)^{-1} \,\mathsf{B}_1^{*} \star_{1,1}^\rhd  ({\textstyle{\sum}_j}\mathbf{V}_j \mathsf{q}_j)^\flat 
  \nonumber \\ &  
 = (\star_{1,0}^\rhd)^{-1} \,\mathsf{B}_1^{*}\star_{1,1}^\rhd {\textstyle{\sum}_j}\mathsf{q}_j^* \{\mathbf{V}_j\cdot{\mathbf{E}_j^\parallel}, \mathbf{V}_j\cdot{\mathbf{E}_j^\perp}\}^\top\nonumber\\
    &
= (\star_{1,0}^\rhd)^{-1} \,\mathsf{B}_1^{*} {\textstyle{\sum}_j} ({F_j}/{T_j^2}) \mathsf{q}_j^* \{-\mathbf{V}_j\cdot{\mathbf{E}_j^\perp, \mathbf{V}_j\cdot{\mathbf{E}_j^\parallel}}\}^\top  \nonumber \\
& = (\star_{1,0}^\rhd)^{-1}  {\textstyle{\sum}_{i,j}} \mathsf{q}_i^* \hat{B}_{ij} ({F_j}/{T_j^2}) \{-\mathbf{V}_j\cdot{\mathbf{E}_j^\perp, \mathbf{V}_j\cdot{\mathbf{E}_j^\parallel}}\}^\top     \nonumber \\
&=   {\textstyle{\sum}_{i,j}}  (\mathsf{q}_i^*/A_i)\hat{B}_{ij}({F_j}/{T_j^2}) \{\mathbf{V}_j\cdot{\mathbf{E}_j^\parallel},\mathbf{V}_j\cdot{\mathbf{E}_j^\perp}\}^\top  \nonumber \\ & 
\defeq\{-\mathrm{div}^c\,\mathsfbf{V},-\mathrm{codiv}^c\,\mathsfbf{V}\}^\top\in \Omega^0_1(\hat{\mathcal{N}}^\rhd). 
\end{align}
\end{subequations}
The magnitude of a divergence field is given by
\begin{align}
\label{eq:magdiv}
    [ (-\mathrm{div}\,\mathsfbf{v})^\sharp, &(-\mathrm{div}\,\mathsfbf{v})^\sharp]_{\hat{\mathcal{V}}}=
    \langle (\star \mathrm{d}\star \mathsfbf{v}^\flat)\wedge \star (\star \mathrm{d}\star \mathsfbf{v}^\flat) \vert\mathsf{1}_v\rangle \nonumber \\
&        =
    \langle (\star_{1,0}^{-1} \mathsf{A}_1^{*\top}\star_{1,1} \mathsfbf{v}^\flat)\wedge (\mathsf{A}_1^{*\top}\star_{1,1} \mathsfbf{v}^\flat )\vert \mathsf{1}_v\rangle 
    \nonumber \\
&= \langle \mathsf{v}^\top [\hat{\mathsf{T}}_e^{-1}\hat{\mathsf{A}} \hat{\mathsf{E}}^{-1}\hat{\mathsf{A}}^\top \hat{\mathsf{T}}_e^{-1})\otimes\mathsf{I}_{\mathcal{P}}]\mathsf{v}\vert\mathsf{1}_v\rangle \nonumber \\
    &={\textstyle{\sum_{j,j'}}}\left[v^\parallel_j\{\hat{\mathsf{T}}_e^{-1}\hat{\mathsf{A}} \hat{\mathsf{E}}^{-1}\hat{\mathsf{A}}^\top \hat{\mathsf{T}}_e^{-1}\}_{j,j'}v^\parallel_{j'}+ v^\perp_j\{\hat{\mathsf{T}}_e^{-1}\hat{\mathsf{A}} \hat{\mathsf{E}}^{-1}\hat{\mathsf{A}}^\top \hat{\mathsf{T}}_e^{-1}    \}_{j,j'}v^\perp_{j'}\right].
\end{align}
Similarly, 
\begin{multline}   
[ (-\mathrm{div}\,\mathsfbf{V})^\sharp, (-\mathrm{div}\,\mathsfbf{V})^\sharp]_{\mathcal{C}} = \\ {\textstyle{\sum_{j,j'}}}\left[V^\parallel_j\{\hat{\mathsf{T}}_l^{-1}\hat{\mathsf{B}}^\top \mathsf{H}^{-1}\hat{\mathsf{B}} \hat{\mathsf{T}}_l^{-1}\}_{j,j'}V^\parallel_{j'}+ V^\perp_j\{\hat{\mathsf{T}}_l^{-1}\hat{\mathsf{B}}^\top \mathsf{H}^{-1}\hat{\mathsf{B}} \hat{\mathsf{T}}_l^{-1}    \}_{j,j'}V^\perp_{j'}\right].
\end{multline}

For $\mathsf{u}\in \mathcal{C}\times\mathcal{P}$ and $\mathsf{U}\in\hat{\mathcal{V}}\times\mathcal{P}$ the adjoint to curl is provided by
\begin{subequations}
    \label{eq:rot}
\begin{align}
    \mathrm{rot}\, \mathsf{u}&=(\star \mathrm{d} \mathsf{u}^\flat)^\sharp 
= (\star\,\mathrm{d} \{\mathsf{u}^{\parallel},\mathsf{u}^{\perp}\}^{\top\flat} )^\sharp =(\star_{1,1}^{-1}\,\mathsf{B}_1^{*\top} ({\textstyle{\sum_i} \{u_i^\parallel,u_i^\perp\}^\top \mathsf{q}_i})^\flat )^\sharp \nonumber \\ 
& = (\star_{1,1}^{-1}\,\mathsf{B}_1^{*\top} {\textstyle{\sum_i}} \{u_i^\parallel,u_i^\perp\}^\top \mathsf{q}_i^\star )^\sharp \nonumber \\
& = (\star_{1,1}^{-1}\, {\textstyle{\sum_{i,j}}}   \{u_i^\parallel  ,u_i^\perp\}^\top \hat{B}_{ij} \mathsf{q}_j^*)^\sharp
= ( {\textstyle{\sum_{i,j}}} \{u_i^\perp,-u_i^\parallel\}^\top \hat{B}_{ij} \mathsf{q}_j^* ({t_j^2}/{F_j})  )^\sharp \nonumber \\
& = {\textstyle{\sum_{i,j}}}  ({u_i^\perp}\hat{B}_{ij}  \mathbf{e}_{j\parallel} {-u_i^\parallel}\hat{B}_{ij}  \mathbf{e}_{j\perp})\mathsf{q}_j({t_j^2}/{F_j}) \defeq \mathrm{rot}^c\,{\mathsf{u}^\perp}+\mathrm{corot}^c\,{\mathsf{u}^\parallel} \in \Gamma(T\mathcal{M}_{\mathcal{E}}), \\
    \mathrm{rot}\, \mathsf{U}&=(\star \mathrm{d} \mathsf{U}^\flat)^\sharp 
= (\star\,\mathrm{d} \{U^{\parallel},U^{\perp}\}^{\top\flat} )^\sharp =((\star_{1,1}^\rhd)^{-1}\,\mathsf{A}_1^{*} ({\textstyle{\sum_k} \{U_k^\parallel,U_k^\perp\}^\top \mathsf{q}_k})^\flat )^\sharp \nonumber \\
& = ((\star_{1,1}^\rhd)^{-1}\,\mathsf{A}_1^{*} {\textstyle{\sum_k}} \{U_k^\parallel,U_k^\perp\}^\top \mathsf{q}_k^\star )^\sharp \nonumber \\
& = ((\star_{1,1}^\rhd)^{-1}\, {\textstyle{\sum_{i,j}}}  \mathsf{q}_j^* \hat{A}_{jk}\{U_k^\parallel  ,U_k^\perp\}^\top)^\sharp
= ( {\textstyle{\sum_{j,k}}}  \mathsf{q}_j^* ({T_j^2}/{F_j}) \hat{A}_{jk} \{U_k^\perp,-U_k^\parallel\}^\top )^\sharp \nonumber \\
& = {\textstyle{\sum_{j,k}}} \mathsf{q}_j({T_j^2}/{F_j}) (\mathbf{E}_{j\parallel}\hat{A}_{jk} {U_k^\perp} {-}\mathbf{E}_{j\perp}\hat{A}_{jk}{U_k^\parallel}  ) \defeq\mathrm{rot}^v\,{\mathsf{U}^\perp}+\mathrm{corot}^v\,{\mathsf{U}^\parallel} \in \Gamma(T\mathcal{M}_{\mathcal{L}}),
\end{align}
\end{subequations}
ensuring that $-\mathrm{div}\,\circ\mathrm{rot}\,\mathsf{U}=\mathsf{0}$ and $-\mathrm{div}\,\circ\mathrm{rot}\,\mathsf{u}=\mathsf{0}$.  The magnitudes of the rot fields are given by
\begin{subequations}
    \begin{align}
    [ \mathrm{rot}\,\mathsf{u}, \mathrm{rot}\,\mathsf{u}]_{\hat{\mathcal{E}}}&=
    \langle (\star \mathrm{d}\mathsf{u}^\flat)\wedge \star (\star \mathrm{d}\mathsf{u}^\flat) \vert\mathsf{1}_e\rangle=
    \langle (\star_{1,1}^{-1} \mathsf{B}_1^{*\top} \mathsf{u}^\flat)\wedge (\mathsf{B}_1^{*\top}\mathsf{u}^\flat )\vert \mathsf{1}_e\rangle \nonumber \\
    & = \langle \mathsf{u}^\top [(\hat{\mathsf{B}} \hat{\mathsf{T}}_e\hat{\mathsf{B}}^\top)\otimes\mathsf{I}_{\mathcal{P}}]\mathsf{u}\vert\mathsf{1}_e\rangle \nonumber \\
    &={\textstyle{\sum_{i,i'}}}\left[u^\parallel_i\{\hat{\mathsf{B}} \hat{\mathsf{T}}_e\hat{\mathsf{B}}^\top\}_{i,i'}u^\parallel_{i'}+ u^\perp_i\{\hat{\mathsf{B}} \hat{\mathsf{T}}_e\hat{\mathsf{B}}^\top\}_{i,i'}u^\perp_{i'}\right], \\
[ \mathrm{rot}\,\mathsf{U}, \mathrm{rot}\,\mathsf{U}]_{\hat{\mathcal{L}}}
    &={\textstyle{\sum_{k,k'}}}\left[U^\parallel_k\{\hat{\mathsf{A}}^\top \hat{\mathsf{T}}_l\hat{\mathsf{A}}\}_{k,k'}U^\parallel_{k'}+ U^\perp_k\{\hat{\mathsf{A}}^\top \hat{\mathsf{T}}_l\hat{\mathsf{A}}\}_{k,k'}U^\perp_{k'}\right].
\end{align}
\end{subequations}

To demonstrate that $-\mathrm{div}$ is adjoint to $\mathrm{grad}$, we write (\ref{eq:adj}a) as
\begin{subequations}
\label{eq:adjs}
    \begin{align}
    \langle \mathsfbf{v}^\flat \wedge (\star_{1,1} \mathsf{A}_1^* \phi) \vert \mathsf{1}_e \rangle 
    &
    ={\textstyle \sum_{j,k}} (F_j/t_j^2)\hat{A}_{jk}(v_j^\parallel\phi_k^\parallel +v_j^\perp\phi_k^\perp)\nonumber \\ & =\langle (\star_{1,0}^{-1} \mathsf{A}_1^{*\top} \star_{1,1}) \mathsfbf{v}^\flat \wedge \star_{1,0} \phi\vert \mathsf{1}_v \rangle, \\
        \langle \mathsfbf{V}^\flat \wedge (\star_{1,1}^\rhd \mathsf{B}_1^{*\top} \Phi) \vert \mathsf{1}_l \rangle 
    &
    ={\textstyle \sum_{i,j}} (F_j/T_j^2)\hat{B}_{ij}(V_j^\parallel\Phi_i^\parallel +V_j^\perp\Phi_i^\perp)\nonumber \\ & =\langle ((\star_{1,0}^\rhd)^{-1} \mathsf{B}_1^{*} \star_{1,1}^\rhd) \mathsfbf{V}^\flat \wedge \star_{1,0}^\rhd \Phi\vert \mathsf{1}_c \rangle. 
\end{align}
Likewise, to demonstrate that rot is adjoint to curl, we write (\ref{eq:adj}b) as
    \begin{align}
        \langle \mathsf{u}^\flat \wedge \mathsf{B}_1^* \mathsfbf{v}^\flat\vert \mathsf{1}_c \rangle
    &
    ={\textstyle \sum_{ij}\hat{B}_{ij}(u_i^\parallel v_j^\perp-u_i^\perp v_j^\parallel)}=\langle (\star_{1,1}^{-1}\mathsf{B}_1^{*\top} \mathsf{u}^\flat)\wedge \star_{1,1} \mathsfbf{v}^\flat \vert \mathsf{1}_e\rangle, \\
    \langle \mathsf{U}^\flat \wedge \mathsf{A}_1^{*\top} \mathsfbf{V}^\flat\vert \mathsf{1}_v \rangle 
    &
    ={\textstyle \sum_{jk}\hat{A}_{jk}(U_k^\parallel V_j^\perp-U_k^\perp V_j^\parallel)}=\langle ((\star_{1,1}^\rhd)^{-1}\mathsf{A}_1^* \mathsf{U}^\flat)\wedge \star_{1,1}^\rhd \mathsfbf{V}^\flat \vert \mathsf{1}_l\rangle.
\end{align}
\end{subequations}
The operators defined in this appendix are summarised in Table~\ref{tab:diffOpDef}.
\begin{table}[]
    \centering
    \setlength{\tabcolsep}{1mm}
    \begin{tabular}{|c|c|c|} 
        \hline 
        {Operator} & \cite{jensen2022} & {Definition}  \\ 
        \hline 
        $-\mathrm{cocurl}^c \,\mathsfbf{b}$ & $-{\mathrm{div}}^c$              & ${\textstyle \sum_{i,j}} \mathsf{q}_i \hat{B}_{ij} (\boldsymbol{\epsilon}_i \mathbf{t}_{j}) \cdot \mathbf{b}_j /A_i$  \\
        $\mathrm{-div}^c \,\mathsfbf{b}$    & $-\widetilde{\mathrm{div}}^c$   & ${\textstyle \sum_{i,j}} \mathsf{q}_i^* \hat{B}_{ij} ({F_j}/ {T_j^2}) \mathbf{T}_j\cdot \mathbf{b}_j/A_i $  \\
        $\mathrm{-div}^v\,\mathsfbf{b}$    & $-\widetilde{\mathrm{div}}^v$   & ${\textstyle \sum_{j,k}} \mathsf{q}_k^* \hat{A}_{jk} ({F_j}/{t_j^2}) \mathbf{t}_j \cdot \mathbf{b}_j /E_k$  \\
        $-\mathrm{cocurl}^v\,\mathsfbf{b}$ & $-{\mathrm{div}}^v$              & ${\textstyle \sum_{j,k}}\mathsf{q}_k  \hat{A}_{jk} (\boldsymbol{\epsilon}_k \mathbf{T}_{j}) \cdot \mathbf{b}_j /E_k $  \\ 
        $\mathrm{curl}^c\,\mathsfbf{b}$   & ${\mathrm{curl}}^c$         & ${\textstyle \sum_{i,j}} \mathsf{q}_i \hat{B}_{ij} \mathbf{t}_j \cdot \mathbf{b}_j /A_i $  \\
        $\mathrm{codiv}^c \,\mathsfbf{b}$  & $\widetilde{\mathrm{CURL}}^c$    & $-{\textstyle \sum_{i,j}}\mathsf{q}_i^* \hat{B}_{ij} ({F_j}/{T_j}^2) (\boldsymbol{\epsilon}_k\mathbf{T}_j) \cdot \mathbf{b}_j/A_i $  \\
        $\mathrm{codiv}^v\,\mathsfbf{b}$  & $\widetilde{\mathrm{curl}}^v$    & $-{\textstyle \sum_{j,k}}\mathsf{q}_k^* \hat{A}_{jk} ({F_j}/{t_j^2}) (\boldsymbol{\epsilon}_i \mathbf{t}_j)\cdot \mathbf{b}_j/E_k$  \\
        $\mathrm{curl}^v\,\mathsfbf{b}$   & ${\mathrm{CURL}}^v$             & ${\textstyle \sum_{j,k}}\mathsf{q}_k \hat{A}_{jk} \mathbf{T}_j \cdot \mathbf{b}_j / E_k $  \\ 
            \hline 
        $\mathrm{grad}^c \,\mathsf{f}$   & $\mathrm{grad}^c$                & ${\textstyle \sum_{i,j}}\mathsf{q}_j \hat{B}_{ij}({\mathbf{T}_j}/{T_j^2}) f_i$  \\
        $\mathrm{cograd}^c \,\mathsf{f}$ & $-\mathrm{CURL}^c$               & ${\textstyle \sum_{i,j}}\mathsf{q}_j  \boldsymbol{\epsilon}_k (\mathbf{T}_j /{T_j^2} )\hat{B}_{ij} f_i$  \\
        $\mathrm{grad}^v\,\phi$   & $\mathrm{grad}^v$                & ${\textstyle \sum_{j,k}}\mathsf{q}_j \hat{A}_{jk}({\mathbf{t}_j}/{t_j^2})\phi_k$  \\
        $\mathrm{cograd}^v\,\phi$    & $-\mathrm{curl}^v$            & ${\textstyle \sum_{j,k}}\mathsf{q}_j \boldsymbol{\epsilon}_i (\mathbf{t}_j/{t_j^2} )\hat{A}_{jk} \phi_k $  \\
        $\mathrm{rot}^c \,\mathsf{f}$    & $\widetilde{\mathrm{curl}}^c$   & ${\textstyle \sum_{i,j}}\mathsf{q}_j \hat{B}_{ij} (\mathbf{t}_j /{F_j})  f_i$  \\
        $-\mathrm{corot}^c \,\mathsf{f}$  & $\widetilde{\mathrm{grad}}^c$    & ${\textstyle \sum_{i,j}}\mathsf{q}_j \hat{B}_{ij} \boldsymbol{\epsilon}_i (\mathbf{t}_j / F_j) f_i$  \\
        $\mathrm{rot}^v\,\phi$    & $\widetilde{\mathrm{CURL}}^v$   & ${\textstyle \sum_{j,k}}\mathsf{q}_j \hat{A}_{jk} (\mathbf{T}_j / F_j) \phi_k $ \\ 
        $-\mathrm{corot}^v\,\phi$  & $\widetilde{\mathrm{grad}}^v$    & ${\textstyle \sum_{j,k}}\mathsf{q}_j \hat{A}_{jk} \boldsymbol{\epsilon}_k (\mathbf{T}_j /F_j)\phi_k $ \\ 
        \hline
    \end{tabular}
    \caption{Definitions of differential operators are given in terms of edge, link and spoke vectors; column 2 shows notation used in \cite{jensen2022}. }
    \label{tab:diffOpDef}
\end{table}

\section{The vertex model}
\label{app:vm}

We derive here a version of the vertex model that ultimately takes a standard form, but which incorporates osmotic effects.  We take the free energy of cell $i$ to be
\begin{multline}
\label{eq:cellenergy}
\mathcal{U}(A_i,L_i,N_i,M_i)=K_A A_0F\left({A_i}/{A_0}\right)+K_L L_0 F\left({L_i}/{L_0}\right)+K_N N_0 F\left({N_i}/{N_0}\right)\\ +K_M M_0 F\left({M_i}/{M_0}\right) +\mathsf{p}_i(A_i-A_0-aN_i)+\mathsf{t}_i(L_i-L_0-\ell M_i),
\end{multline}
where $F(\theta)$ is a convex function satisfying $F(1)=F'(1)=0$, $F''(1)=1$ and $K_A$, $K_L$, $K_N$, $K_M$ are positive constants.  $A_0$ and $L_0$ are a reference area and reference perimeter; taking $L_0\lesssim 3.72\sqrt{A}_0$ prevents the area and perimeter contributions to the energy from both achieving minimal values and contributes to mechanical rigidity.  A common choice for $F$ is $F(\theta)=\tfrac{1}{2}(\theta-1)^2$.  In addition to $A_i$ and $L_i$, we allow the energy to depend on two chemical species with molecular number $N_i$ (occupying the cell's apical face) and $M_i$ (occupying the apical cortex), with concentrations $n_i=N_i/A_i$ and $m_i=M_i/L_i$.  Intracellular gradients are neglected.  $\mathsf{p}_i$ and $\mathsf{t}_i$ in (\ref{eq:cellenergy}) are Lagrange multipliers enforcing the steric relationships $A_i=A_0+aN_i$ and $L_i=L_0+\ell M_i$, where $a$ and $\ell$ are a molecular area and length.  (Volumetric constraints of this kind are used in models of hydrogels, \hbox{e.g.} \cite{hennessy2020}.)  The parameter range of interest is one in which energetic and steric contributions balance, namely
\begin{equation}
\label{eq:enst}
    K_AA_0\sim K_LL_o\sim K_N N_0\sim K_M M_0, \quad A_0\sim aN_0\sim L_0^2 \sim  (\ell M_0)^2,
\end{equation}
where $\sim$ denotes `scales like.'  The first derivatives of $\mathcal{U}$ define a pressure, tension and chemical potentials
\begin{subequations}
\label{eq:firstderivs}
\begin{align}
P_i&=\mathsf{p}_i+K_A F'(A_i/A_0), & \mu_i&=K_NF'(N_i/N_0)-a\mathsf{p}_i, \\
T_i&=\mathsf{t}_i+K_L F'(L_i/L_0), & \nu_i& =K_MF'(M_i/M_0)-\ell \mathsf{t}_i.
\end{align}
\end{subequations}
In order to derive evolution equations for vertex locations $\mathsfbf{r}\equiv \sum_k \mathbf{r}_k\mathsf{q}_k$ and chemical numbers $\mathsf{N}\equiv \sum_i N_i\mathsf{q}_i^*$, $\mathsf{M}\equiv \sum_i M_i\mathsf{q}_i^*$, we first define some relevant operators.

We define $\mathcal{N}^\Diamond$ as the network $\mathcal{N}$ on a flat manifold $\mathcal{M}$ supplemented with links between edge centroids, as illustrated in Fig.~\ref{fig:Figure4ForceNetwork}(c).  Such links are defined by \cite{cowley2024} as
\begin{equation}
\label{eq:sik}
    \mathbf{s}_{ik}=\tfrac{1}{2}{\textstyle \sum_j} B_{ij}\mathbf{t}_j \vert{A}_{jk}\vert=-{\textstyle\sum_j} B_{ij} A_{jk}\mathbf{c}_j.
\end{equation}
They can be associated with cell area changes because, in the vertex model, the pressure exerted by cell $i$ on vertex $k$ acts along $-\boldsymbol{\epsilon}_i \mathbf{s}_{ik}$.  The connection is emphasised by taking a time derivative of ${\boldsymbol\epsilon}_iA_i=\sum_jB_{ij}\mathbf{t}_j \otimes \mathbf{c}_j$, giving
\begin{align}
    \boldsymbol{\epsilon}_i\dot{A}_i
    &={\textstyle \sum_j}B_{ij}(\dot{\mathbf{t}}_j \otimes \mathbf{c}_j+\mathbf{t}_j \otimes \dot{\mathbf{c}}_j)
    ={\textstyle \sum_{j,k}}B_{ij}(A_{jk}\dot{\mathbf{r}}_k \otimes \mathbf{c}_j+\tfrac{1}{2}\vert A_{jk}\vert \mathbf{t}_j \otimes \dot{\mathbf{r}}_k) \nonumber \\
    &={\textstyle \sum_i}(\mathbf{s}_{ik} \otimes \dot{\mathbf{r}}_k-\dot{\mathbf{r}}_k \otimes \mathbf{s}_{ik}).
\end{align}
Multiplication by $\boldsymbol{\epsilon}_i$ and taking the trace gives 
\begin{equation}
\label{eq:areachange}
    \dot{A}_i= -{\textstyle \sum_k}\boldsymbol{\epsilon}_i\mathbf{s}_{ik}\cdot \dot{\mathbf{r}}_k.
\end{equation}
The links define the area $D_k$ of triangles enclosing each vertex via the relationship \cite{cowley2024}
\begin{equation}
\boldsymbol{\epsilon}_i\mathbf{s}_{ik} \cdot \mathbf{s}_{i'k}=2D_k {\textstyle \sum_j A_{jk} B_{ij} \vert {B}_{i'j}\vert}.
\end{equation}
Complementing (\ref{eq:sik}), noting that tension acts along edges, we define 
\begin{equation}
\label{eq:uik}
    \mathbf{u}_{ik}={\textstyle{\sum_{j}}} \vert B_{ij}\vert A_{jk}\hat{\mathbf{t}}_j
\end{equation}
and complementing (\ref{eq:areachange}) we observe that 
\begin{equation}
\label{eq:perimeterchange}
  {\textstyle{\sum_{j,k}}} \mathbf{u}_{ik} \cdot \dot{\mathbf{r}}_k={\textstyle{\sum_{k}}} \vert B_{ij}\vert \hat{\mathbf{t}_j}\cdot \dot{\mathbf{t}}_j={\textstyle{\sum_{k}}} \vert B_{ij}\vert  \dot{t}_j=\dot{L}_i.
\end{equation}
It is evident from (\ref{eq:areachange}) and (\ref{eq:perimeterchange}) that
\begin{equation}
    \frac{\partial A_i}{\partial \mathbf{r}_k}=-\boldsymbol{\epsilon}_i\mathbf{s}_{ik}, \quad \frac{\partial L_i}{\partial \mathbf{r}_k}=\mathbf{u}_{ik}.
\end{equation}

In \cite{cowley2024}, we defined the operators (with minor notational changes)
\begin{subequations}
\label{eq:cwops}
\begin{align}
\mathrm{grad}_A\,&={\textstyle{\sum_{i,k}}}D_k^{-1}\boldsymbol{\epsilon}_i \mathbf{s}_{ik} \mathsf{q}_k\otimes\mathsf{q}_i^*, &
\mathrm{curl}_A\,&={\textstyle{\sum_{i,k}}}A_i^{-1} (\mathbf{s}_{ik} \cdot ) \mathsf{q}_i\otimes\mathsf{q}_k, \\
-\mathrm{div}_A\,&={\textstyle{\sum_{i,k}}}A_i^{-1}(\boldsymbol{\epsilon}_i \mathbf{s}_{ik}\cdot) \mathsf{q}_i^*\otimes\mathsf{q}_k, &
\mathrm{rot}_A\,&={\textstyle{\sum_{i,k}}}D_k^{-1} \mathbf{s}_{ik} \mathsf{q}_k\otimes\mathsf{q}_i, 
\end{align}
and Laplacians
\begin{align}
\mathsf{L}_A&={\textstyle\sum_{i,i',k}}A_i^{-1} \frac{\mathbf{s}_{ik}\cdot\mathbf{s}_{i'k}}{D_k} \mathsf{q}_i^*\otimes \mathsf{q}_{i'}^*, \\\mathsfbf{L}_A&={\textstyle\sum_{i,k,k'}}A_i^{-1} \left(\frac{\mathbf{s}_{ik}\otimes\mathbf{s}_{ik'}}{D_k}+\frac{(\boldsymbol{\epsilon}_i\mathbf{s}_{ik})\otimes(\boldsymbol{\epsilon}_i\mathbf{s}_{ik'})}{D_k}\right) \mathsf{q}_k\otimes \mathsf{q}_{k'},
\end{align}
\end{subequations}
and demonstrated that $\mathrm{curl}_A\,\circ\mathrm{grad}_A=\mathsf{0}$ and $-\mathrm{div}_A\,\circ\mathrm{rot}_A=\mathsf{0}$.  The operators in (\ref{eq:cwops}) are adjoint under inner products 
\begin{subequations}
\label{eq:cvmadj}
    \begin{align}
    [\mathsfbf{v},\mathrm{grad}_A\phi]_{\mathcal{V}}\equiv {\textstyle{\sum}_{i,k}\phi_i\boldsymbol{\epsilon}\mathbf{s}_{ik}\cdot \mathbf{v}_k}&=[-\mathrm{div}_A\mathsfbf{v},\phi]_{\mathcal{C}A},\\
    [\mathsfbf{v},\mathrm{rot}_A\phi]_{\mathcal{V}}\equiv {\textstyle{\sum}_{i,k}\phi_i \mathbf{s}_{ik}\cdot \mathbf{v}_k}&=[\mathrm{curl}_A\mathsfbf{v},\phi]_{\mathcal{C}A}.
\end{align}
\end{subequations}
Similarly, following \cite{cowley2024} we define
\begin{align}
\label{eq:cwop2}
    \mathrm{grad}_L=-{\textstyle\sum_{i,k}}D_k^{-1}\mathbf{u}_{ik}\mathsf{q}_k\otimes\mathsf{q}_i^*, \quad 
    -\mathrm{div}_L=-{\textstyle\sum_{i,k}}L_i^{-1}(\mathbf{u}_{ik}\cdot) \mathsf{q}_i^*\otimes\mathsf{q}_k,
\end{align}
satisfying $[\mathsfbf{v},\mathrm{grad}_L\phi]_{\mathcal{V}}=-\sum_{i,k}\mathbf{v}\cdot\mathbf{u}_{ik}=[-\mathrm{div}_L\mathsfbf{v},\phi]_{\mathcal{C}L}$.  (Inner products labelled $\mathcal{CA}$ and $\mathcal{CL}$ are sums over cells weighted by area and perimeter respectively.)  Eq.~(\ref{eq:areachange}) and (\ref{eq:perimeterchange}) imply that 
\begin{align}
\label{eq:areachange2}
\dot{A}_i&=A_i\{\mathrm{div}_A\,\dot{\mathsfbf{r}}\}_i, &
\dot{L}_i&=L_i\{\mathrm{div}_L\,\dot{\mathsfbf{r}}\}_i.
\end{align}

With (\ref{eq:areachange2}) at our disposal, we can write mass conservation equations for the chemicals as 
\begin{subequations}
\label{eq:mcon}
\begin{align}
\dot{n}_i+ {n}_i\,\{\mathrm{div}_A\,\dot{\mathsfbf{r}}\}_i&=\dot{n}_i+(\dot{A}_i/{A}_i){n}_i \equiv \dot{N_i}/A_i= -\{\mathrm{div}_A\,\mathsfbf{J}_{A}\}_i, \\
\dot{m}_i+ {m}_i\,\{\mathrm{div}_L\,\dot{\mathsfbf{r}}\}_i&=\dot{m}_i+(\dot{L}_i/{L}_i){m}_i \equiv \dot{M_i}/L_i= -\{\mathrm{div}_L\,\mathsfbf{J}_{L}\}_i,
\end{align}
\end{subequations}
for some fluxes $\mathsfbf{J}_A=\sum_k n_k \mathbf{u}_k\mathsf{q}_k^*$ and $\mathsfbf{J}_L=\sum_k m_k \mathbf{v}_k\mathsf{q}_k^*$.  The time derivative in (\ref{eq:mcon}) is Lagrangian (for fixed $i$).  Here we follow \cite{doi2011} in introducing fields $\mathsfbf{u}\in\Gamma(T\mathcal{M}_{\mathcal{V}})$ and $\mathsfbf{v}\in\Gamma(T\mathcal{M}_{\mathcal{V}})$ that transport chemicals between cells; $n_k$ and $m_k$ are concentrations projected onto vertices.  

Treating $\mathsfbf{r}(t)$, $\mathsf{N}(t)$ and $\mathsf{M}(t)$ as independent variables, and using (\ref{eq:mcon}), changes of the total energy $U=\sum_i \mathcal{U}(A_i,L_i,N_i,M_i)$ satisfy, using (\ref{eq:firstderivs}) and the chain rule,
\begin{align}
    \dot{U}&={\textstyle \sum_i}(P_i \dot{A}_i+T_i\dot{L}_i+\mu_i\dot{N}_i+\nu_i \dot{M}_i) \nonumber \\
    & ={\textstyle \sum_i}\left(P_i \dot{A}_i+T_i\dot{L}_i-\mu_iA_i \{\mathrm{div}_A\, \mathsfbf{J}_A \}_i -\nu_i L_i \{\mathrm{div}_L\, \mathsfbf{J}_L\}_i\right)\nonumber \\
    &={\textstyle \sum_{i,k}}\left[P_i \frac{\partial {A}_i}{\partial \mathbf{r}_k} +T_i\frac{\partial L_i}{\partial \mathbf{r}_k}\right]\cdot \dot {\mathbf{r}}_k-[\boldsymbol{\mu}, \mathrm{div}_A\,\mathsfbf{J}_A]_{\mathcal{C}A} -[\boldsymbol{\nu}, \mathrm{div}_L\,\mathsfbf{J}_L]_{\mathcal{C}L}\nonumber \\
    &=-\left[\dot{\mathsfbf{r}}, \mathrm{grad}_A \mathsf{P}\right]_{\mathcal{V}} -\left[\dot{\mathsfbf{r}}, \mathrm{grad}_L \mathsf{T}\right]_{\mathcal{V}}
     +[\mathsfbf{J}_A,\mathrm{grad}_A\,\boldsymbol{\mu}]_\mathcal{V} + [\mathsfbf{J}_L,\mathrm{grad}_L\,\boldsymbol{\nu}]_\mathcal{V}, 
\end{align}
using (\ref{eq:cwops}) and (\ref{eq:cwop2}) and imposing no-flux conditions at the monolayer periphery.  We define a dissipation rate $\Phi= \eta [ \dot{\mathsfbf{r}},\dot{\mathsfbf{r}}]_{\mathcal{V}} + \xi [\mathsfbf{u}, \mathsfbf{u}]_{\mathcal{V}}+ \omega [\mathsfbf{v}, \mathsfbf{v}]_{\mathcal{V}}$ where $\xi>0$, $\eta>0$ and $\omega>0$ are weightings applied at vertices, and construct a Rayleighan $\mathcal{R}=\tfrac{1}{2}\Phi+\dot{U}$.  Following \cite{doi2011}, we enforce $\partial\mathcal{R}/\partial \dot{\mathsfbf{r}}=0$, $\partial \mathcal{R}/\partial \mathsfbf{u}=0$ and $\partial \mathcal{R}/\partial \mathsfbf{v}=0$ to give
\begin{subequations}
\label{eq:intvm}
\begin{align}
\eta\dot{\mathsfbf{r}}&=- \mathrm{grad}_A \mathsf{P} - \mathrm{grad}_L \mathsf{T}, \\
\xi\mathbf{u}_k&=-n_k \,\{\mathrm{grad}_A \,\boldsymbol{\mu}\}_k, \\
\omega\mathbf{v}_k&=-m_k \,\{\mathrm{grad}_L \,\boldsymbol{\nu}\}_k,
\end{align}
\end{subequations}
ensuring that $\dot{U}=-\eta [\dot{\mathsfbf{r}},\dot{\mathsfbf{r}}]_\mathcal{V} -\xi [{\mathsfbf{u}},{\mathsfbf{u}}]_\mathcal{V}
-\omega [{\mathsfbf{v}},{\mathsfbf{v}}]_\mathcal{V}\leq 0$.  We expect $n_k$ and $m_k$ to be averages of neighbouring cells, so that 
\begin{equation}
    n_k={\textstyle\sum_i}C_{ik}n_i/({\textstyle\sum_i}C_{ik}), \quad
    m_k={\textstyle\sum_i}C_{ik}m_i/({\textstyle\sum_i}C_{ik})
\end{equation}
where $C_{ik}=\tfrac{1}{2}\sum_j \vert A_{jk} \vert ~\vert B_{ij}\vert$ is the face-vertex adjacency matrix.  Hence, taking $F$ to be quadratic, we recover the coupled system
\begin{subequations}
\label{eq:fullvm}
    \begin{align}
\eta \dot{\mathsfbf{r}}&=- \mathrm{grad}_A [\mathsf{p} +K_A (\mathsf{A}/A_0-\mathsf{1}_c)] -\mathrm{grad}_L [\mathsf{t}+ K_L(\mathsf{L}/L_0-\mathsf{1}_c)] , \\ 
\dot{N}_i&=A_i \left\{\mathrm{div}_A\,\left( {\textstyle\sum_k}
 \frac{n_k^2}{\xi}\left[\frac{K_N}{N_0} \mathrm{grad}_A\, \mathsf{N} - a\,\mathrm{grad}_A\,\mathsf{p}\right]_k \mathsf{q}_k^* \right)\right\}_i, \\
 \dot{M}_i&=L_i \left\{\mathrm{div}_L\,\left({\textstyle\sum_k}
 \frac{m_k^2}{\omega}\left[\frac{K_M}{M_0} \mathrm{grad}_L\, \mathsf{M} - \ell \,\mathrm{grad}_L\,\mathsf{t}\right]_k \mathsf{q}_k^* \right)\right\}_i,  \\
 A_i&=A_0+aN_i, \qquad L_i=L_0+\ell M_i,
\end{align}
\end{subequations}
for $i=1,\dots,N_c$.  The (osmotic) pressure $\mathsf{p}$ and tension $\mathsf{t}$ couple mechanical and chemical processes.  The diffusion of species $\mathsf{N}$ and $\mathsf{M}$ in (\ref{eq:fullvm}b,c) are regulated by distinct Laplacian operators.  

To nondimensionalise (\ref{eq:fullvm}) we set 
\begin{align*}
a&=A_0\tilde{A}, & 
L&=\sqrt{A_0}\tilde{L}, & 
\mathsf{p}&=K_A\tilde{\mathsf{p}}, & \mathsf{t}&=K_L\tilde{\mathsf{t}}, \\
\mathsfbf{r}(t)&=\sqrt{A_0}\tilde{\mathsfbf{r}}(\tilde{t}), &
t&=(\eta A_0/K_A)\tilde{t}, \\
\mathsf{N}(t)&=N_0\tilde{\mathsf{N}}(\tilde{t}), &
\mathsf{M}(t)&=M_0\tilde{\mathsf{M}}(\tilde{t}), &
\mathsf{n}&=(N_0/A_0)\,\tilde{\mathsf{n}}, &
\mathsf{m}&=(M_0/\sqrt{A_0}) \,\tilde{\mathsf{m}}, &
\end{align*}
and define parameters
\begin{align*}
    \tilde{L}_0&=L_0/\sqrt{A_0},& 
    \tilde{\Gamma}&=K_L /(K_A \sqrt{A_0}),& 
    \theta_N&=N_0 K_N/(A_0 K_A),\\
    \theta_{M}&=M_0 K_M/(A_0 K_A),&
    \tilde{a}&=a N_0/A_0,&
    \tilde{\ell}&=\ell M_0/\sqrt{A_0},\\
    \tilde{\xi}&=\xi/\eta,&
    \tilde{\omega}&=\omega/\eta,
\end{align*}
giving
\begin{subequations}
\label{eq:fullvm1}
    \begin{align}
\dot{\tilde{\mathsfbf{r}}}&=- \mathrm{grad}_A [\tilde{\mathsf{p}} +(\tilde{\mathsf{A}}-\mathsf{1}_c)] -\tilde{\Gamma}\,\mathrm{grad}_L [\tilde{\mathsf{t}}+ (\tilde{\mathsf{L}}/\tilde{L}_0-\mathsf{1}_c)] , \\ 
\dot{\tilde{N}}_i&=\tilde{A}_i \left\{\mathrm{div}_A\,\left({\textstyle\sum_k}
 \frac{\tilde{n}_k^2}{\tilde{\xi}}\left[ \theta_N \,\mathrm{grad}_A\, \tilde{\mathsf{N}} - \tilde{a}\,\mathrm{grad}_A\,\tilde{\mathsf{p}}\right]_k \mathsf{q}_k^*\right)\right\}_i, \\
 \dot{\tilde{M}}_i&=\tilde{L}_i \left\{\mathrm{div}_L\,\left({\textstyle\sum_k}
 \frac{\tilde{m}_k^2}{\tilde{\omega}}\left[ \theta_M \,\mathrm{grad}_L\, \tilde{\mathsf{M}} - \tilde{\ell} \,\mathrm{grad}_L\,\tilde{\mathsf{t}}\right]_k \mathsf{q}_k^* \right)\right\}_i,  \\
 \tilde{\mathsf{A}}&=\mathsf{1}_c+\tilde{a} \tilde{\mathsf{N}},\qquad \tilde{\mathsf{L}}=\tilde{L}_0\mathsf{1}_c+\tilde{\ell} \tilde{\mathsf{M}}.
\end{align}
\end{subequations}

We now suppose that the molecular mobilities are large ($\xi\ll \eta$, $\omega\ll \eta$), \hbox{i.e.} the dissipation is dominated by movement of vertices.  In this limit, concentrations equilibrate faster than the cells change shape and $\theta_N \tilde{\mathsf{N}}=\tilde{a}\tilde{\mathsf{p}}$, $\theta_M\tilde{\mathsf{M}}=\tilde{\ell} \tilde{\mathsf{t}}$.  It follows that
\begin{equation}
\dot{\tilde{\mathsfbf{r}}}=- \left(1+\frac{\theta_N}{\tilde{a}^2}\right)\mathrm{grad}_A (\tilde{\mathsf{A}}-\mathsf{1}_c) -\left(\frac{\theta_M}{\tilde{\ell}^2}+\frac{\tilde{\Gamma}}{\tilde{L}_0}\right)\,\mathrm{grad}_L  (\tilde{\mathsf{L}}-\tilde{L}_0\mathsf{1}_c). 
\end{equation}
Defining 
\begin{equation}
\label{eq:gamma}
    \Gamma=\left(\frac{K_L}{K_A L_0}+\frac{K_M}{K_A \ell^2 M_0}\right)\bigg/\left(1+\frac{K_N A_0}{K_A a^2 N_0}\right)
\end{equation}
and rescaling time on $(1+\theta_N/\tilde{a}^2)^{-1}$, we recover the standard implementation of the vertex model (\ref{eq:vmvan}) parametrized by $\tilde{L_0}$ and $\Gamma$.  $\Gamma$ is of order unity when the balances (\ref{eq:enst}) hold.  The parameters $K_M/(K_A\ell^2 M_0)$ and $K_NA_0/(K_A a^2 N_0)$ capture stiffening of the cortex and apical face by packing of chemicals $M$ and $N$ in the respective domains.

The model outlined above assigns the same quadratic free energy $F$ to the four contributing variables $A_i$, $L_i$, $N_i$, $M_i$ in (\ref{eq:cellenergy}).  A variant of this model might use $F(\theta)=\theta(\log\theta -1)$ for one or more of the chemical components, reflecting the entropic origins of steric repulsion; passing from (\ref{eq:mcon}, \ref{eq:intvm}) to (\ref{eq:fullvm}), such a choice of $F$ would then introduce additional nonlinearities, not considered here.  The model could also be generalized to relax the hard steric constraints imposed by the Lagrange multipliers $\mathsf{p}_i$ and $\mathsf{t}_i$ in (\ref{eq:cellenergy}).  We have also assumed that the species $M_i$ and $N_i$, that contribute stiffness to the cell via crowding effects, also diffuse directly between cells; the model could be adapted to incorporate different diffusible signalling molecules that regulate the abundance of $M_i$ and $N_i$.

\section{Validation}
\label{app:valid}

Table~\ref{tab:integrals} shows that the solvability condition (\ref{eq:solvd}) is violated, which we attribute to non-orthogonality.  To address this, we subtract $[\{\mathsf{1}_c,\mathsf{1}_c\}^\top,(-\mathrm{div}\,\mathsfbf{V})^\sharp]_\mathcal{C}$ from the left-hand-side of (\ref{eq:poid}a), enforcing (\ref{eq:solvd}) and enabling Moore--Penrose pseudoinversion.  Now $\mathsf{B}^\top\mathsf{1}_c$ defines the chain identifying peripheral edges, and is therefore non-zero.  Replacing $\hat{\mathsf{B}}$ with $\mathsf{B}$ in $\mathsf{L}_{\mathcal{C}}$ in (\ref{eq:lvctf}a) defines a non-singular Laplacian $\mathsf{L}_{\mathcal{C}}^{\mathrm{D}}$ (that implicitly imposes Dirichlet rather than Neumann conditions).  We then invert $\mathsf{L}_{\mathcal{C}}^{\mathrm{D}}\psi =\mathsf{1}_c$ numerically and add $[\{\mathsf{1}_c,\mathsf{1}_c\}^\top,(-\mathrm{div}\,\mathsfbf{V})^\sharp]_\mathcal{C} \psi$ to the solution of the pseudoinversion.  This yields a Helmholtz--Hodge representation of $\mathsfbf{h}$ 
in all but the peripheral cells, where $\mathsf{L}$ and $\mathsf{L}^{\mathrm{D}}_{\mathcal{C}}$ differ.  The operators $\mathsf{L}_{\mathcal{V}}$ and $\mathcal{L}_{\mathcal{T}}$ do not require a solvability condition and the solvability condition $\mathsf{L}_{\mathcal{F}}$ is satisfied (Table~\ref{tab:integrals}).

We evaluated Laplacians of computed potentials and recovered the derivatives of $\mathsfbf{h}$ to within machine precision, except in peripheral cells when evaluating $\mathrm{div}^c\,\mathsfbf{h}$ and $\mathrm{codiv}^c\,\mathsfbf{h}$; this imperfection is a consequence of the non-zero integrals in Table~\ref{tab:integrals}.

\section{Stress induced by the harmonic field}
\label{app:harmonicstress}

\begin{figure}
    \centering
    \includegraphics[width=0.95\textwidth]{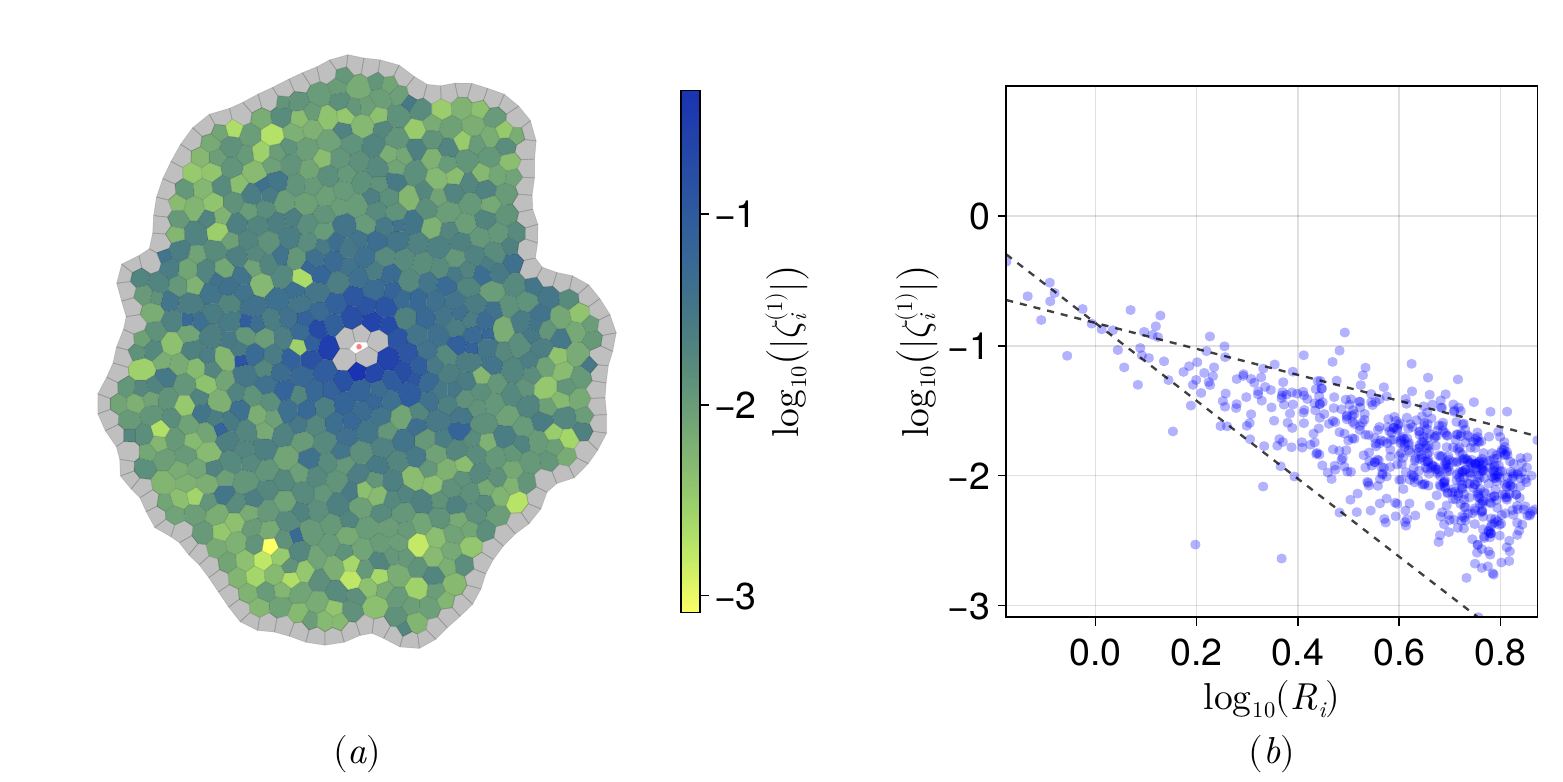}
    \caption{
    (a) Cell shear-stress magnitude $\vert \zeta^{(1)}_i\vert$ induced over internal cells $i^i$ by the harmonic field (\ref{eq:shearstress}).  This shear stress is undefined over peripheral cells (shown in black).  (b) Scatterplot of $\vert \zeta^{(1)}_i\vert$ versus $R_i=\vert\mathbf{R}_i\vert$ ($i=1,\dots,N_c$), the distance of the centre of cell $i$ from the centre of the ablated cell, shown with a red dot in (a). Dashed lines have gradients $-1$ and $-3$.  
    }
    \label{fig:FigureG1EdgeLaplacianShearStressCellRemoved}
\end{figure}

Recalling (\ref{eq:cellstress}), we can evaluate the stress field in cell $i$ associated with the harmonic contribution $\mathsfbf{x}^{(m)}$ to $\mathbf{h}_j$ in (\ref{eq:edgex}) as
\begin{equation}
\boldsymbol{\sigma}_i^{(m)}=
z^{\perp(m)} A_i^{-1} {\textstyle{\sum}_j}B_{ij} w_j^{(m)}\mathbf{e}_j^\parallel \otimes \mathbf{e}_{j\parallel}
\end{equation}
for some $z^{\perp(m)}$.  We have set $z^{\parallel(m)}=0$ as this would contribute an asymmetric couple stress in cells, which is not observed in simulations. $\sum_j B_{ij} w_j^{(m)}\hat{\mathbf{t}}_j\otimes \hat{\mathbf{t}}_j$
is symmetric and traceless, because, from (\ref{eq:zerodivcurl}), $\mathsf{B}\mathsf{w}^{(m)}=\mathsf{0}$.  Therefore the harmonic field does not contribute to the isotropic component of the stress, but instead contributes a shear stress of magnitude
\begin{equation}
    \zeta_i^{(m)}=z^{\perp(m)} \sqrt{-\mathrm{det} \left( A_i^{-1}{\textstyle \sum_j } B_{ij} w_j^{(m)} \hat{\mathbf{t}}_j\otimes \hat{\mathbf{t}}_j\right)},
    \label{eq:shearstress}
\end{equation}
for some $z^{\perp(m)}$.  Each ablation induces a global field that decays with distance from the ablation.  Fig.~\ref{fig:FigureG1EdgeLaplacianShearStressCellRemoved} shows the field $\zeta^{(1)}$ for the monolayer shown in Fig.~\ref{fig:Figure6harmonicFieldCellRemoved}, with $z^{\perp(m)}=1$. 

\end{appendices}

\bibliography{abbreviatedNew}
\end{document}